\documentclass[10pt, journal, onecolumn, draftcls]{IEEEtran}

\usepackage{cite}
\usepackage{amsmath,amssymb,amsfonts}
\usepackage{graphicx}
\usepackage{textcomp}

\usepackage{acronym}

\usepackage{amsmath}
\usepackage{amssymb}
\usepackage{amsfonts}
\usepackage[cmintegrals]{newtxmath}

\usepackage{hhline}
\usepackage{color}
\usepackage{epsfig}
\usepackage{epstopdf}
\usepackage{psfrag}
\usepackage[caption=false,font=footnotesize]{subfig}
\usepackage[table]{xcolor}

\acrodef{$P_{EM}$}{probability of emulation, or false alarm}
\acrodef{$P_{FA}$}{probability of false alarm}
\acrodef{$P_{MD}$}{probability of missed detection}
\acrodef{$P_{D}$}{probability of detection}

\acrodef{ACF}{autocorrelation function}
\acrodef{ACG}{automatic	gain control}
\acrodef{ACI}{adjacent channel interference}
\acrodef{ACK}{acknowledge}
\acrodef{AcR}{autocorrelation receiver}
\acrodef{ADC}{analog-to-digital converter}
\acrodef{AF}{amplify \& forward}
\acrodef{AFL}{anchor-free localization}
\acrodef{AGNSS}{assisted-GNSS}
\acrodef{AGPS}{assisted GPS}
\acrodef{AI}{automatic identification}
\acrodef{AIC}{Akaike information criterion}
\acrodef{AOA}{angle-of-arrival}
\acrodef{AOD}{angle-of-departure}
\acrodef{AOT}{approximate optimum threshold}
\acrodef{AP}{access point}
\acrodef{API}{application programming interface}
\acrodef{ASK}{amplitude shift keying}
\acrodef{ASNR}{accumulated signal-to-noise ratio}
\acrodef{AUB}{asymptotic union bound}
\acrodef{AWGN}{additive white Gaussian noise}
\acrodef{BAN}{body area network}
\acrodef{BAV}{balanced antipodal Vivaldi}
\acrodef{BCH}{Bose Chaudhuri Hocquenghem}
\acrodef{BEP}{bit error probability}
\acrodef{BER}{bit error rate}
\acrodef{BF}{brute force}
\acrodef{BFC}{block fading channel}
\acrodef{BIC}{Bayesian information criterion}
\acrodef{BLUE}{best linear unbiased estimator}
\acrodef{BPAM}{binary pulse amplitude modulation}
\acrodef{BPF}{bandpass filter}
\acrodef{BPPM}{binary pulse position modulation}
\acrodef{bps}{bits per second}
\acrodef{BPSK}{binary phase shift keying}
\acrodef{BPZF}{band-pass zonal filter}
\acrodef{BS}{base station}
\acrodef{BSC}{binary symmetric channel}
\acrodef{BTB}{Bellini-Tartara bound}
\acrodef{c.c.d.f.}{complementary cumulative distribution function}
\acrodef{c.d.f.}{cumulative distribution function}
\acrodef{CAD}{computer-aided design}
\acrodef{CAIC}{consistent Akaike information criterion}
\acrodef{CAP}{continuous aperture phased}
\acrodef{CCF}{cross correlation function}
\acrodef{CCI}{co-channel interference}
\acrodef{CD}{cooperative diversity}
\acrodef{CDMA}{code division multiple access}
\acrodef{CEOT}{channel ensemble optimum threshold}
\acrodef{CEP}{codeword error probability}
\acrodef{CFAR}{constant	 false alarm rate}
\acrodef{ch.f.}{characteristic function}
\acrodef{CH}{cluster head}
\acrodef{CIR}{channel impulse response}
\acrodef{CL}{centroid localization}
\acrodef{CM}{channel model}
\acrodef{CNR}{clutter-to-noise ratio}
\acrodef{CP}{ciclic prefix}
\acrodef{CPR}{channel pulse response}
\acrodef{CR}{channel response}
\acrodef{CRB}{Cram\'{e}r-Rao bound}
\acrodef{CRC}{cyclic redundancy check}
\acrodef{CRLB}{Cram\'{e}r-Rao lower bound}
\acrodef{CS}{clock skew}
\acrodef{CSCG}{circularly symmetric complex Gaussian}
\acrodef{CSI}{channel state information}
\acrodef{CSMA}{carrier sense multiple access}
\acrodef{CSS}{chirp spread spectrum}
\acrodef{CTS}{clear-to-send}
\acrodef{CW}{continuous wave}
\acrodef{DAA}{detect and avoid}
\acrodef{DAB}{digital audio broadcasting}
\acrodef{DBB}{digital base band}
\acrodef{DBPSK}{differential binary phase shift keying}
\acrodef{DCM}{dual-carrier modulation}
\acrodef{DDP}{detected direct path}
\acrodef{DF}{detect \& forward}
\acrodef{DFMS}{monopole dual feed stripline antenna}
\acrodef{DGPS}{differential GPS}
\acrodef{DLL}{delay-locked loop}
\acrodef{DoD}{Department of Defense}
\acrodef{DoF}{degrees of freedom}
\acrodef{DP}{direct path}
\acrodef{DR}{detection rate}
\acrodef{DRT}{distance ratio test}
\acrodef{DS-SS}{direct-sequence spread-spectrum}
\acrodef{DS}{delay spread}
\acrodef{DTR}{differential transmitted-reference}
\acrodef{DVB-H}{digital video broadcasting\,--\,handheld}
\acrodef{DVB-T}{digital video broadcasting\,--\,terrestrial}
\acrodef{e.m.}{electromagnetic}
\acrodef{ECC}{European Community Commission}
\acrodef{ED}{energy detector}
\acrodef{EDR}{energy detector receiver}
\acrodef{EFIM}{equivalent Fisher information matrix}
\acrodef{EIRP}{effective radiated isotropic power}
\acrodef{EKF}{extended Kalman filter}
\acrodef{ELP}{equivalent low-pass}
\acrodef{EM}{electromagnetic}
\acrodef{EMCB}{extended Miller Chang bound}
\acrodef{EME}{minimum eigenvalue ratio detector}
\acrodef{ENP}{estimated noise power}
\acrodef{ESA}{European Space Agency}
\acrodef{EU}{European Union}
\acrodef{FAR}{false alarm rate}
\acrodef{FCC}{Federal Communications Commission}
\acrodef{FDMA}{frequency division multiple access}
\acrodef{FDMA}{frequency division multiple access}
\acrodef{FEC}{forward error correction}
\acrodef{FEC}{forward error correction}
\acrodef{FFD}{full function device}
\acrodef{FFR}{full function reader}
\acrodef{FF}{far-field}
\acrodef{FFT}{fast Fourier transform}
\acrodef{FG}{factor graph}
\acrodef{FH-SS}{frequency-hopping spread-spectrum}
\acrodef{FH}{frequency-hopping}
\acrodef{FIM}{Fisher information matrix}
\acrodef{FLL}{Frequency-locked loop}
\acrodef{FS}{frame synchronization}
\acrodef{GA}{Gaussian approximation}
\acrodef{GD}{gradient descent}
\acrodef{GDOP}{geometric dilution of precision}
\acrodef{GLR}{generalized likelihood ratio}
\acrodef{GLRT}{generalized likelihood ratio test}
\acrodef{GML}{generalized maximum likelihood}
\acrodef{GPRS}{general packet radio service}
\acrodef{GPS}{global positioning system}
\acrodef{HAP}{high altitude platform}
\acrodef{HCRB}{hybrid Cram\'{e}r-Rao bound}
\acrodef{HDSA}{high-definition situation-aware}
\acrodef{Hi-RADIAL}{High-accuracy RAdio Detection, Identification, And Localization}
\acrodef{HMM}{hidden Markov model}
\acrodef{HPA}{high-power amplifier}
\acrodef{HPBW}{half power beam width}
\acrodef{HW}{hardware}
\acrodef{i.i.d.}{independent, identically distributed}
\acrodef{ICT}{information and communication technologies}
\acrodef{IE}{informative element}
\acrodef{IEEE}{Institute of Electrical and Electronics Engineers}
\acrodef{IF}{intermediate frequency}
\acrodef{IFFT}{inverse fast Fourier transform}
\acrodef{IMF}{ideal matched filter}
\acrodef{IMU}{inertial measurement unit}
\acrodef{INR}{interference-to-noise ratio}
\acrodef{INS}{inertial navigation system}
\acrodef{IoT}{Internet of things}
\acrodef{IIoT}{industrial Internet of things}
\acrodef{INS}{inertial navigation system}
\acrodef{IR-UWB}{impulse radio UWB}
\acrodef{IR}{impulse radio}
\acrodef{IRI}{inter-reader interference}
\acrodef{IRS}{intelligent reflecting surface} 
\acrodef{ISI}{inter-symbol interference} 
\acrodef{isi}{intra-symbol interference} 
\acrodef{ISM}{industrial, scientific and medical}
\acrodef{ISNR}{interference-plus-signal-to-noise-ratio}
\acrodef{IT}{interference temperature}
\acrodef{ITC}{information theoretic criteria}
\acrodef{JBSF}{jump back and search forward}
\acrodef{JF}{just forward}
\acrodef{KF}{Kalman filter}
\acrodef{LDC}{low duty cycle}
\acrodef{LDPC}{low density parity check}
\acrodef{LEO}{localization error outage}
\acrodef{LIS}{large intelligent surface}
\acrodef{LLR}{log-likelihood ratio}
\acrodef{LLRT}{log-likelihood ratio test}
\acrodef{LRT}{likelihood ratio test}
\acrodef{LNA}{low-noise amplifier}
\acrodef{LOS}{line-of-sight}
\acrodef{LRT}{likelihood ratio test}
\acrodef{LS}{least square}
\acrodef{LS}{least squares}
\acrodef{M-PSK}{$M$-ary phase shift keying}
\acrodef{M-QAM}{$M$-ary quadrature amplitude modulation}
\acrodef{m.g.f.}{moment generating function}
\acrodef{MAC}{medium access control}
\acrodef{MAE}{mean absolute error}
\acrodef{MAI}{multiple access interference}
\acrodef{MAN}{metropolitan area network}
\acrodef{MAP}{maximum a posteriori}
\acrodef{MB-OFDM}{multi-band OFDM}
\acrodef{MB-UWB}{multi-band UWB}
\acrodef{MB}{multi-band}
\acrodef{MC}{multi-carrier}
\acrodef{MCB}{Miller Chang bound}
\acrodef{MCRB}{modified Cram\'{e}r-Rao bound}
\acrodef{MDD}{minimum distance distribution}
\acrodef{MDL}{minimum description length}
\acrodef{MF}{matched filter}
\acrodef{MGF}{moment generating function}
\acrodef{MI}{mutual information}
\acrodef{MIMO}{multiple-input multiple-output}
\acrodef{MISO}{multiple-input single-output}
\acrodef{ML}{maximum likelihood}
\acrodef{MM}{min-max}
\acrodef{MME}{maximum-minimum eigenvalue ratio detector}
\acrodef{MMSE}{minimum mean-square error}
\acrodef{MPC}{multipath component}
\acrodef{MRC}{maximal ratio combiner}
\acrodef{MS}{mobile station}
\acrodef{MSB}{most significant bit}
\acrodef{MSE}{mean square error}
\acrodef{MSE}{mean squared error}
\acrodef{MSK}{minimum shift keying}
\acrodef{MUI}{multi-user interference}
\acrodef{MUR}{multistatic radar}
\acrodef{MVU}{minimum variance unbiased}
\acrodef{MZZB}{modified Ziv-Zakai bound}
\acrodef{NB}{narrowband}
\acrodef{NBI}{narrowband interference}
\acrodef{NEO}{navigation error outage}
\acrodef{NFER}{near-Þeld electromagnetic ranging}
\acrodef{NF}{near-field}
\acrodef{NFF}{near-field focused}
\acrodef{NL}{nonlinear}
\acrodef{NLOS}{non-line-of-sight}
\acrodef{NP}{Neyman-Pearson}
\acrodef{NTIA}{National Telecommunications and Information Administration}
\acrodef{NTP}{network time protocol}
\acrodef{OAM}{orbital angular momentum} 
\acrodef{OC}{optimum combining}
\acrodef{OFDM}{orthogonal frequency division multiplexing}
\acrodef{OOK}{on-off keying}
\acrodef{OP}{outage probability}
\acrodef{OT}{optimum threshold}
\acrodef{P-Max}{$P$-Max}  
\acrodef{p.d.f.}{probability density function}
\acrodef{p.m.f.}{probability mass function}
\acrodef{PA}{power amplifier}
\acrodef{PAM}{pulse amplitude modulation}
\acrodef{PAN}{personal area network}
\acrodef{PAR}{peak-to-average ratio}
\acrodef{PD}{probability of detection}
\acrodef{PDP}{power delay profile}
\acrodef{PE}{probability of emulation}
\acrodef{PEB}{position error bound}
\acrodef{PEP}{packet error probability}
\acrodef{PF}{particle filter}
\acrodef{PFA}{probability of false alarm}
\acrodef{PHY}{physical layer}
\acrodef{PL}{path-loss}
\acrodef{PLL}{phase-locked loop}
\acrodef{PMD}{probability of missed detection}
\acrodef{PN}{pseudo-noise}
\acrodef{ppm}{part-per-million}
\acrodef{PPM}{pulse position modulation}
\acrodef{PR}{pseudo-random}
\acrodef{PRake}{partial rake}
\acrodef{PRF}{pulse repetition frequency}
\acrodef{PRP}{pulse repetition period}
\acrodef{PSD}{power spectral density}
\acrodef{PSEP}{pairwise synchronization error probability}
\acrodef{PSK}{phase shift keying}
\acrodef{PSWF}{prolate spheroidal wave function}
\acrodef{PU}{primary user}
\acrodef{QAM}{quadrature amplitude modulation}
\acrodef{QoS}{quality of service}
\acrodef{QPSK}{quadrature phase shift keying}
\acrodef{R.V.}{random variable}
\acrodef{RADAR}{radar}
\acrodef{RCS}{radar cross section}
\acrodef{RDL}{"random data limit"}
\acrodef{REM}{radio environment map}
\acrodef{REO}{ranging error outage}
\acrodef{RF}{radio-frequency}
\acrodef{RFID}{radio-frequency identification}
\acrodef{RFR}{reduced function reader}
\acrodef{RFT}{reduced function tag}
\acrodef{RII}{ranging information intensity}
\acrodef{RIS}{reconfigurable intelligent surface}
\acrodef{rms}{root mean square}
\acrodef{RMSE}{root-mean-square error}
\acrodef{ROC}{receiver operating characteristic}
\acrodef{RRC}{root raised cosine}
\acrodef{RSN}{radar sensor network}
\acrodef{RSS}{received signal strength}
\acrodef{RSSI}{received signal strength indicator}
\acrodef{RTLS}{real time locating systems}
\acrodef{RTT}{round-trip time}
\acrodef{S-V}{Saleh-Valenzuela}
\acrodef{SA}{simulated annealing}
\acrodef{SaG}{stop-and-go}
\acrodef{SBS}{serial backward search}
\acrodef{SBSMC}{serial backward search for multiple clusters}
\acrodef{SCM}{supply chain management}
\acrodef{SCR}{signal-to-clutter ratio}
\acrodef{SEP}{symbol error probability}
\acrodef{SIS}{small intelligent surface}
\acrodef{SFD}{start frame delimiter}
\acrodef{SIMO}{single-input multiple-output}
\acrodef{SINR}{signal-to-interference plus noise ratio}
\acrodef{SIR}{signal-to-interference ratio}
\acrodef{SISO}{single-input single-output}
\acrodef{SNR}{signal-to-noise ratio}
\acrodef{SoC}{system on chip}
\acrodef{SoO}{signal of opportunity}
\acrodef{SoP}{system on package}
\acrodef{SOT}{sub-optimum threshold}
\acrodef{SPAWN}{sum-product algorithm over a wireless network}
\acrodef{SPEB}{squared position error bound}
\acrodef{SPMF}{single-path matched filter}
\acrodef{SQNR}{signal-to-quantization-noise ratio}
\acrodef{SS}{spread spectrum}
\acrodef{ST}{simple thresholding}
\acrodef{SU}{secondary user}
\acrodef{SVD}{singular value decomposition}
\acrodef{SW}{software}
\acrodef{SW}{sync word}
\acrodef{TDE}{time delay estimation}
\acrodef{TDL}{tapped delay line}
\acrodef{TDMA}{time division multiple access}
\acrodef{TDOA}{time difference-of-arrival}
\acrodef{TH}{time-hopping}
\acrodef{TNR}{threshold-to-noise ratio}
\acrodef{TOA}{Time-of-arrival}
\acrodef{TOF}{time-of-flight}
\acrodef{TPC}{transmit power control}
\acrodef{TR}{transmitted-reference}
\acrodef{TS}{tabu search}
\acrodef{UAV}{unmanned aerial vehicle}
\acrodef{UB}{union bound}
\acrodef{UDP}{undetected direct path}
\acrodef{UHF}{ultra-high frequency}
\acrodef{ULP}{user location protocol}
\acrodef{UMP}{uniformly most powerful}
\acrodef{UMPI}{uniformly most powerful invariant}
\acrodef{UT}{user terminal}
\acrodef{UTC}{coordinated universal time}
\acrodef{UTM}{universal transverse Mercator}
\acrodef{UTRA}{UMTS terrestrial radio access}
\acrodef{UAV}{unmanned aerial vehicle}
\acrodef{UUV}{unmanned underwater vehicle}
\acrodef{UWB}{ultrawide-band}
\acrodef{UWBcap}[UWB]{Ultrawide band}
\acrodef{VFIL}{virtual force iterative localization}
\acrodef{VGA}{variable-gain amplifier}
\acrodef{VNA}{vector network analyzer}
\acrodef{WAF}{wall attenuation factor}
\acrodef{WB}{wideband}
\acrodef{WBI}{wideband interference}
\acrodef{WCL}{weighted centroid localization}
\acrodef{WED}{wall extra delay}
\acrodef{WiMAX} {worldwide interoperability for microwave access}
\acrodef{WLAN}{wireless local area network}
\acrodef{WLS}{weighted least squares}
\acrodef{WMAN}{wireless metropolitan area network}
\acrodef{WPAN}{wireless personal area networks}
\acrodef{WRAPI}{wireless research application programming interface}
\acrodef{WSN}{wireless sensor network}
\acrodef{WSR}{wireless sensor radar}
\acrodef{WSS}{wide-sense stationary}
\acrodef{WWB}{Weiss-Weinstein bound}
\acrodef{WWLB}{Weiss-Weinstein lower bound}
\acrodef{ZZB}{Ziv-Zakai bound}
\acrodef{ZZLB}{Ziv-Zakai lower bound}

 \DeclareMathOperator{\arccot}{arccot} 

\newcommand{\degree}{\ensuremath{^\circ}}

\newcommand{\rect}[1] {\text{rect} \left ({#1} \right )}
\newcommand{\sinc}[1] {\text{sinc} \left ({#1} \right )}

\newcommand{\lr}{L_R}
\newcommand{\lt}{L_T}

\newcommand{\rr}{{\bf r}}
\newcommand{\rt}{{\bf s}}

\newcommand{\boldr} {{\bf r}}

\newcommand{\yc}{y_{\mathrm{c}}}

\begin{document}
\title{Communication Modes with Large Intelligent Surfaces in the Near Field}

\author{Nicol\'o Decarli,~\IEEEmembership{Member,~IEEE}, Davide~Dardari,~\IEEEmembership{Senior~Member,~IEEE}


\thanks{
   N. Decarli is with the National Research Council - Institute of Electronics, Computer and Telecommunication Engineering (CNR-IEIIT), and WiLab-CNIT, Bologna (BO), Italy (e-mail:  {nicolo.decarli}@ieiit.cnr.it).
 D.~Dardari is with the 
   Dipartimento di Ingegneria dell'Energia Elettrica e dell'Informazione ``Guglielmo Marconi"  (DEI), and WiLab-CNIT, 
   University of Bologna, Cesena Campus, 
   Cesena (FC), Italy, (e-mail: davide.dardari@unibo.it). 
   \IEEEcompsocthanksitem 
    }

\thanks{
This work was sponsored, in part, by Theory Lab, Central Research Institute, 2012 Labs, Huawei Technologies Co., Ltd.
}

}

\maketitle

\begin{abstract}
This paper proposes a practical method for the definition of communication modes when antennas operate in the near-field region, by realizing ad-hoc beams exploiting the focusing capability of large antennas. The beamspace modeling proposed to define the communication modes is then exploited to derive expressions for their number (i.e., the degrees of freedom) in a generic setup, beyond the traditional paraxial scenario, together with closed-form definitions for the basis set at the transmitting and receiving antennas for several cases of interest, such as for the communication between a large antenna and a small antenna. Numerical results show that quasi-optimal communication can be obtained starting from focusing functions. This translates into the possibility of a significant enhancement of the channel capacity even in line-of-sight channel condition, without the need of implementing optimal but complex phase/amplitude profiles on transmitting/receiving antennas as well as resorting to intensive numerical solutions. Traditional results valid under paraxial approximation are revised in light of the proposed modeling, showing that similar conclusions can be obtained from different perspectives.
\end{abstract}


\maketitle



\section{Introduction}

\IEEEPARstart{T}{he} even-increasing demand for high-speed wireless communication is requiring a shift towards high frequency (e.g., millimeter-wave and terahertz) where large bandwidth is available \cite{San:19,DeLEtAl:J21}. In this context, the introduction of new technologies such as intelligent surfaces made of metamaterials has been proposed to realize passive \acp{IRS} and active \acp{LIS} used as highly-flexible antennas \cite{HuRusEdf:18,Tre:15,GonMinChaMac:2017,HolKueGorOHaBooSmi:12,Fink:14,WuZhaZheYouZha:20,DiRenzoJSAC:20}. %
In fact, metamaterials enable the manipulation of the \ac{EM} field or the local control of amplitude and phase reflecting behavior at an unprecedented level, thus enabling the design of specific characteristics in terms of radiated \ac{EM} field \cite{HuRusEdf:18}. 

The use of millimeter-wave and terahertz technologies jointly with large antennas poses new challenges and opportunities at the same time, since traditional models based on the assumption of far-field \ac{EM} propagation fail \cite{DarDec:J20,BjoSan:20}. 
Recently, \cite{Dar:J20} discussed how multiple \emph{communication modes} can be obtained using LISs. Communication modes correspond to orthogonal channels between a couple of antennas, thus capable of enhancing significantly the channel capacity thanks to parallel data multiplexing. They can be realized by designing ad-hoc orthogonal current distributions at the transmitting antenna side, and by correlating the \ac{EM} field at the receiving antenna side with other properly-designed orthogonal functions \cite{Mil:J19}. Despite the concept of communication modes is well known especially in optics and in guided \ac{EM} propagation, its exploitation for radio communication received only a little attention so far. This is mainly because, for the time being, radio communication systems have been realized with relatively small antennas and using low frequencies; in such a case, only one communication mode is generally available, if considering a single polarization, corresponding to a plane wave traveling from the transmitting to the receiving antenna. Therefore, the \ac{LOS} link capacity is  limited, and multi-path propagation is usually exploited by \ac{MIMO} antenna arrays to enhance the number of parallel channels (modes), namely the \ac{DoF} of communication. Unfortunately, due to the trend of adopting even increasing frequencies, multi-path tends to become sparse and \ac{LOS} propagation (e.g., using pencil-like beams) becomes predominant, thus limiting the possibility of exploiting multi-path propagation for increasing the \ac{DoF}. For this reason, there is a great interest in exploiting new methods for enhancing the communication \ac{DoF}, even in \ac{LOS}, especially considering the characteristics of propagation in the near field arising from the use of large antennas and high frequency bands. Driven by these motivations, this paper focuses the discussion on multi-mode communications between intelligent antennas, namely \acp{LIS}, considering a \ac{LOS} propagation channel among them and without accounting for additional reflecting elements such as \acp{IRS}.\footnote{Notice that, in the case of reflectarrays, \ac{LIS}-based antennas can be realized as reflecting surfaces.}

\subsection{State of the Art}

When the paraxial approximation holds (i.e., antennas are parallel, oriented towards their maximum of the radiation intensity, namely the boresight direction, and small with respect to the link distance), analytical results for the number of communication modes are known, and the corresponding exciting functions at transmitting and receiving antenna side can be written in closed form \cite{Mil:J00}. In particular, the number of communication modes and the relative coupling intensity are related to the geometry of the scenario (dimension and shape of the antennas, link distance) and to the operating frequency \cite{Mil:J00}. Unfortunately, as discussed in \cite{Dar:J20}, the results known for the number of communication modes (i.e., the \ac{DoF}) cannot be adopted when the link distance becomes comparable to the antenna size, condition that can happen with the use of \acp{LIS} (or very large antenna arrays) and \acp{IRS}. In such a case, ad-hoc models must be considered also for describing the path loss between antennas, since traditional methods (e.g., the well-known Friis formula) do not capture properly the propagation phenomenon in the near-field region \cite{BjoSan:20, Dar:J20}. Moreover, if the assumptions of the paraxial approximation are not fulfilled, no closed-form solutions for the definition of communication modes are known, at the best of authors' knowledge.
In the general case, the derivation of the communication modes can be realized by resorting to numerical solutions, which can be not practical for antennas of large size, especially for run-time operations, and do not allow to get insights about the effects of the different system parameters. Thus, new and easy methods for the definition of the communication modes must be investigated.

In the last years, practical methods for obtaining multiple communication channels also in \ac{LOS} have been proposed. For example, \cite{BraBehSay:J13} introduces the \ac{CAP}-\ac{MIMO} technique capable of generating several orthogonal beams in the angular domain using aperture antennas with high flexibility in shaping the \ac{EM} wavefront. If the operating frequency is high, for example in the millimeter-wave band or even in the terahertz band, very narrow beams can be realized \cite{HeaEtAl:J18}, thus a single aperture antenna can capture some of them, then allowing multi-mode communication. Unfortunately, the number of communication modes is limited when antennas are small compared to the link distance (e.g., in the far field). Other studies showed the potential of realizing multiple channels when antenna arrays are located at close distance each other (short-range \ac{MIMO}) or when very large arrays are used (extra-large MIMO), even in \ac{LOS} thus without the need of exploiting multi-path propagation \cite{JiaIng:J05,BohOrtOie:J09,NisTomHir:J11}. However, a classical point-wise definition of antenna arrays consisting of a finite number of antenna elements, equi-spaced at $\lambda/2$ to make mutual coupling negligible, has been considered. This translates into a sampling of the continuous-space \ac{EM} channel and continuous signals (propagating waves) according to a specific placement of the array's elements \cite{HanFu:06}, which cannot capture the ultimate limits offered by the wireless channel and exploitable by metamaterials, as recently envisioned under the holographic communication concept \cite{DarDec:J20}.

Other practical methods proposed in the last years are based on the generation of \ac{OAM} beams, but they assume frequently large distance among the antennas and paraxial conditions \cite{CheEtAl:J20,HuEtAl:J19,TriParZghOoiAlo:J19,MurEtAl:J17}. Moreover, it has been shown that the maximum number of communication modes is limited also using these techniques\cite{Mil:J19}. Despite the method for realizing multiple communication modes, the traditional results valid under paraxial approximation cannot be adopted for more general configurations, and the number of communication modes related to traditional expressions is, frequently, overestimated \cite{Dar:J20}.

\subsection{Contributions}

In this paper, starting from the analytical formulation leading to the optimum strategy for establishing the communication modes between a couple of antennas, we define a  practical method for approximating them. 

The proposed method  exploits the multi-focusing capability of large antennas \cite{Han:J85,NepBuf:J17,LiuWu:J19} in the near-field region. In particular, focusing will be proposed to create quasi-orthogonal beams on the receiving antenna. It will be shown that, using focusing in the near field, the number of communication modes is approximatively equal to that using optimal beams obtained from the exact numerical solution of the problem. Moreover, it is shown that the number of communication modes corresponds to the maximum number of diffraction-limited focused-beams that can be realized on a screen (i.e., a receiving antenna) by using a certain aperture (i.e., a transmitting antenna) as commonly assumed in optics under paraxial approximation and at large distance \cite{Mil:J19}. 
Closed-form solutions will be provided for the case of communication between a \ac{LIS} and a smaller antenna, that is, a \ac{SIS}, which is of great interest as practical case of communication between a fixed large antenna used at base station and a mobile user with, for example, a handheld terminal.

The main contributions of the paper can be summarized as follows:
\begin{itemize}
\item{Proposal of a method based on diffraction theory and use of focusing capability of large antennas for generating multiple quasi-orthogonal beams, corresponding to practical communication modes, in the near-field region.}
\item{Derivation of closed-form approximate solutions for the definition of communication modes (basis set) in the case of communication between a \ac{LIS} and a \ac{SIS}, considering both uplink and downlink scenarios.}
\item{Presentation of novel closed-form expressions for the number of communication modes capable of providing the correct number for any configuration (not only under paraxial approximation) and even in the near-field region.}
\item{Review of traditional results valid under paraxial approximation, showing that they can be obtained as particular cases of the proposed method.}
\item{Discussion of the conditions allowing for the exploitation of  multiple communication modes, in relation to the regions of space around a transmitting/receiving antenna.}
\end{itemize}

The remainder of the paper is organized as follows. Sec.~\ref{sec:scenarioproblem} describes the scenario we consider and briefly reviews the theoretical foundations behind the communication through multiple communication modes. Then, Sec.~\ref{sec:basisdef} introduces the approximate method we propose, capable of defining multiple communication modes in the near-field region; in the same section, particular cases of interest are discussed and easy closed-form expressions for the available number of communication modes are provided. Sec.~\ref{sec:zones} discusses the conditions for exploiting multiple communication modes. Numerical results are presented in Sec.~\ref{Sec:results}, and Sec.~\ref{sec:conclusion} concludes the paper.

\bigskip

\section{Scenario and Problem Formulation}\label{sec:scenarioproblem}

\subsection{Scenario Considered}
\label{sec:scenario}

\begin{figure}[t!]
\centering
\includegraphics[width=0.5\columnwidth,keepaspectratio=true]{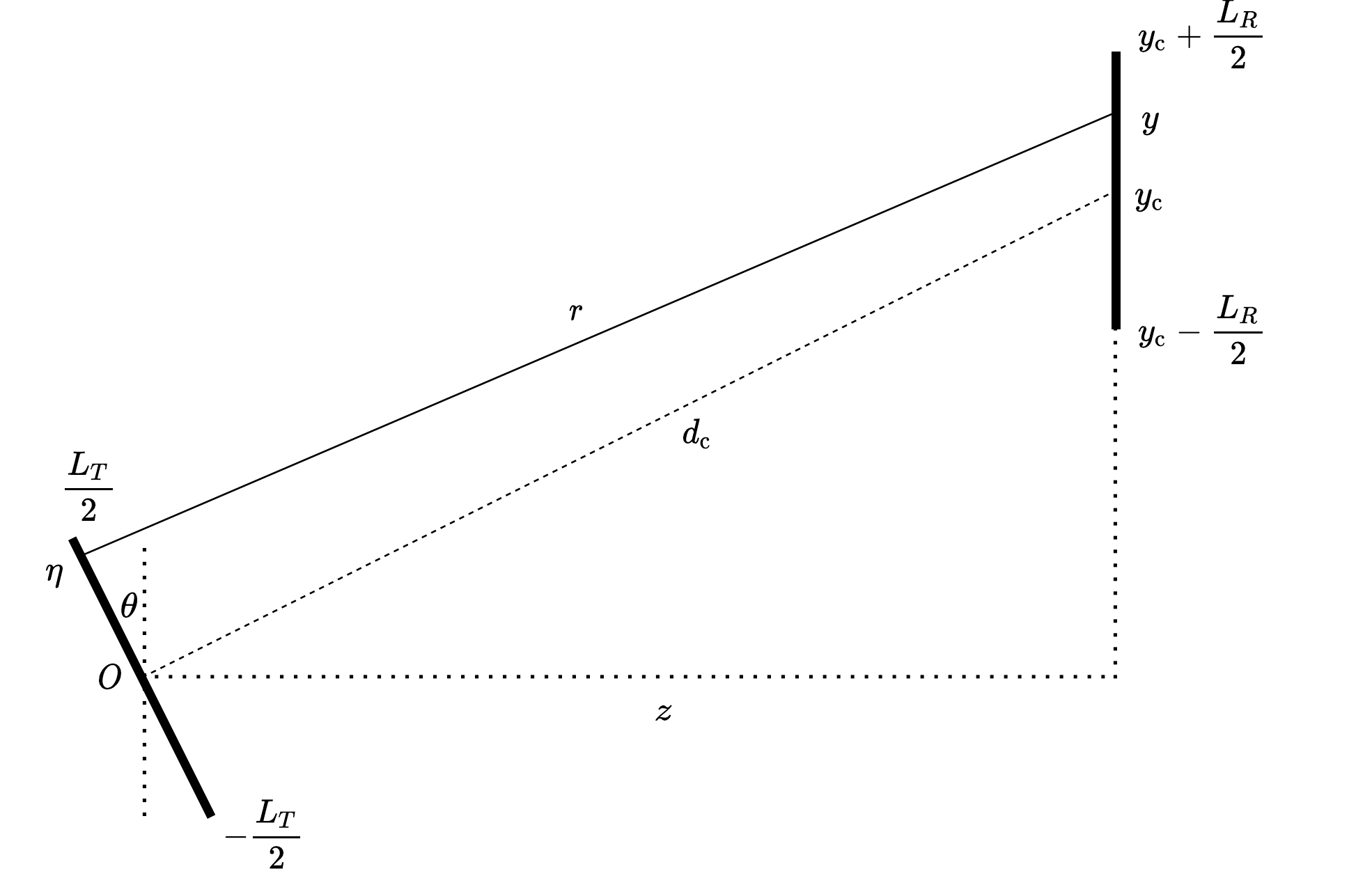}
\caption{The considered scenario.}
\label{fig:scenario}
\end{figure}

We consider two linear antennas (i.e., segments in space), according to Fig.~\ref{fig:scenario}. The receiving antenna has a length $\lr$ and is oriented along the $y$ axis, with center in $y=\yc$. The transmitting antenna has a length $\lt$ and is rotated of an angle $\theta$ with respect to the $y$ axis, considering a positive angle for counterclockwise rotation. The origin of the reference system is in the center of the transmitting antenna, with the $z$ axis oriented in the horizontal direction and the $y$ axis oriented in the vertical direction. The distance between the centers of the two antennas is $d_{\mathrm{c}}$, and the horizontal projection of such a distance on the $z$ axis is $z$. For further convenience, we consider $\eta$ the coordinate along the transmitting antenna, with upward positive direction when $\theta=0$, and $y$ the vertical coordinate along the receiving antenna.
The two antennas are assumed lying on the same plane (2D analysis), and any relative orientation and displacement among them are considered with $y_\mathrm{c}$ and $\theta$. This assumption can serve as a benchmark for more complicated scenarios and to understand the key concepts behind the realization of multiple communication modes that will be discussed. Notice that, recently, practical schemes exploiting linear antennas, namely \textit{radio stripes}, have been proposed and have several interesting practical implications \cite{ShaBjoLar:C20}. Moreover, the linear antenna model we are considering can serve as a benchmark for practical implementations related to classical uniform linear arrays.

We denote with $\rt$ in $\mathcal{S}_T$ and $\rr$ in $\mathcal{S}_R$ a couple of vectors pointing from the origin $O$ towards a point on the transmitting and receiving antennas, being $\mathcal{S}_T$ and $\mathcal{S}_R$, the sets of points belonging to the transmitting and receiving antennas, respectively. 
The distance between two points on the transmitting and receiving antennas is indicated with $r$, so that 
\begin{equation}\label{eq:distance}
r=||\rr-\rt||=\sqrt{(z+\eta\sin\theta)^2+(y-\eta\cos\theta)^2} \, 
\end{equation}
where $||\cdot||$ is the Euclidean norm.
The explicit dependence of $r$ on $z, y, \eta, \theta$ is omitted to simplify the notation.

In the following, we consider the antennas as ideal apertures in free space, where there is the complete freedom of drawing any current distribution on their surface (holographic capability). In fact, an analysis based on dense discrete antenna elements would require a careful characterization of the mutual coupling among the elements which is, in most of cases, feasible only through \ac{EM} simulations, and it would be sticked to a specific technology. Instead, considering  the antenna as a continuum composed of an infinite number of infinitesimal antennas, each producing an outgoing spherical wave with a given initial amplitude and phase, allows investigating the ultimate limits of the wireless communication without accounting for a specific technology or implementation-related aspects.  
Notice that such an assumption corresponds to supposing the antenna as an ideal \ac{EM} source obeying the Huygens-Fresnel principle, where each infinitesimal antenna is a Huygens' source.
Although the \ac{EM} field is a vector quantity, here it is treated, as first approximation, as a complex-valued scalar quantity to simplify the discussion. This is a common approach, frequently adopted in optics, which can be consider a reasonable approximation until all the elements of the discussion (dimension of antennas, distance...) are relatively large with respect to the wavelength \cite{God:B05}. Of course, by considering the field as a vector, additional communication modes could be obtained by exploiting, for example, polarization diversity \cite{Mil:J19,PooBroTse:05,HanFu:06}.

\subsection{Communication Modes: General Formulation}\label{eq:general}

Consider a monochromatic source at frequency $f_0$, that is, a function $\phi(\rt)$. This function can describe the current density distribution (amplitude and phase) on the transmitting antenna \cite{Dar:J20}. At the receiving antenna side, the source function produces a wave (electric field), that is a function $\psi(\rr)$, which can be obtained, in free-space conditions, as a solution of the inhomogeneous Helmholtz equation, and it is given by\footnote{We neglect the reactive near field, whose effect vanishes for distance larger than a few wavelengths from the transmitting antenna.} \cite{Mil:J00}
\begin{equation}\label{eq:distribution}
\psi(\rr)=\int_{\mathcal{S}_T}G(\rr,\rt)\, \phi(\rt)\,d\rt
\end{equation}
where $G(\boldr_1,\boldr_2)$ denotes the \emph{Green function} between points represented by vectors $\boldr_1$ and $\boldr_2$, which is
\begin{equation}\label{eq:greenf}
G(\boldr_1,\boldr_2)=\frac{\exp{(-\jmath \kappa ||\boldr_1-\boldr_2||})}{4\pi||\boldr_1-\boldr_2||}
\end{equation}
with $\kappa=2\pi/\lambda$ indicating the wavenumber, $\lambda=c/f_0$ the wavelength, $c$ standing for the speed of light, and $\jmath=\sqrt{-1}$. According to \eqref{eq:distribution}, at receiver side we can see the effect of the transmitting antenna (source) as a sum of infinitesimally small contributions (Huygens' sources) producing outgoing spherical waves centered on $\rt$ over $\mathcal{S}_T$ of initial amplitude/phase given by $\phi(\rt)$.

By expanding $\phi(\rr)$ and $\psi(\rr)$ using orthonormal basis sets  $\phi_n(\rt)$ and $\psi_m(\rr)$ complete in $\mathcal{S}_T$ and $\mathcal{S}_R$, respectively,  we have
\begin{align}
\phi(\rt)=\sum_n a_n \, \phi_n(\rt) \\
\psi(\rr)=\sum_m b_m \, \psi_m(\rr)
\end{align}
where $a_n$, $b_m$ are the series expansion coefficients, and the orthonormal conditions ensure that
\begin{align}
\int_{\mathcal{S}_T} \phi_m(\rt) \, \phi_n^*(\rt)\,d\rt=\delta_{mn} \label{eq:orthoTX}\\
\int_{\mathcal{S}_R} \psi_m(\rr) \, \psi_n^*(\rr)\,d\rr=\delta_{mn}  \label{eq:orthoRX}
\end{align}
with $\delta_{mn}$ indicating the Kronecker delta function,
so that it is possible to write
\begin{align}
b_m = \sum_n \xi_{mn} a_n
\end{align}
or, in matrix notation,\footnote{In the sense of Hilbert spaces, thus also considering vectors and matrices of infinite dimension \cite{Mil:J19}.}
\begin{align}\label{eq:matrixcomm}
\bold{B} = \bold{\Gamma} \bold{A}
\end{align}
where $\bold{\Gamma}=\{ \xi_{mn}\}$ is the communication operator between transmitting and receiving antennas, $\bold{B}=\{ b_m\}$ and $\bold{A}=\{ a_n\}$ and
\begin{align}\label{eq:commoperator}
\xi_{mn} = \int_{\mathcal{S}_R} \int_{\mathcal{S}_T} \psi_m^*(\rr) \, G(\rr,\rt) \, \phi_n(\rt) \, d\rt d\rr \, 
\end{align}
is the \textit{coupling intensity}. 
For a given geometry (locations in space of transmitting and receiving antennas, of finite size) it is possible to show that the sum of all the coupling coefficients $\xi_{mn}$ is limited (sum rule) \cite{Mil:J00}, that is
\begin{align}\label{eq:SumRule}
\gamma_{RT} = \sum_{m n } \xi_{mn}^2 = \frac{1}{(4\pi)^2}\int_{\mathcal{S}_R} \int_{\mathcal{S}_T} \frac{1}{|| \rr-\rt ||^2} \, d\rt d\rr \, . 
\end{align}
This sum is intimately related to the path loss of the channel \cite{Dar:J20}.

In general, each basis function at the transmitting antenna is coupled with more basis  functions at the receiving antenna. If the orthonormal basis set is chosen so that the operator $\Gamma$ is a diagonal matrix, a one-to-one correspondence among the $n$th {TX} and the $n$th {RX} basis functions is established.    
Formally, we have $\bold{\Gamma}=\operatorname{diag}\{\xi_n\}$ so that the $n$th basis function $\phi_n(\rt)$ produces an effect $\xi_n \psi_n(\rr)$ on the receiving antenna space, where $\xi_n$ is the largest possible coupling coefficient, and $\xi_1$, $\xi_2$,... are in decreasing order on the main diagonal of $\bold{\Gamma}$. For a well-coupled communication mode (i.e., large $\xi_n$), the wave generated by the transmitting antenna impinges the receiving antenna. Differently, for a loosely-coupled communication mode (i.e., small $\xi_n$), the wave generated by the transmitting antenna is mostly spread away from the receiving antenna \cite{Mil:J19}.

These basis sets (communication modes) can be found by solving a coupled eigenfunction problem, specifically \cite{Mil:J00}
\begin{align}
\xi_n^2\, \phi_n(\rt) = \int_{\mathcal{S}_T} K_T(\rt,\rt')\, \phi_n(\rt')\, d\rt' \label{eq:autoTX} \\
\xi_n^2\, \psi_n(\rr) = \int_{\mathcal{S}_R} K_R(\rr,\rr')\, \psi_n(\rr')\, d\rr' \label{eq:autoRX}
\end{align}
where the kernels $K_T(\rt',\rt)$ and $K_R(\rr,\rr')$ are given by
\begin{align}
K_T(\rt',\rt) = \int_{\mathcal{S}_R} G^*(\rr,\rt)\, G(\rr,\rt')\, d\rr \label{eq:kernel1} \\
K_R(\rr,\rr') = \int_{\mathcal{S}_T} G(\rr,\rt)\, G^*(\rr',\rt)\, d\rt\, .\label{eq:kernel2}
\end{align}
More in particular, by solving \eqref{eq:autoTX} and \eqref{eq:autoRX}, the eigenfunctions (basis set) ensure the most accurate approximation of the Green function \eqref{eq:greenf} for any cardinality of the basis set itself.\footnote{See \cite{Dar:J20} for a more detailed discussion.}

From a practical point of view, for finite-size antennas, it is possible to show that the number of significantly non-zero eigenvalues on the main diagonal of $\bold{\Gamma}$ is limited. In other words, there is a physical limit on the number of communication modes (namely \ac{DoF}) with significant coupling, after which the coupling intensity falls off rapidly to zero. Such a behavior is well known for the eigenvalues of problems having as eigenfunctions \acp{PSWF} \cite{Slepian:83}, as obtained by resolving \eqref{eq:autoTX} and \eqref{eq:autoRX} in the case of linear antennas of small size with respect to the link distance under paraxial conditions (i.e., parallel antennas, with $\yc=0$, $\theta=0$ and $\lt, \lr\ll z$) \cite{Mil:J00}. Then, it is possible to consider a finite number $N$ of bases  both at transmitting and receiving antenna side, and hence \eqref{eq:matrixcomm} assumes the usual meaning of vector-matrix product.
Under this assumption, it is well known that  \cite{Mil:J00,BraBehSay:J13}
\begin{equation}\label{eq:Ngeneral}
N\approx\frac{\lt \lr}{\lambda z} \, .
\end{equation}
When operating in the far field (i.e., large $\lambda z$) and in \ac{LOS} between the antennas, only a single communication mode can be exploited ($N=1$), corresponding to a plane wave traveling from the transmitting to the receiving antenna.\footnote{Equation \eqref{eq:Ngeneral} should be intended as the closest integer. The limit for large $z$ in \eqref{eq:Ngeneral} is 0, however a single communication mode can be exploited provided that the receiver noise is low if compared to the relative coupling intensity (i.e., with respect to the received power). Moreover, it is known that $N$ in \eqref{eq:Ngeneral} is a conservative estimate, and the number of well-coupled modes is generally between $N$ and $N+1$ \cite{BraBehSay:J13}.} Notice that \eqref{eq:Ngeneral} is verified only according to the geometric paraxial approximation assumption, without any requirement for the operating frequency; thus, with the increasing of the operating frequency (i.e., smaller wavelength), multiple communication modes become feasible, even at  relatively large distance among the antennas. In such a case, communication can take part within the so-called near-field region. In particular, for an antenna of size $D$, the distance corresponding to the boundary between the far-field region and near-field region of the antenna itself, called Fraunhofer distance, is defined as $r_{\text{ff}}={2D^2}/{\lambda}$ \cite{Bal:B15}. 

\subsection{Communication Modes as Beams}

A possibility for realizing multiple communication modes is represented by the generation of orthogonal beams with the transmitting antenna.
This is the case of beamspace \ac{MIMO} \cite{BraBehSay:J13}. The different beams are obtained by adopting arrays implementing constant gradients for the phase excitation on the transmitting antenna elements (i.e., beam steering), by selecting properly the beam-pointing angular directions in order to ensure orthogonality. When mapping this technique to a continuous antenna, as in the holographic MIMO case we are considering, a constant phase profile (i.e., a constant transmitting functions $\phi(\eta)$)  translates into a beam directed on the antenna boresight (i.e., the \textit{optical axis} of the transmitting antenna, making a parallelism with optics). More generally,  a linear phase profile, that is a transmitting function $\phi(\eta)$ in the form
\begin{equation}\label{eq:BeamSteering}
\phi(\eta)=\rect{\frac{\eta}{\lt}}e^{-\jmath\frac{2\pi }{\lambda } \eta \sin{\varphi} \,\, }
\end{equation}
with $\rect{x}=1$ for $x\in[-0.5,\, 0.5]$ and $0$ otherwise, realizes, in the far-field region of the transmitting antenna, a beam in the  direction (angle) $\varphi$ with respect to the antenna boresight; \eqref{eq:BeamSteering} generalizes the well-known results for beam steering with uniform linear arrays \cite{Bal:B15} in the holographic case  considering a continuous  antenna. In fact, we can see \eqref{eq:BeamSteering} as the array factor related to the Huygens' sources on the transmitting antenna; $\varphi=0$ is obtained with all the sources in-phase (broadside array, corresponding to a main lobe perpendicular to the direction of the array), while $\varphi=\pm\pi/2$ is obtained by imposing a phase difference equal to the separation of adjacent sources scaled by $\lambda$ (end-fire array, corresponding to a main lobe along the direction of the array).
\begin{figure}
\centering
     \subfloat[]{\includegraphics[width=0.3\columnwidth,keepaspectratio=true]{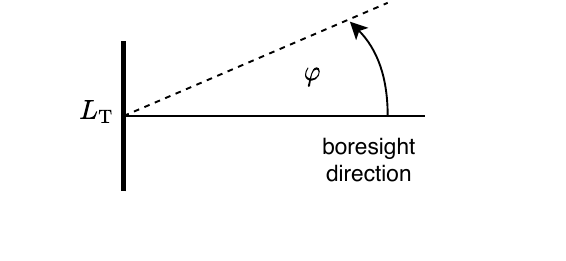}}
     \subfloat[]{\includegraphics[width=0.2\columnwidth,keepaspectratio=true]{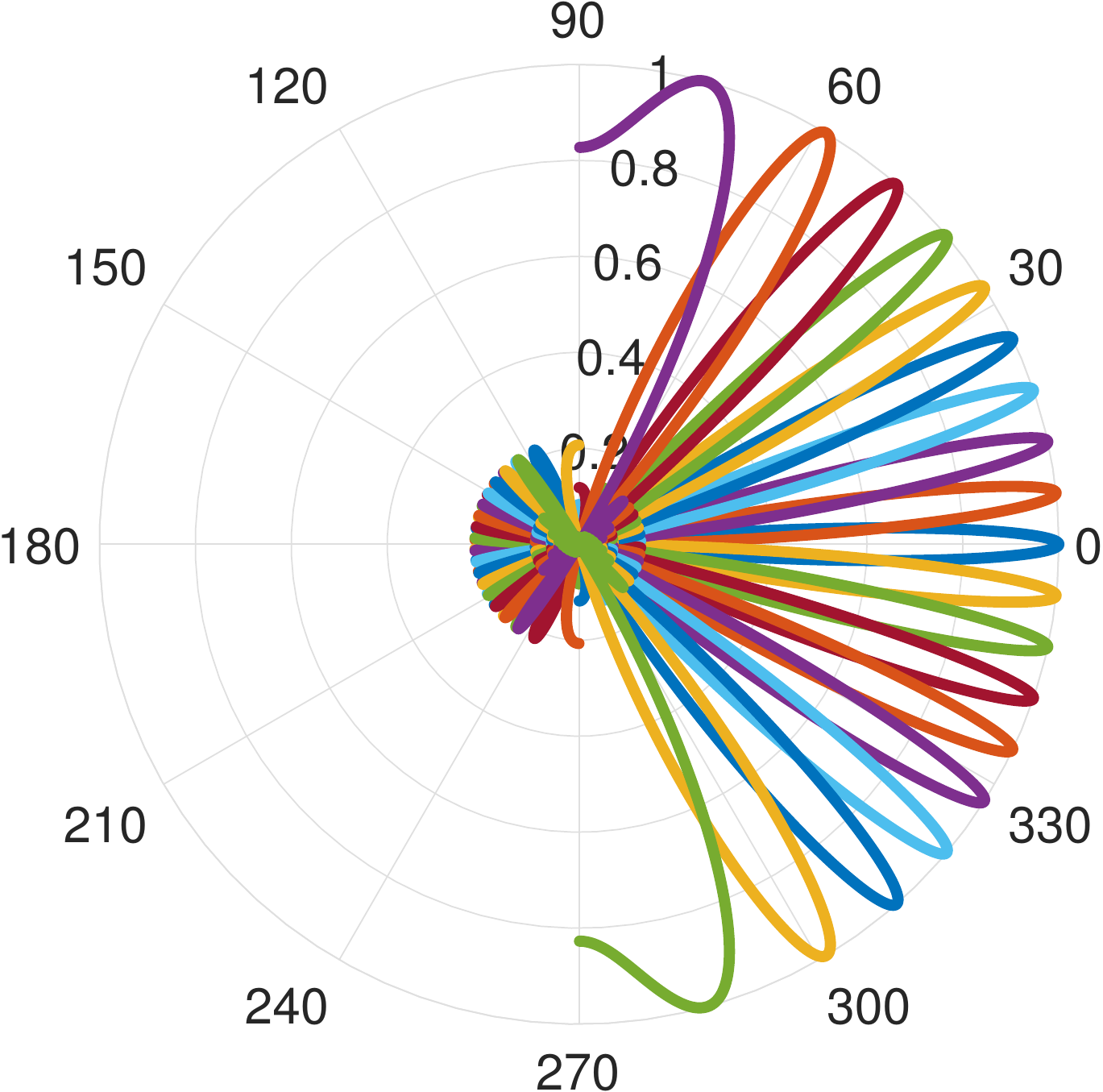}}
\caption{Far-field orthogonal beams generated by a linear antenna for $f_0=28\,$GHz and $\lt=10\,$cm.}
\label{fig:FarFieldBeams}
\end{figure}  
The beamspace technique can be adopted only when the antennas are small with respect to to the propagation distance; unfortunately, such a condition may correspond to a very limited number of communication modes, due to the small coupling among the antennas  \cite{BraBehSay:J13}. 
In fact, a transmitting antenna can realize, in the far field, up to $N_{\text{max}}=2\lt/\lambda$ orthogonal beams in a hemisphere (i.e., for $-\pi/2 \leq \varphi \leq \pi/2$), as depicted in Fig.~\ref{fig:FarFieldBeams}.\footnote{\label{footnote:Border}In Fig.~\ref{fig:FarFieldBeams}, $N_{\text{max}}=2\lt/\lambda=18.7$ and, in fact, 19 beams are shown.
} However, only a fraction $N\ll N_{\text{max}}$ will intercept the receiving antenna of size $\lr$ thus coupling properly with the transmitting antenna\cite{BraBehSay:J13}.

\begin{figure}
\centering
\includegraphics[width=0.5\columnwidth,keepaspectratio=true]{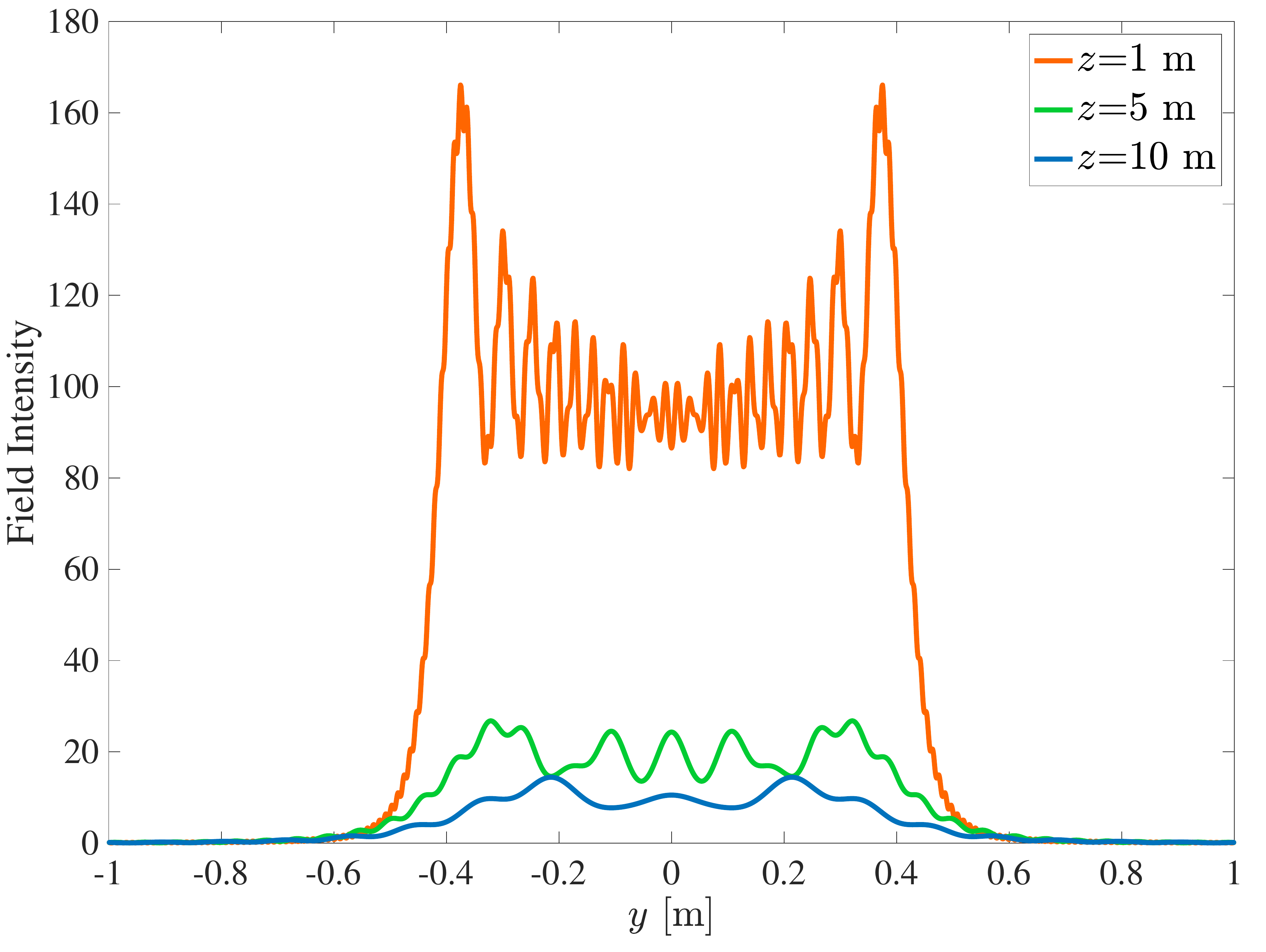}
\caption{ Beams obtained at a receiving antenna placed in the near-field region of a transmitting antenna of size $\lt=1\,$m, for $\yc=0\,$m, $\theta=0\degree$, $f_0=28\,$GHz.}
\label{fig:FresnelPatterns}
\end{figure}

When the distance between the transmitting antenna and the receiving antenna decreases, propagation takes part in the near-field region. 
Diffraction theory developed in optics describes the beams obtained in such propagation condition using the Fresnel diffraction patterns (analytically described by the Fresnel integrals) \cite{God:B05}. Assuming the beam steering transmitting function in \eqref{eq:BeamSteering} with $\varphi=0$, the corresponding beam on the receiving antenna, assumed centered in the boresight direction, is reported in Fig.~\ref{fig:FresnelPatterns};  in particular, the different beams shown in the figure are obtained for different values of the distance between the transmitting and receiving antennas considering $f_0=28\,$GHz and a transmitting \ac{LIS} of $\lt=1\,$m. %
It can be noticed how, in this case, it is easy to have very wide beams even at practical operating distances\footnote{Notice that, at the distance considered, the beam width is close to the transmitting antenna size $\lt$.} ($1\div10\,$m in the example of Fig.~\ref{fig:FresnelPatterns}); therefore, a mobile terminal equipped with a relatively small antenna could not intercept much of the beam, thus making difficult to exploit multiple communication modes (very low coupling due to a wave mainly dispersed away from the receiving antenna). In this case, classical beam steering techniques may be extremely inefficient, and one should resort to the exact solution of the coupled eigenfunction problem in \eqref{eq:autoTX}-\eqref{eq:autoRX}.

For a generic setup, even in the near field, finding the solution of the coupled eigenfunction problem in \eqref{eq:autoTX}-\eqref{eq:autoRX} requires extensive and sometimes prohibitive simulations if large antennas are considered. In particular, a discretization into a fine mesh of the transmitting and receiving antenna regions can be realized, then solving numerically the eigenfunction problem applying \ac{SVD} which might lead to huge matrices. For this reason, there is a great interest in finding simple ways to define the communication modes. A practical method extending the beamspace techniques also in the near field will be presented in Sec.~\ref{sec:basisdef}.

\bigskip
\section{Derivation of Approximate Basis Sets}\label{sec:basisdef}

\subsection{General Case}\label{sec:BasisGen}

In this section, we propose a method for the definition of a practical basis set, thus approximating optimal solutions without the need of evaluating numerically \eqref{eq:autoTX} and \eqref{eq:autoRX}. Moreover, the proposed method will be specified in particular cases of interest leading to analytical results, thus providing insights into the problem, not otherwise available with numerical solutions.

Let us consider the kernel $K_R(\rr,\rr')$ in \eqref{eq:kernel2}, by declining it in the scenario described in Fig.~\ref{fig:scenario}. 
In this case, we can write $G(\rr,\rt)$ as a function of $r=||\rr-\rt||$ given by \eqref{eq:distance}. In order to simplify the notation, in the following we will denote the Green function $G(\rr,\rt)$ in \eqref{eq:greenf} with $G(r)$, and the kernel $K_R(\rr,\rr')$ in \eqref{eq:kernel2} with $K_R(y,y')$.

Defining $r'=\sqrt{(z+\eta\sin\theta)^2+(y'-\eta\cos\theta)^2}$, we have
\begin{align}\label{eq:kernelJ}
K_R(y,y') = &\int_{-\lt/2}^{\lt/2} G(r)\,G^*(r')\, d\eta\, \\
 =& \int_{-\lt/2}^{\lt/2}  \frac{e^{-\jmath\kappa\sqrt{(z+\eta\sin\theta)^2+(y-\eta\cos\theta)^2}}}{4\pi\sqrt{(z+\eta\sin\theta)^2+(y-\eta\cos\theta)^2}}\, \frac{e^{\jmath\kappa\sqrt{(z+\eta\sin\theta)^2+(y'-\eta\cos\theta)^2}}}{4\pi\sqrt{(z+\eta\sin\theta)^2+(y'-\eta\cos\theta)^2}} \, d\eta\,  .\nonumber
\end{align}
For practical operating distances among the antennas, in~\eqref{eq:kernelJ} we approximate the denominators with $d_\mathrm{c}$, while such an approximation cannot be adopted for the numerators due to the interference effects produced by the phase terms. Then,
\begin{align}\label{eq:kernelJ1}
K_R(y,y') \approx  \\\frac{1}{\left(4\pi d_\mathrm{c}\right)^2} & \int_{-\lt/2}^{\lt/2} {\exp{\left(-\jmath\kappa\sqrt{(z+\eta\sin\theta)^2+(y-\eta\cos\theta)^2}\right)}}  \,\,{\exp{\left(\jmath\kappa\sqrt{(z+\eta\sin\theta)^2+(y'-\eta\cos\theta)^2}\right)}} \, d\eta\, .\nonumber
\end{align}
On the right hand side of the integral in~\eqref{eq:kernelJ1}, we recognize the phase profile required at the transmitting antenna (i.e., function $\phi(\eta)$) to focus its energy on the point $y'$ at the receiving antenna (focal point), that is, to have all the components (outgoing spherical waves produced by Huygens' sources on the  transmitting antenna) summing up in-phase at the point of coordinates $(z,y')$, thus compensating for the phase delay introduced by the propagation. Using this phase profile corresponding to the focusing operation, we have an increase of the \ac{EM} power density in a size-limited region of the space close to the focal point, if this is located relatively close to the transmitting antenna. In fact, differently from the beam steering, which consists in concentrating the energy on a specific direction in the far field (corresponding to focusing at infinite distance), operating in the near field  allows concentrating the field on a specific point in the space \cite{NepBuf:J17}, as it will be further discussed in the Sec.~\ref{sec:zones} (beam with finite depth). With reference to Fig.~\ref{fig:scenario}, we define the focusing function\footnote{Here, differently from \ac{NFF}  antenna arrays \cite{NepBuf:J17}, we have a continuous phase profile with $\eta$ (holographic assumption, with the \ac{LIS} considered as an ideal aperture antenna).} 
 at the transmitting antenna required to focus the energy at the point $y'$ on the receiving antenna as 
\begin{equation}\label{eq:focusingfunc}
F_T(\eta)\rvert_{y'}=\rect{\frac{\eta}{\lt}}e^{\jmath\frac{2\pi }{\lambda}\sqrt{(z+\eta\sin\theta)^2+(y'-\eta\cos\theta)^2}} \, .
\end{equation}
%
%
At the receiving antenna side the field is given by \eqref{eq:distribution}, that is
\begin{align}\label{eq:kernelJ2}
&\psi(y)\bigr\rvert_{y'} =  \int_{-\lt/2}^{\lt/2}  G(r) \, F_T(\eta)\rvert_{y'} \, d\eta\, 
\end{align}
where we used the notation $\psi(y)\rvert_{y'}$ to identify the field distribution in the case we used a phase profile at transmitting antenna side to focus the energy towards $y'$ on the receiving antenna. 
\begin{figure}
\centering
\includegraphics[width=0.5\columnwidth,keepaspectratio=true]{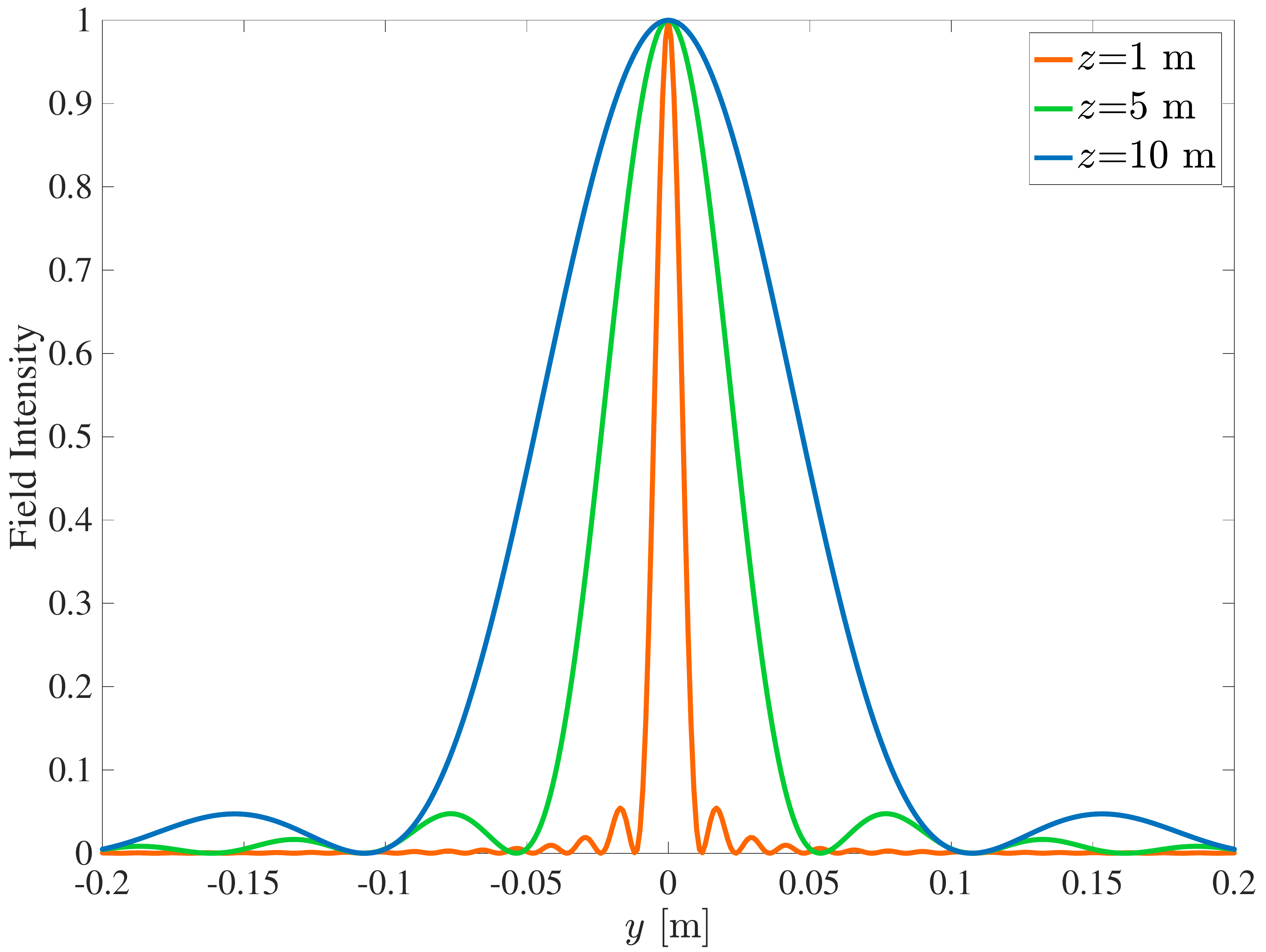}
\caption{Beams (normalized) obtained at a receiving antenna placed in the near-field region of a transmitting antenna of size $\lt=1\,$m, when a phase profile corresponding to the focusing operation is implemented, for $\yc=0\,$m, $\theta=0\degree$, $f_0=28\,$GHz.}
\label{fig:FresnelFocusingPatterns}
\end{figure}  
Fig.~\ref{fig:FresnelFocusingPatterns} reports the beams obtained on the receiving antenna (i.e., functions $|\psi(y)|$ given by \eqref{eq:kernelJ2}) using focusing according to \eqref{eq:focusingfunc} for $y'=0$, $\theta=0$, and the same distance set considered in Fig.~\ref{fig:FresnelPatterns}. It is immediate to notice that the focusing function changes completely the beams corresponding to the Fresnel diffraction patterns of Fig.~\ref{fig:FresnelPatterns}, obtained with constant phase profiles at the transmitting antenna side, into much more concentrated functions.\footnote{Notice the different scale on the horizontal axis of Fig.~\ref{fig:FresnelFocusingPatterns} with respect to that of Fig.~\ref{fig:FresnelPatterns}.} Then, even a mobile user equipped with a relatively small antenna could intercept the beam, resulting in a good coupling among transmitting and receiving antennas. This behavior suggests the possibility of implementing a beamspace-like multi-mode communication using focusing functions in the form \eqref{eq:focusingfunc} instead of beam steering functions in the form \eqref{eq:BeamSteering} when operating in the near field.

We notice that \eqref{eq:kernelJ2} is formally identical to the kernel \eqref{eq:kernelJ1}, except for a multiplicative factor. Moreover, it is possible to write the kernel as the product of a focusing function towards $y'$ and the complex conjugate of a focusing function towards a generic $y$, that is
\begin{align}\label{eq:kernelJ3}
K_R(y,y') \approx \frac{1}{\left(4\pi d_{\mathrm{c}}\right)^2} \int_{-\lt/2}^{\lt/2} \left(F_T(\eta)\rvert_{y} \right)^* F_T(\eta)\rvert_{y'} \, d\eta\, .
\end{align}
If \eqref{eq:kernelJ3} is zero for any couple of focusing functions towards $y'$ and $y$, for some $y'\neq y$, we obtain the same condition \eqref{eq:orthoTX} that we have to meet for orthogonal basis sets. More specifically, by assuming that the first basis function at transmitting side is defined as the focusing function towards $y'$, we ask how far we have to move along the direction $y$ (i.e., on the receiving antenna) to have orthogonality with respect to such a focusing function. Once $y$ is found,  the second basis function at the transmitting side will be a focusing function towards that specific $y$. The process can be repeated until it is not possible to find a new $y$  inside the receiving antenna capable of fulfilling the orthogonality condition. The orthonormal version can be  obtained by dividing each basis function by a scale factor $\sqrt{\lt}$. Thus, focusing functions in the form \eqref{eq:focusingfunc}, with proper choices of the points $y'$, can form approximate basis sets at the transmitting antenna side.

\begin{figure}[t]
\centering
\includegraphics[width=0.5\columnwidth,keepaspectratio=true]{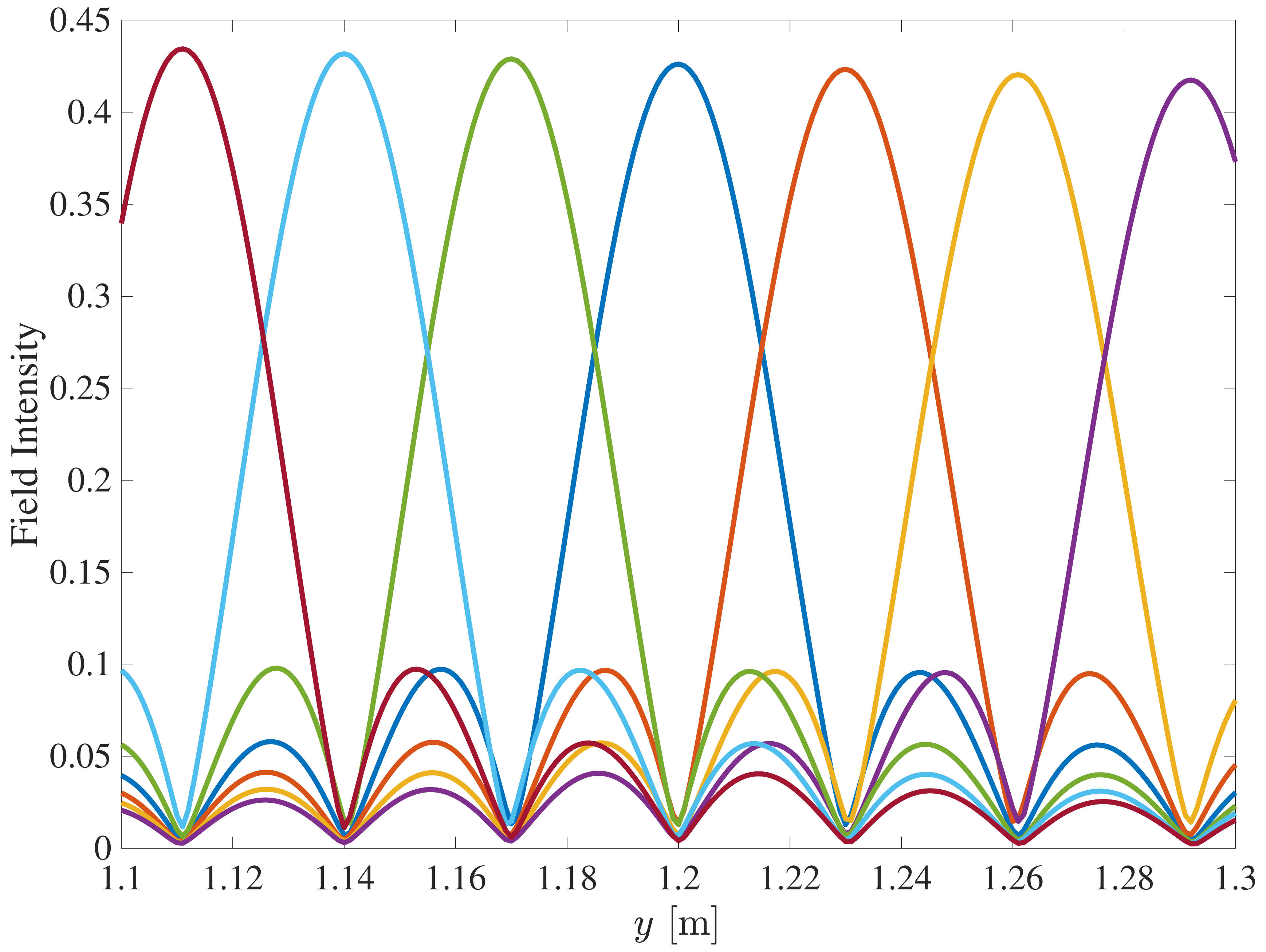}
\caption{The $N=7$ beams obtained using focusing according to \eqref{eq:focusingfunc} in the near-field region for $\yc=1.2\,$m, $\theta=20\degree$, $z=2\,$m, $f_0=28\,$GHz and a \ac{LIS} with $\lt=1\,$m (non-paraxial case). A receiving antenna of size $\lr=20\,$cm is assumed.}
\label{fig:ModesEx1}
\end{figure}  

From a practical point of view, we have to construct the functions starting from a particular point at the receiving antenna. A viable way consists in assuming a first basis function at the transmitting antenna side as the focusing function towards the center $y_\mathrm{c}$ of the receiving antenna, thus using $(z,\yc)$ as focal point, that is
\begin{equation}
\phi_0(\eta)=\frac{1}{\sqrt{\lt}}F_T(\eta)\rvert_{\yc}\, .
\end{equation}
The next step is to move along $y$ in positive direction until a null (or a local minimum from a practical point of view) of the kernel $K_R(y,\yc)$ in \eqref{eq:kernelJ1} (i.e., an orthogonal focusing function) is found, corresponding to point $y_1$; a new basis is then constructed as a focusing function towards that point. The process is iterated with all the other nulls of $K_R(y,\yc)$ until the focal point falls outside the receiving antenna.
Then, the same process is repeated for the negative direction starting from the receiving antenna center.%

\begin{figure}[t]
     \centering
     \subfloat[][Transmitter functions set]{\includegraphics[width=0.25\columnwidth,keepaspectratio=true]{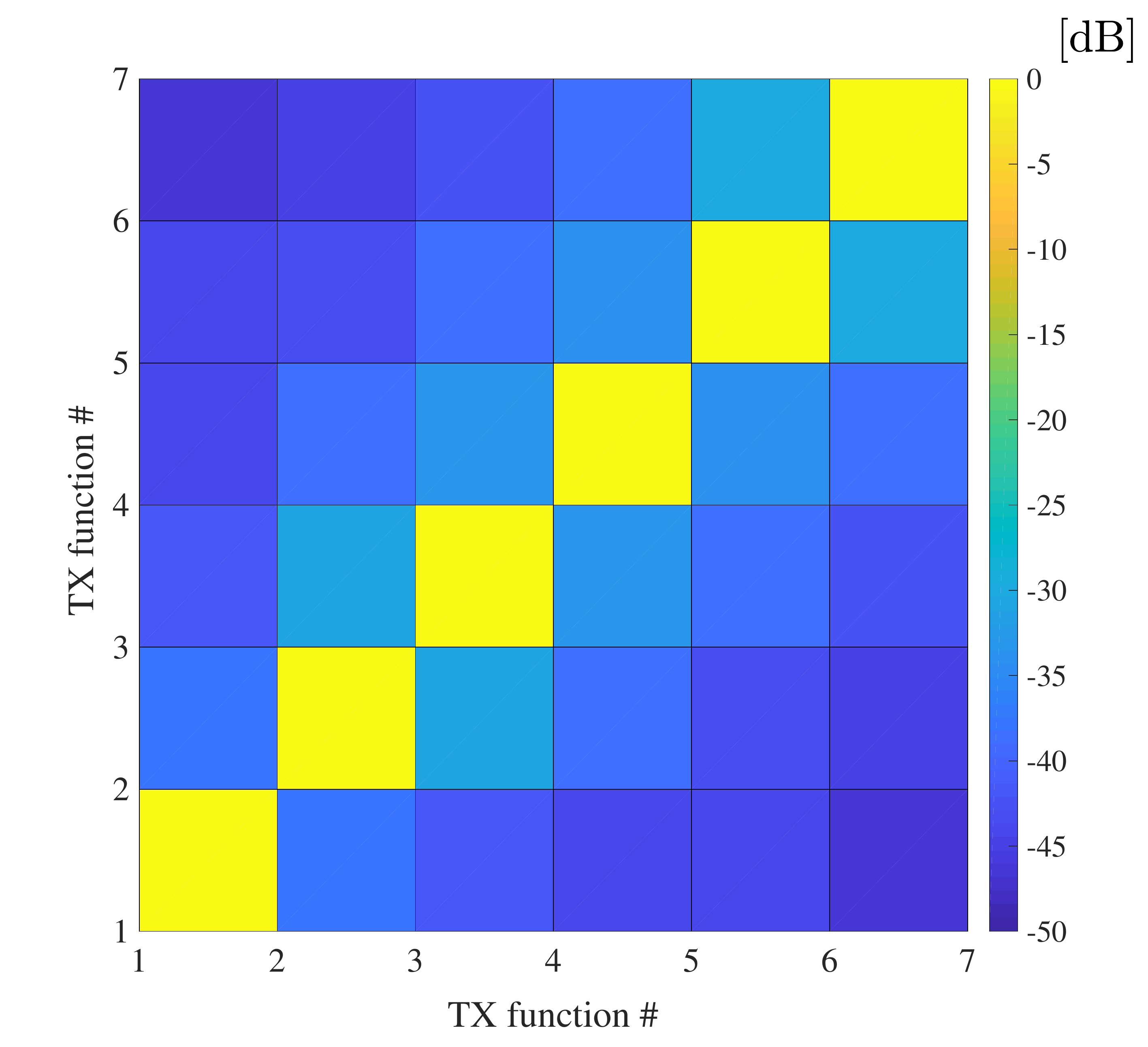}}
     \subfloat[][Receiver functions set]{\includegraphics[width=0.25\columnwidth,keepaspectratio=true]{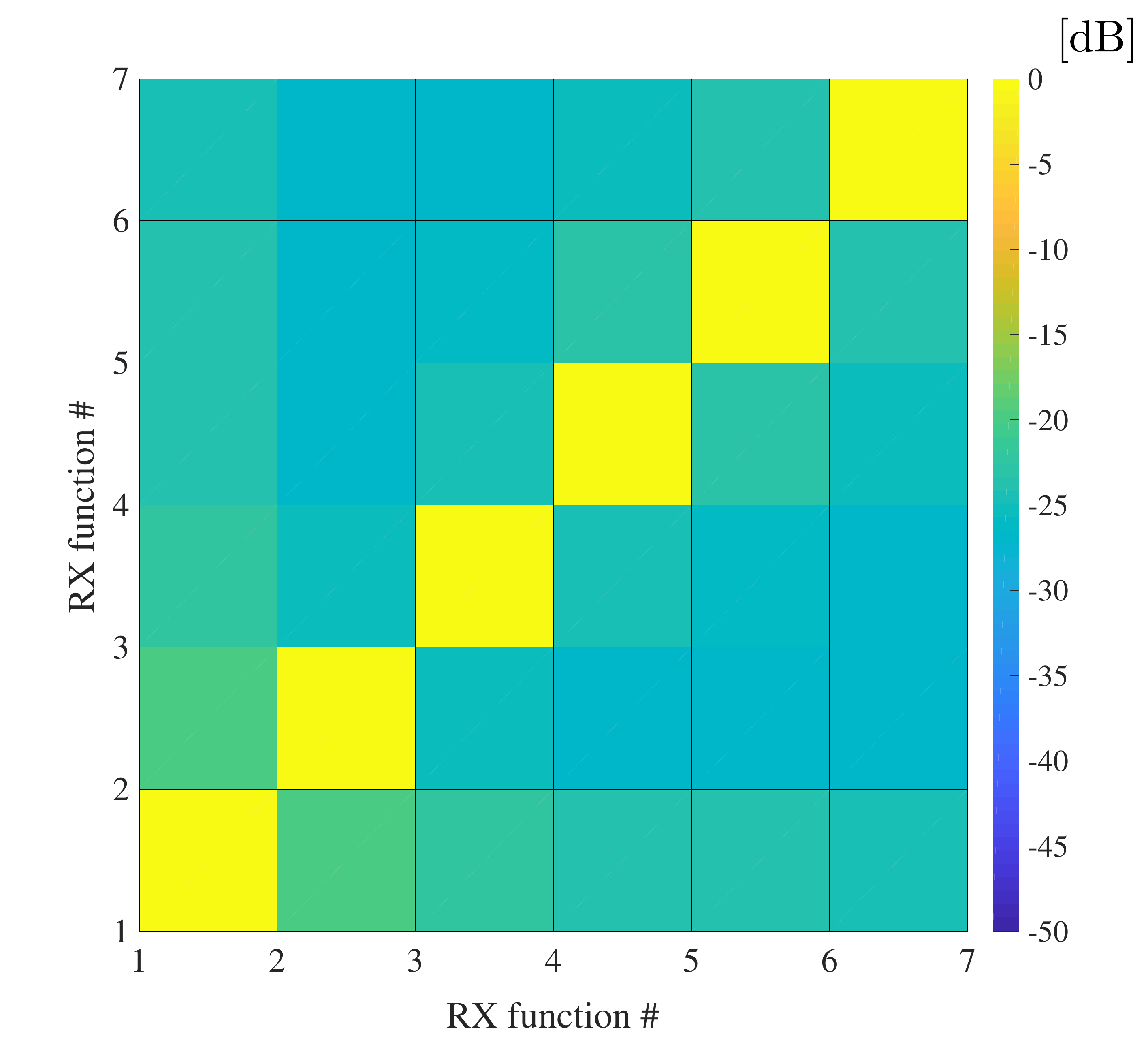}}
     \caption{Numerical check of orthogonality for the functions at transmitter and receiver side (worst-case $-23\,$dB at the transmitting antenna side, and $-16\,$dB at the receiving antenna side).}
     \label{fig:ModesEx1Ortho}
\end{figure}

The method defines basis functions at the transmitting antenna. The effect at the receiving antenna is given, for every focusing function, by \eqref{eq:kernelJ2}. 
An example of the beams realized at the receiving antenna with the proposed approach is reported in Fig.~\ref{fig:ModesEx1}. In particular, a receiving antenna of size $\lr=20\,$cm is assumed, with the other geometrical parameters reported in the caption (non-paraxial case). It can be noticed that $N=7$ beams according to \eqref{eq:kernelJ2} can be realized. It is worth to underline that finding orthogonal functions does not ensure that these can be used to form (orthogonal) communication modes, that is, solutions of the eigenfunction problem \eqref{eq:autoTX}-\eqref{eq:autoRX}. Orthogonality among the functions at transmitting and receiving antenna side was checked numerically and it is graphically reported in Fig.~\ref{fig:ModesEx1Ortho}. It can be noticed a good orthogonality even at the receiving antenna side (around $-20\,$dB) for the considered example. Thus, $N=7$ practically-orthogonal communication modes in the form of focused beams can be realized, thus capable of boosting the capacity between the transmitting antenna equipped with the \ac{LIS} and the user equipped with a smaller antenna here considered.
Moreover, the close orthogonality ensures that the basis functions constructed starting from focusing functions, even though not optimal in general, provide almost the same number of communication modes when using the optimal basis set solution of \eqref{eq:autoTX}-\eqref{eq:autoRX}, as it will be discussed more in detail later.

Notice that, with the choice made on the specific focusing functions, each beam (i.e., each basis function at receiving antenna side) is centered in the first null corresponding to the adjacent beam (the adjacent basis function at the receiving antenna side, see Fig.~\ref{fig:ModesEx1}); this corresponds to the Rayleigh criterion for resolution of optical images \cite{God:B05}, also known as Rayleigh spacing (usually denoted in the angular domain, that is in the far field in \cite{SayBeh:C10}). 
Thus, the number of communication modes equals the maximum number of diffraction-limited focused-beams that can be obtained at the receiving antenna side, produced by an aperture of size equal to that of the transmitting antenna, without mutual interference. 
As happens in diffraction, larger apertures (i.e., larger antennas) produce narrower beams (i.e., more concentrated basis functions) at the receiving antenna side; thus, the wider the transmitting and receiving antennas are, the higher is the number of communication modes that can be established, as expected.

Focusing functions in the near field allow realizing a beamspace-MIMO-like approach similarly to what done in the angular domain (far field) with beam steering. Here, orthogonality cannot be considered in the angular domain since, in the near field, the transmitting antenna beam pattern changes with the distance due to the spherical wavefront propagation. Differently, orthogonality is ensured on the receiving antenna region. 
It has to be remarked that, from an implementation point of view, the realization of beams with focusing requires the knowledge of relative position and orientation among the antennas; this can be realized with high accuracy by exploiting the information carried by the spherical wavefront \cite{GuiDar:J21,ElzEtAl:J20}.

Summarizing:
\begin{itemize}
\item The approach described allows defining orthogonal functions at the transmitting antenna using focusing functions (phase profiles) towards the nulls of the kernel $K_R(y,\yc)$.
\item It can be verified numerically that the focusing functions produce practically-orthogonal beams at the receiving antennas in several cases of interest, as it will be discussed in depth in Sec.~\ref{Sec:results}. Thus, we are approximately defining multiple communication modes.
\end{itemize}
The functions proposed are not the optimum, which is obtained by solving \eqref{eq:autoTX}-\eqref{eq:autoRX} leading to perfectly orthogonal functions, thus diagonalizing $\mathbf{\Gamma}$. However, the good level of orthogonality verified numerically allows defining practical communication modes, and it will be shown that their numbers equal that of optimum solutions. 
   
In the following sections:
\begin{itemize}
\item The proposed approach will be declined to particular but practical configurations leading to analytical solutions both at transmitting and receiving antenna side.
\item Traditional results obtained as approximate solutions in the paraxial case will be revised in light of the proposed method.
\end{itemize}
In particular, we now derive easy closed-form expressions for the basis functions at transmitting and receiving antenna side and for the number of communication modes considering the communication between an antenna of small size, and a large intelligent antenna.

\subsection{Communication Modes between a SIS and LIS: Uplink}\label{sec:SISLIS}

\begin{figure}[t!]
\centering
\includegraphics[width=0.5\columnwidth,keepaspectratio=true]{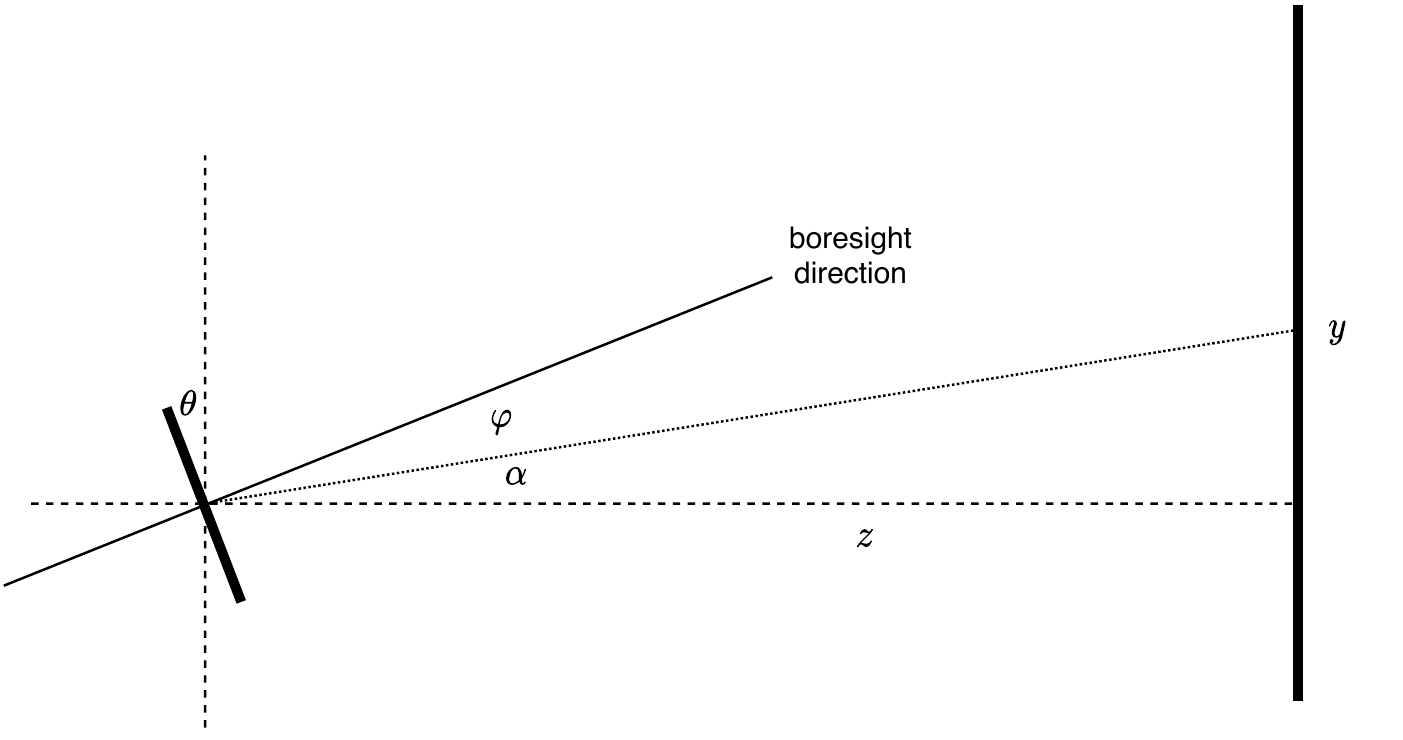}
\caption{Geometry of communication between a SIS (left) and a LIS (right), uplink case.}
\label{fig:geometrySISLIS}
\end{figure}

As first, we consider a small transmitting antenna (i.e., uplink scenario). In this case, we can adopt the Maclaurin series expansion at the first term for the distance $r$ as a function of the variable $\eta$, in the numerators (phase terms) of \eqref{eq:kernelJ}-\eqref{eq:kernelJ3}, around the antenna center $\eta=0$, that is
\begin{equation}\label{eq:SISapprox}
\frac{2\pi}{\lambda}r(\eta)  \approx  \frac{2\pi}{\lambda} \left[ r(0) +\eta \frac{\partial}{\partial \eta} r(\eta)\biggr\rvert_{\eta=0}  \right] \, .
\end{equation}
Using \eqref{eq:SISapprox}, it is possible to write the focusing function $F_T(\eta)\rvert_{y}$ in \eqref{eq:focusingfunc} as
\begin{equation}\label{eq:focusingSIS}
F_T(\eta)\rvert_{y}=\rect{\frac{\eta}{\lt}}e^{\jmath\frac{2\pi }{\lambda } \frac{\sin\theta-\gamma\cos\theta}{\sqrt{1+\gamma^2}} \eta}
\end{equation}
whose derivation is reported in Appendix~\ref{App:SISLIS}, where $\gamma=y/z$.
It is immediate to see that we have obtained a phase profile of the same form of \eqref{eq:BeamSteering}, that is, linear with $\eta$. 	Thus, this corresponds to a beam steering phase profile allowing to concentrate the  energy on a specific direction in the far field, that is, towards an angle $\varphi$ with respect to the boresight of the (small) transmitting antenna. In particular, the angle $\varphi$ obeys to
\begin{equation}\label{eq:SISangle}
\rho=-\sin{\varphi}= \frac{\sin\theta-\gamma\cos\theta}{\sqrt{1+\gamma^2}} \, 
\end{equation}
where the explicit dependence of $\rho$ from $y$, $z$ and $\theta$ is omitted to simplify the notation.
By writing $\gamma=y/z=\tan\alpha$, with simple trigonometric manipulations, we have that $\varphi$ in \eqref{eq:SISangle} corresponds to an angle $\varphi=\alpha-\theta$, which is expected, since the focusing function was constructed to focus the  energy towards a point $(z,y)$ on the receiving antenna, also considering a generic transmitting antenna orientation $\theta$ (see Fig.~\ref{fig:geometrySISLIS}). Thus, beam steering in the direction of the focal point is realized. In this case, the focusing behavior degenerates into a beam steering operation, since the small dimension of the transmitting antenna does not provide the focusing capability (or equivalently, the near-field region of the transmitting antenna, where focusing is feasible, is small, as it will discussed more in detail in Sec.~\ref{sec:zones}). 
%

We can now exploit approximation \eqref{eq:SISapprox} to rewrite the kernel \eqref{eq:kernelJ3} or equivalently the received function \eqref{eq:kernelJ2}, which are the same except for a multiplicative factor, considering the focusing towards $(z, \yc)$, obtaining
\begin{align}\label{eq:SISkernelJ}
K_R(y,\yc) \approx &\frac{1}{\left(4\pi d_{\mathrm{c}}\right)^2} \int_{-\infty}^{\infty} \rect{\frac{\eta}{\lt}} e^{-\jmath\frac{2\pi }{\lambda } (\rho-\rho_{\mathrm{c}}) \eta}
\, d\eta\, 
\end{align}
where $\rho_{\mathrm{c}}$ is obtained from \eqref{eq:SISangle} by considering the steering direction towards $\yc$ at the receiving antenna. Then, \eqref{eq:SISkernelJ} corresponds to the Fourier transform of $\rect{\eta/\lt}$ evaluated at $(\rho-\rho_{\mathrm{c}})/\lambda$, so that
\begin{align}\label{eq:SISkernelJ2}
K_R(y,\yc) \approx &\frac{\lt}{\left(4\pi d_{\mathrm{c}}\right)^2} \, \sinc{\frac{\lt}{\lambda}(\rho-\rho_{\mathrm{c}})} \, 
\end{align}
where $\sinc{x}=\sin(\pi x)/(\pi x)$.
The kernel, or equivalently the beam at the receiving antenna side, is a space-varying sinc function of $y$. Under the hypothesis of steering towards $\yc$, it is then possible to find orthogonal steering functions for values of $y$ corresponding to integer values of $\frac{\lt}{\lambda}(\rho-\rho_{\mathrm{c}})$, that is
\begin{align}\label{eq:SISnulls}
y_n\!=\! z \tan{\left[\arcsin{\left(-\frac{\lambda}{\lt}n-\rho_{\mathrm{c}}\right)}\!+\!\theta\right]} \, , \quad |n|=1, 2, \ldots \,\, .
\end{align}
Thus, by assuming a first beam directed towards $(z, \yc)$, from \eqref{eq:SISnulls} the allowed set of integers $n$ must satisfy
\begin{align}\label{eq:condkappa}
-1\leq  \frac{\sin\theta-\frac{\yc}{z}\cos\theta}{\sqrt{1+{\left(\frac{\yc}{z}\right)}^2}} +\frac{\lambda}{\lt} n \leq 1 \, 
\end{align}
that for $\theta=0$ and $\yc=0$ simplifies to 
\begin{align}
-\frac{\lt}{\lambda}\leq  n \leq \frac{\lt}{\lambda} \, 
\end{align}
corresponding to a maximum number of $2\lt/\lambda$ beams in the hemisphere, which is the same result discussed Sec.~\ref{eq:general}. Among the indexes $n$ of \eqref{eq:condkappa}, it is necessary to evaluate which of them correspond to beams intercepting the receiving \ac{LIS}, thus leading to good coupling among the antennas. Formally, by defining $\mathcal{I}$ the set of indexes corresponding to well-coupled beams, we have
\begin{align}\label{eq:Nexact}
 \mathcal{I}=\{n\} : \yc-\frac{\lr}{2} < y_n < \yc+\frac{\lr}{2} 
\end{align}
and $N=|\mathcal{I}|$, where $|\cdot|$ stands for the cardinality of a set.

Equation~\eqref{eq:SISnulls} does not give immediate insights in its general form. However, considering parallel antennas ($\theta=0$) and exploiting $\tan\arcsin x = x/\sqrt{1-x^2}$, we can write
\begin{align}\label{eq:SISnulls2}
y_n&\!=\! z \frac{1}{\sqrt{1-\left(\frac{\frac{\yc}{z}}{\sqrt{1+\left(\frac{\yc}{z}\right)^2}}\!-\!\frac{\lambda}{\lt}n\right)^2}} \left(\frac{\frac{\yc}{z}}{\sqrt{1+\left(\frac{\yc}{z}\right)^2}}-\frac{\lambda}{\lt}n\right)
\end{align}
for $n\in\mathcal{I}$, where it is evident the stretching of the sinc lobes while moving along $y$. From \eqref{eq:SISnulls2} we can observe several facts:
\begin{itemize}
\item The distance among the nulls increases with $z$. This is expected, since larger distance produces wider beams, thus the density of beams decreases.
\item The distance among the nulls increases with $\lambda$. Again, using smaller wavelength is beneficial to concentrate the energy towards a smaller area on the receiving antenna, thus increasing the number of beams.
\item The distance among the nulls decreases with $\lt$. Also, this fact is expected, since larger antennas can concentrate more efficiently the energy towards a smaller area on the receiving antenna, thus increasing the number of beams.
\end{itemize}

Coming back to the general case, we have at the transmitting antenna side
\begin{equation}\label{eq:basisTX_SISLIS}
\phi_n (\eta)= \frac{1}{\sqrt{\lt}}F_T(\eta)\rvert_{y_n}=\frac{1}{\sqrt{\lt}}\rect{\frac{\eta}{\lt}}e^{\jmath\frac{2\pi }{\lambda } \frac{\sin\theta-\gamma_n\cos\theta}{\sqrt{1+\gamma_n^2}} \eta} \,\,\, 
\end{equation}
for $n\in \mathcal{I}$ and with $\gamma_n=y_n/z$ and, at the receiving antenna side
\begin{equation}\label{eq:basisRX_SISLIS}
\psi_n (y)\!=\! K \sinc{\!\frac{\lt}{\lambda}\!\!\left(\!\frac{\sin\theta\!-\!\frac{y}{z}\cos\theta}{\sqrt{1+\left(\frac{y}{z}\right)^2}}-\frac{\sin\theta-\gamma_n\cos\theta}{\sqrt{1+\left(\gamma_n\right)^2\!}}\!\right)}\, 
\end{equation}
where $K$ is a normalization constant needed to obtain an orthonormal basis.\footnote{Its value is of interest only for the calculation of the coupling intensity of each communication mode.}

This result extends the beamspace-{MIMO} approach for an uplink case between a transmitting \ac{SIS} and a receiving \ac{LIS}, beyond the traditional paraxial approximation and considering generic orientations among the antennas. At the transmitting antenna side, due to the small size of the antenna, the \ac{SIS} operates beam steering according to \eqref{eq:basisTX_SISLIS} towards specific angles $\varphi_n=\alpha_n-\theta$ with respect to the transmitting antenna boresight, with $\alpha_n=\arctan{\gamma_n}$. At the receiving antenna side, the \ac{LIS} correlates the impinging \ac{EM} field with the $n$th basis function, that is, a space-varying sinc function according to \eqref{eq:basisRX_SISLIS}, in order to retrieve the information associated to the $n$th communication mode.

\subsubsection{Number of Communication Modes}

\begin{figure}[t!]
\centering
\includegraphics[width=0.5\columnwidth,keepaspectratio=true]{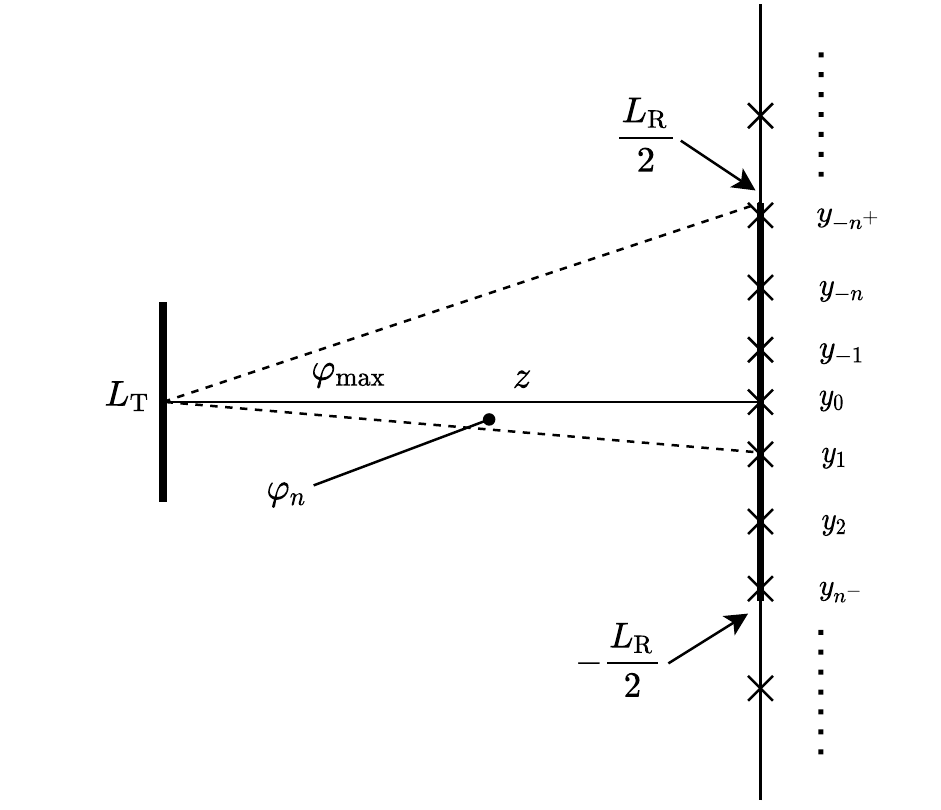}
\caption{Relationship among geometry and nulls in the paraxial case. Transmitting SIS of size $\lt$ (left), receiving LIS of size $\lr$ (right).}
\label{fig:paraxialN}
\end{figure}

As  specified in \eqref{eq:Nexact}, the number of communication modes for a generic configuration corresponds to the cardinality of the set $\mathcal{I}$, that is, the number of nulls of the kernel $K_R(y,\yc)$ following inside the receiving antenna of size $\lr$. We now derive simple closed-form expressions for cases of particular interest.

In the case of parallel antennas ($\theta=0$, $\yc=0$) we have $\rho_{\mathrm{c}}=0$ and \eqref{eq:SISnulls} gives
\begin{align}\label{eq:ykparr}
y_n= z \tan{\left[\arcsin{\left(-\frac{\lambda}{\lt}n\right)}\right]} \, , \quad\quad |n|=1, 2, \ldots \,\, .
\end{align}
The number of nulls $n^+$ in the positive half plane of the receiving antenna ($0<y_n<\lr/2$), according to Fig.~\ref{fig:paraxialN}, corresponding to indices $n<0$, is
\begin{align}\label{eq:condParr}
n^+ = n : y_n < \frac{\lr}{2} \, , \quad\quad n=-1, -2, \ldots \,\, .
\end{align}
Due to the symmetry of the configuration, the number of nulls $n^-$ in the negative half plane of the receiving antenna ($-\lr/2<y_n<0$) is the same. Then, considering also the beam directed towards $y=0$, the number of communication modes is
\begin{align}\label{eq:NParr}
N=1+n^+ + n^- = 1 + \frac{2\lt\lr}{\lambda\sqrt{4z^2+\lr^2}} 
\end{align}
which is obtained by solving \eqref{eq:condParr} with $y_n$ given by \eqref{eq:ykparr}, as showed in Appendix~\ref{App:NParr}.
For a very large receiving antenna (limit for $\lr\rightarrow\infty$) we have\footnote{It should be $N=2\lt/\lambda+1$ however, considering a high frequency so that $\lt\gg\lambda$, here and in the following we neglect the  factor ``+1'' for what the limits are concerned. This can be considered, again, a border effect, as also discussed in footnote~\ref{footnote:Border}.} $N=2\lt/\lambda$, which depends only on the size of the transmitting antenna (i.e., the smallest of the two), which is consistent with the results already found in \cite{Dar:J20}. In fact, an infinite-size receiving antenna, parallel to the transmitting one, would be able to intercept all the beams generated by the \ac{SIS} (showed for example in Fig.~\ref{fig:FarFieldBeams}), leading to the maximum number of communication modes.
Differently, when $z$ is very large, the number of communication modes tends to 1 considering the central beam for $n=0$, which is the only possible for very large distance, as commonly assumed by traditional radio links in the far field. Finally, for $z\rightarrow0$, again the limit $N=2\lt/\lambda$ arises, regardless the size of the receiving antenna. 
Equation \eqref{eq:NParr} generalizes the traditional result \eqref{eq:Ngeneral} to the case of large antennas, thus also for the case of antenna size comparable with the link distance, which it was not accounted by \eqref{eq:Ngeneral}; in particular, it is shown that the number of communication modes cannot increase to arbitrary large values, as erroneously indicated by \eqref{eq:Ngeneral}, but intrinsic limits arise.

In the case of perpendicular antennas ($\theta=\pi/2$, $\yc=0$) we have $\rho_{\mathrm{c}}=1$ and \eqref{eq:SISnulls} gives
\begin{align}\label{eq:ykperp}
y_n= &z \tan{\left[\arcsin{\left(-\frac{\lambda}{\lt}n -1\right)} +\frac{\pi}{2}\right]} = 
  z \cot{\left[\arcsin{\left(\frac{\lambda}{\lt}n +1\right)} \right]}  \, , \quad n=-1, -2, \ldots \,\, .
\end{align}
The number of nulls $n^+$ in the positive half plane of the receiving antenna ($0<y_n<\lr/2$), corresponding to indices $n<0$, is given again by \eqref{eq:condParr}; however, in this case, $n^-=0$, since every beam in the upper hemisphere is always coupled with a beam in the lower hemisphere.\footnote{The antenna was considered as an ideal aperture capable of generating waves in every direction, thus avoiding the need of considering boundary conditions deriving from ground planes. For the symmetry of the Green function, a given phase profile produces a symmetric effect on the two hemispheres around the aperture.} Thus, considering also the beam directed towards $y=0$, we have
\begin{align}\label{eq:NPerp}
N=1+n^+ = 1+\frac{\lt\left[\sqrt{4z^2+\lr^2}-2z\right]}{\lambda\sqrt{4z^2+\lr^2}}
\end{align}
which is obtained by solving \eqref{eq:condParr} with $y_n$ given by \eqref{eq:ykperp}, as showed in Appendix~\ref{App:NPerp}.
For a very large receiving antenna (limit for $\lr\rightarrow\infty$) we have $N=\lt/\lambda$, which depends only on the size of the transmitting antenna (i.e., the smallest of the two), which is reasonable since only a half of the orthogonal beams (showed for example in Fig.~\ref{fig:FarFieldBeams}) can be intercepted by an infinite-size receiving antenna, perpendicular to the transmitting one.
Again, the same limit arises for $z\rightarrow0$, regardless the size of the receiving antenna. 

Expressions for a generic $\theta$ can be found in closed form and are derived in Appendix~\ref{App:NGen}. In particular, we have
\begin{align}\label{eq:Ngeneric}
N\!=\!1\!+\!
    \begin{cases}
      \frac{2\lt}{\lambda}\sin{\left(\varphi_\text{max}\right)}\cos\theta &\theta\leq\frac{\pi}{2}-\frac{1}{2}\varphi_\text{max}\\
      \frac{\lt}{\lambda}\left[\sin{\left(\varphi_\text{max}-\theta\right)}+\sin\theta\right] &\theta\geq\frac{\pi}{2}-\frac{1}{2}\varphi_\text{max}\\
    \end{cases} 
\end{align}
where $\varphi_\text{max}=\arctan{\lr/2z}$ (see Fig.~	\ref{fig:paraxialN}) for $0\leq\theta\leq\pi/2$. When $\theta=0$ and $\theta=\pi/2$, expressions \eqref{eq:NParr} and \eqref{eq:NPerp} are obtained.
For $\lr\rightarrow\infty$ or $z\rightarrow0$, the limit for a generic $\theta$ is given by
\begin{align}\label{eq:Nbound}
N=1+
    \begin{cases}
      \frac{2\lt}{\lambda}\cos\theta &\theta\leq\frac{\pi}{4}\\
      \frac{\lt}{\lambda}\left(\cos\theta+\sin\theta\right) &  \theta\geq\frac{\pi}{4}\, .
    \end{cases} 
\end{align}

\subsubsection{Paraxial Case}

We discuss now more in detail the paraxial case.
When $\theta=0$ and $\yc=0$, the basis set at transmitting antenna side is
\begin{equation}\label{eq:basisTX_SISLISpx}
\phi_n (\eta)= \frac{1}{\sqrt{\lt}}F_T(\eta)\rvert_{y_n}=\frac{1}{\sqrt{\lt}}\rect{\frac{\eta}{\lt}}e^{-\jmath\frac{2\pi }{\lambda } \frac{\frac{y_n}{z}}{\sqrt{1+\left(\frac{y_n}{z}\right)^2}} \eta} \,\,\, 
\end{equation}
with $y_n$ given by \eqref{eq:ykparr} or, in other form,
\begin{equation}\label{eq:ykparass2}
y_n = - z \frac{\lambda}{\lt} \frac{1}{ \sqrt{1-\frac{\lambda^2}{\lt^2}n^2}   } n \,\,\, 
\end{equation}
and, at the receiving antenna side
\begin{align}\label{eq:basisRX_SISLISpx}
\psi_n (y) &= K\, \sinc{\!\frac{\lt}{\lambda}\!\!\left(\!\frac{y_n}{\sqrt{z^2+y_n^2}}-\frac{y}{\sqrt{z^2+{y}^2}}\right)}\, \nonumber \\
&= K\, \sinc{\!-\frac{\lt}{\lambda}\frac{y}{\sqrt{z^2+{y}^2}}-n}\, .
\end{align}
It is also possible to analyze this case in the angular domain. In fact, by using $\varphi_n=\arctan \gamma_n =\arctan(y_n/z)$ (see Fig.~\ref{fig:paraxialN}) we have from \eqref{eq:ykparr}
\begin{equation}\label{eq:phinparax}
\varphi_n = \arcsin{-\frac{\lambda}{\lt}n} \,\,\, .
\end{equation}
Thus, in a half hemisphere, it is 
\begin{align}
n^{\text{max}} = n : 0 < \varphi_n < \frac{\pi}{2}  = \frac{\lt}{\lambda} \,, \quad\quad n=-1, -2, \ldots
\end{align}
leading to the same result on the maximum number of beams introduced in Sec.~\ref{eq:general}.
At the transmitting antenna side, we obtain 
\begin{align}
\phi_n (\eta) &= \frac{1}{\sqrt{\lt}}F_T(\eta)\rvert_{y_n}=\frac{1}{\sqrt{\lt}}\rect{\frac{\eta}{\lt}}e^{-\jmath\frac{2\pi }{\lambda } \frac{{\tan{\varphi_n}}}{\sqrt{1+{\tan^2{\varphi_n}}}} \eta} \nonumber \\ 
&= \frac{1}{\sqrt{\lt}}\rect{\frac{\eta}{\lt}}e^{-\jmath\frac{2\pi }{\lambda } \eta \sin{\varphi_n} }\nonumber \\ 
\end{align}
which highlights the beam steering phase profile towards angles $\varphi_n$. Thus, by considering \eqref{eq:phinparax}, it is
\begin{equation}
\phi_n (\eta)= \frac{1}{\sqrt{\lt}}\rect{\frac{\eta}{\lt}}e^{\jmath\frac{2\pi }{\lt } n \eta}
\end{equation}
and the orthogonality of the basis functions is evident as integer multiple of a phasor.
The corresponding beams can be written in the angular domain as $\psi_n(\varphi)$, where $\varphi=\arctan(y/z)$, obtaining from \eqref{eq:basisRX_SISLISpx}
\begin{align}\label{eqFarFieldPattern}
\psi_n (\varphi) &\propto \sinc{\!\frac{\lt}{\lambda}\!\!\left(\frac{\tan\varphi_n}{\sqrt{1+\tan^2\varphi_n}}-\frac{\tan\varphi}{\sqrt{1+\tan^2\varphi}}\right)}\nonumber \\
&= \sinc{\!-\frac{\lt}{\lambda}\sin\varphi-n} \, 
\end{align}
which, for $n=0$, is the far field antenna pattern of an ideal linear aperture \cite{Bal:B15}. Then, exploiting the angular domain and the paraxial geometry, the maximum number of communication modes can be found by considering the largest angle sustained by the receiving antenna, which is $\varphi_\text{max}=\arctan{\lr/2z}$ (see Fig.~\ref{fig:paraxialN}). Thus, it is
\begin{align}
n^+ = n : \varphi_n < \varphi_\text{max} 
\end{align}
whose solution leads to the same result \eqref{eq:NParr}. Notice that, since the \ac{LIS} is considered in the far field of the \ac{SIS}, whose angular pattern results independent on the distance, the knowledge of the orientation only among the antennas is required to define the communication modes at the transmitter side, instead of the complete definition of the geometry.

\subsection{Communication Modes between a LIS and a SIS: Downlink}\label{sec:LISSIS}

When communicating between a transmitting \ac{LIS} and a receiving \ac{SIS}, that is in the downlink case (dual case with respect to the discussion of Sec.~\ref{sec:SISLIS}), the number of communication modes is of course the same, due to the reciprocity of the radio link, and the same basis functions can be adopted. 

However, the use of the approach based on focusing at the transmitting side may bring some advantages from the practical point of view and the possibility to exploit the method described in Sec.~\ref{sec:BasisGen} both in uplink and downlink.\footnote{This translates also in the possibility of using phase-tapering only at the transmitting antenna side.} For this reason, in this section we investigate the downlink case, where focusing is realized with the transmitting \ac{LIS} towards the receiving \ac{SIS}. In this case, the receiving \ac{SIS} will be likely in the near-field region of the transmitting \ac{LIS} (see discussion in Sec.~\ref{sec:zones}), so focusing will be realized toward the \ac{SIS} and we cannot adopt the Maclaurin expansion \eqref{eq:SISapprox} due to the large size of the transmitting antenna.
However, for a parallel \ac{SIS} on the \ac{LIS}'s boresight ($\theta=0$, $\yc=0$), it is possible to exploit the Fresnel approximation \cite{God:B05}. Specifically, when adopting such an approximation, it is possible to expand the numerators (phase terms) $\sqrt{z^2+\left({y-\eta}\right)^2}$ in \eqref{eq:kernelJ}-\eqref{eq:kernelJ3} using the Maclaurin series at the first term with respect to $(y-\eta)/z$, as
\begin{equation}\label{eq:fresnel}
z\sqrt{1+\left(\frac{y-\eta}{z}\right)^2} \approx z\left[1+\frac{1}{2}\left(\frac{y-\eta}{z}\right)^2\right] \, .
\end{equation}
Using \eqref{eq:fresnel}, the focusing function $F_T(\eta)\rvert_{y}$ can be written as
\begin{equation}\label{eq:focusingFresnel}
F_T(\eta)\rvert_{y}=\rect{\frac{\eta}{\lt}}e^{\jmath\frac{\pi }{\lambda z} \eta^2} e^{-\jmath\frac{2 \pi y}{\lambda z} \eta} 
\end{equation}
where, again, all the phase terms independent of $\eta$ have been discarded, since the addition of a constant phase shift would not change the focusing behavior. In particular, we have that the focusing function is composed of a quadratic term and a linear term.\footnote{Properties of focused continuous apertures have been studied since a long time, see for example \cite{She:J62}.} When the focusing point is along the boresight direction (i.e., at $y=0$), the only quadratic phase term is retained, and the focusing function is simply $F_T(\eta)=F_T(\eta)\rvert_{y=0}=\rect{\frac{\eta}{\lt}}e^{\jmath\frac{\pi }{\lambda z} \eta^2}$, which is consistent with standard definitions (see, for example \cite{Mil:J00, HanFu:06}). Thus, when a receiving \ac{SIS} is located in the near field of the \ac{LIS} and with its center along the boresight direction (paraxial case), we have that
\begin{equation}
\phi_0(\eta)=\frac{1}{\sqrt{\lt}} F_T(\eta)=\frac{1}{\sqrt{\lt}}  \rect{\frac{\eta}{\lt}}e^{\jmath\frac{\pi }{\lambda z} \eta^2} \, .
\end{equation}
A second orthonormal function can be  constructed starting from \eqref{eq:kernelJ3}. We obtain
\begin{align}
&\int_{-\lt/2}^{\lt/2} \left(F_T(\eta)\rvert_{y} \right)^* F_T(\eta) \, d\eta\, =  \int_{-\lt/2}^{\lt/2} e^{\jmath\frac{2\pi y}{\lambda z}\eta}\, d\eta \propto \sin{\frac{\pi y \lt}{\lambda z}} \nonumber
\end{align}
which gives nulls (orthogonal condition) for
\begin{equation}\label{eq:nullsTXFresnel}
y_n=n\frac{\lambda z}{\lt}\, ,\quad\quad |n|=1,2,\ldots \, .
\end{equation}
Under the Fresnel approximation, focusing functions in the form \eqref{eq:focusingFresnel} establish, with proper choices of $y$ as in \eqref{eq:nullsTXFresnel} and when normalized by $\sqrt{\lt}$, an orthonormal basis set, whose $n$th element is
\begin{equation}\label{eq:TXbasisFresnel}
\phi_n (\eta)= \frac{1}{\sqrt{\lt}}  \rect{\frac{\eta}{\lt}} e^{\jmath\frac{\pi }{\lambda z} \eta^2} e^{-\jmath\frac{2 \pi }{\lt} n \eta} \, , \quad n\in \mathcal{I}\,
\end{equation}
where $\mathcal{I}$ is the set of points $y_n$ in \eqref{eq:nullsTXFresnel} following inside the SIS.

We now compute the field at the receiving antenna given by basis functions in the form \eqref{eq:TXbasisFresnel}. By exploiting \eqref{eq:distribution} and approximation \eqref{eq:fresnel}, we have
\begin{align}\label{eq:RXfuncFresnel}
& \psi(y)\rvert_{y_n} = \nonumber \\
&\frac{1}{4\pi z \sqrt{\lt}}  e^{-\jmath\frac{\pi}{\lambda}\left(2z+\frac{y^2}{z}\right)} \int_{-\infty}^{\infty}  \rect{\frac{\eta}{\lt}} e^{\jmath{2\pi}{}\frac{y}{\lambda z}\eta} e^{-\jmath\frac{2 \pi }{\lt} n\eta}\, d\eta = \nonumber \\
&\frac{1}{4\pi z \sqrt{\lt}}  e^{-\jmath\frac{\pi}{\lambda}\left(2z+\frac{y^2}{z}\right)} \mathcal{F} \left\{\rect{\frac{\eta}{\lt}} e^{-\jmath\frac{2 \pi }{\lt} n \eta} e^{2 \jmath2\pi f_Y \eta} \right\}\biggr\rvert_{f_{Y}=\frac{y}{\lambda z}}  \nonumber \\
& = \frac{\sqrt{\lt}}{4\pi z } e^{-\jmath\frac{\pi}{\lambda}\left(2z+\frac{y^2}{z}\right)} \sinc{-\frac{\lt}{\lambda z}y-n}
\end{align}
where $\mathcal{F}(\cdot)$ indicates the Fourier transform and where now we have $d_c=z$ due to the paraxial approximation. 
Since the exponential term on the left-hand side of \eqref{eq:RXfuncFresnel} can be considered approximatively constant when $y\ll z$ (small receiving antenna), it  is possible to notice that \eqref{eq:RXfuncFresnel} defines a set of orthogonal functions at the receiving antenna side. The $n$th function is centered at the point $y_n$ given by \eqref{eq:nullsTXFresnel} where the corresponding focusing function was pointing, and the same corresponds to the nulls of the adjacent functions. Then we have
\begin{equation}\label{eq:RXbasisFresnel}
\psi_n (y)= K\,  \sinc{-\frac{\lt}{\lambda z}y-n}  \, , \quad n\in \mathcal{I}\, .
\end{equation}
Notice that two adjacent basis functions at the receiving antenna side are spaced by $\frac{\lambda z}{\lt}$. Thus, considering a receiving antenna of size $\lr$, centered at $\yc=0$, we have a maximum number of communication modes given by \eqref{eq:Ngeneral}. After such a number $N$, the focal points would fall outside the receiving antenna, thus making impossible to define a further communication mode with a significant level of coupling. This is consistent with traditional results based on paraxial approximation (see \cite{Mil:J19}), but it has been found starting with the specific choice of focusing functions as basis set at transmitting antenna side, instead of resorting to the exact solution of the eigenfunction problem.\footnote{Thus, the same limitations of the traditional formulation arise, in the case the transmitting antenna becomes very large (number of communication modes overestimated).}

Interestingly, by writing $\tan\varphi=y/z$, it is possible to write the basis functions at the receiving antenna side \eqref{eq:RXbasisFresnel} in the angular domain, obtaining
\begin{align}\label{eq:RXbasisFresnelangle}
\psi_n (\varphi) &\propto  \sinc{-\frac{\lt}{\lambda }\tan\varphi-n}  \approx  \sinc{-\frac{\lt}{\lambda }\sin\varphi-n}  \, ,\quad n\in \mathcal{I}\,
\end{align}
where the approximation holds for the condition $y\ll z$, that is for small angles. 
With \eqref{eq:RXbasisFresnelangle} it is possible to notice that, using focusing, the same angular pattern of a (small) antenna of size $\lt$ is obtained, as in \eqref{eqFarFieldPattern}, regardless we are observing the angular pattern in the near field. Basically, focusing modifies the beams related to the Fresnel diffraction patterns typical of the near field (e.g., that of Fig.~\ref{fig:FarFieldBeams}), thus an angular pattern changing with the distance from the antenna, into an angular pattern corresponding to the same obtained in the far field (e.g., that of Fig.~\ref{fig:FarFieldBeams}). As the beam width in the far field has an inverse relationship with the antenna size $\lt$, using focusing the width of the focal spot has an inverse relationship with the antenna size $\lt$ (wider antenna, more concentrated focal spot).

\smallskip
\section{Antenna Operating Zones and Communication Modes}\label{sec:zones}
Traditionally, the space around a transmitting antenna is divided into several regions, depending on the characteristics of the field emitted by the antenna itself. The closest area to the antenna corresponds to the reactive near field, where reactive field components from the source antenna cannot be neglected \cite{Bal:B15}. Then, by increasing the distance from the antenna, we have the radiating near field (Fresnel region), and the far field (Fraunhofer region), conventionally separated by the Fraunhofer distance $r_{\text{ff}}$ introduced in Sec.~\ref{eq:general}. 
In the far field:
\begin{itemize}
\item The angular pattern does not depend on the distance;
\item Rays\footnote{We define a ray as a line that is perpendicular to the wavefront, and that points in the direction of the energy radiated by the antenna (i.e., the receiving antenna).} can be considered almost parallel;
\item The wavefront can be considered approximately plane.
\end{itemize}
In the near field:
\begin{itemize}
\item The angular pattern does depend on the distance\footnote{As discussed in Sec.~\ref{sec:LISSIS} the angular patter does not change with the distance when a focusing phase profile is implemented at the transmitting antenna side.};
\item Rays cannot be considered parallel;
\item The wavefront is approximately spherical.
\end{itemize}

The focusing capability exploited in Sec.~\ref{sec:basisdef} for the practical definition of the communication modes is known to be a feature which can be exploited when the focal point is located in the Fresnel region \cite{NepBuf:J17}. Differently from the traditional definition of $r_{\text{ff}}$, which is a function of the antenna size and of the operating frequency, in the derivation of Sec.~\ref{sec:basisdef} several approximations regarded the relation among the size of the antenna and the link distance, regardless the operating frequency. Moreover, when a \ac{SIS} and a \ac{LIS} are communicating, as assumed in Sec~\ref{sec:SISLIS} and Sec.~\ref{sec:LISSIS}, the boundaries between the respective near-field and far-field regions will be severely different for the two antennas, since scaling with the square of the antenna size. Thus, the aim of this section is that of discussing the relationship among the traditional definitions of the regions of space around antennas, and the number of communication modes that can be realized.
As a rule of thumb, a transmitting antenna will adopt a beam steering phase profile (i.e., linear phase to concentrate the energy towards a given angle $\varphi$) when transmitting to a receiver located in its far-field region (i.e., above $r_{\text{ff}}$) and a focusing phase profile (i.e., a phase profile to concentrate the energy towards a specific point in the space $(z,y)$)  when transmitting to a receiver located in its near-field region (i.e., below $r_{\text{ff}}$). The focusing capability, thus the possibility of establishing multiple communication modes, decreases quickly in the case of:
\begin{itemize}
\item Large distance between the transmitter and receiver.
\item Large angle with respect to the boresight direction.
\item Small antennas.
\item Low frequency.
\end{itemize}
%
%
Of course, $r_{\text{ff}}$ is only a conventional boundary not fixed in space, and the characteristics of the \ac{EM} field generated by an antenna change smoothly with the distance. 
Depending on the relative dimension of these zones for both the transmitting and receiving antennas, a different number of communication modes will be realized.

Let us now consider the paraxial case, from which some insights can be  drawn in an easier way. By expressing the size of the transmitting and receiving antenna as a function of their respective boundaries between near field and far field, that is
\begin{align}
\lt=\sqrt{\frac{\lambda\, r_{\text{ff}}^{(T)}}{2}}\, ,  \quad\quad \lr=\sqrt{\frac{\lambda\, r_{\text{ff}}^{(R)}}{2}}
\end{align}
it is possible to write relation \eqref{eq:Ngeneral} as
\begin{align}
N=\frac{\sqrt{r_{\text{ff}}^{(T)}}\sqrt{r_{\text{ff}}^{(R)}}}{2z}\, .
\end{align}
If two antennas of the same size are used (e.g., for a point-to-point radio link), so that $r_{\text{ff}}^{(T)}=r_{\text{ff}}^{(R)}=r_{\text{ff}}$, it is immediate to observe that a single communication mode ($N=1$) is obtained at least up to $z=r_{\text{ff}}/2$. Thus, for antennas of the same size, placing them within their boundary between near field and far field is a necessary condition to establish multiple communication modes. In general, $N>1$ can be realized when $z<\lt\lr/\lambda$ or, equivalently, multiple communication modes can be established only if antennas are placed at a distance much smaller than their boundary between near field and far field.
As presented in Sec.~\ref{sec:SISLIS}, a transmitting \ac{SIS} can also establish multiple communication modes with a receiving \ac{LIS}, by considering a beam steering phase profile (i.e., linear phase); in fact, despite the receiving \ac{LIS} is likely in the far field of the transmitting \ac{SIS}, the transmitting \ac{SIS} is likely in the near field of the \ac{LIS}. Then, to  correctly determine the number of communication modes that can be established by a couple of antennas of any dimension, it is fundamental to consider the operating zones of both transmitting and receiving antennas, although traditionally the near field region is referred to the the transmitting one. %

Regardless the condition for exploiting multiple communication modes, Sec.~\ref{sec:SISLIS} has highlighted how, when one antenna is much larger than the other, the traditional relation \eqref{eq:Ngeneral} is not valid anymore, and a much more accurate expression is \eqref{eq:NParr}. This is reasonable, since the number of communication modes cannot grow to any possible value if the distance $z$ is reduced or very large \acp{LIS} are adopted. Differently, the two relations are equivalent when the distance is large, and in particular when $\lr\ll z$.  By comparing \eqref{eq:Ngeneral} and \eqref{eq:NParr}, we have that, for example,  
at $z=\lr/2$ the actual number of communication modes is a half of that predicted by \eqref{eq:Ngeneral}, while at $z=\lr/(2\sqrt{2})$ it is only one third of that. This condition depends on geometrical aspects only and it is not a function of the operating frequency.

\begin{figure}
\centering
\hspace*{-1cm}                                                           
\includegraphics[width=0.5\columnwidth,keepaspectratio=true]{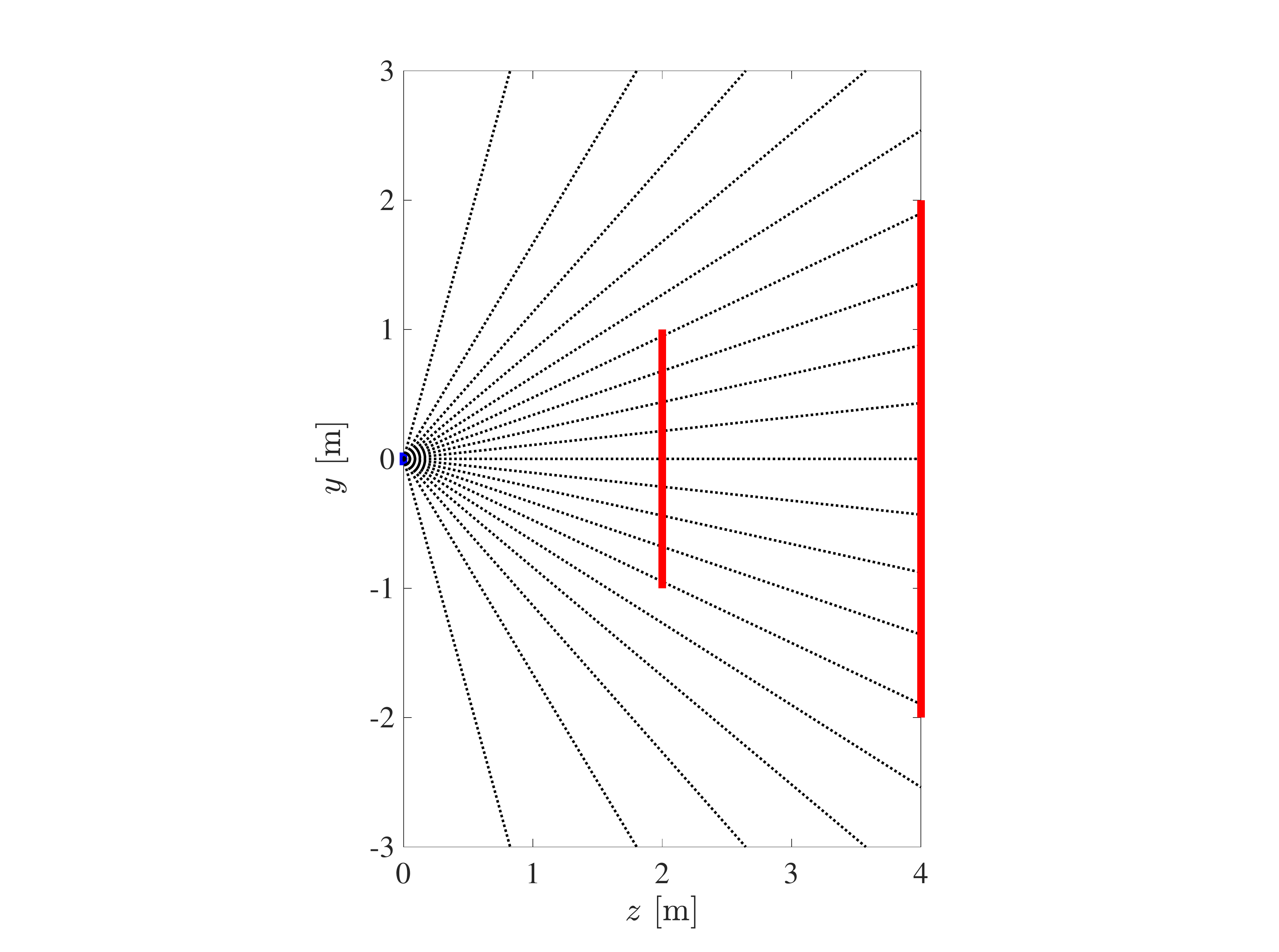}
\caption{Directions $\varphi_n$ of the orthogonal beams for a SIS of size $\lt=10\,$cm (on the left at $(z,y)=(0,0)$), in the same configuration of Fig.~\ref{fig:FarFieldBeams}. Two LISs characterized with the same ratio $F=z/\lr=1$ are shown (in red), highlighting that the same number $N=9$ of communication modes can be realized (intercepted beams).}
\label{fig:BeamPoints}
\end{figure}

Analyzing the expression for the number of communication modes here derived, it can be noticed that, in \eqref{eq:NParr}, the number of communication modes is not a function of $\lr$ and $z$ themselves, but of their ratio $\lr/z$. Specifically, it is possible to write \eqref{eq:NParr} as
\begin{equation}\label{eq:NparrF}
N=1+\frac{2\lt}{\lambda\sqrt{1+4F^2}}
\end{equation} 
with $F=z/\lr$. The same number of communication modes can be obtained by keeping such a ratio constant. This fact can be easily explained in the angular domain. In fact, we have $\lr/(2z)=\tan\varphi_{\text{max}}$ according to Fig.~\ref{fig:paraxialN}. Since the \ac{LIS} was supposed located in the far field of the \ac{SIS} (hypothesis of small transmitting antenna), the angular pattern of the \ac{SIS} does not change with $z$, so that the same number of angular beams can be intercepted by keeping the ratio $F$ constant. As example, Fig.~\ref{fig:BeamPoints} reports the directions $\varphi_n$ of the orthogonal beams of Fig.~\ref{fig:paraxialN} for the same setting of Fig.~\ref{fig:FarFieldBeams} ($N=19$) and two receiving antennas with $F=1$, leading to the same number $N=9$ of communication modes, as predicted by \eqref{eq:NparrF}.
Similarly, for perpendicular antennas, we have
\begin{equation}\label{eq:NperpF}
N=1+\frac{\lt \left(\sqrt{1+4F^2}-2F\right)}{\lambda\sqrt{1+4F^2}}\, .
\end{equation} 

Summarizing:
\begin{enumerate}
\item Operating in the near field is a necessary condition to establish multiple communication modes; in particular, operating in the near field should be intended as the inclusion of at least one antenna in the near field of the other one. The design variables for exploiting multi-mode communication are both the size of antennas and the operating frequency.
\item When the link distance becomes comparable with the size $D$ of the largest antenna (e.g., $z<D$), ad-hoc models for evaluating the number of communication modes must be considered and traditional results, for example, \eqref{eq:Ngeneral}, fail.
\item For very large \acp{LIS} or very small distance among the antennas, the number of communication modes depends mainly on the size of the smallest antenna only; in the general case, the number of communication modes is a function of the ratio $F=z/D$, where $D$ is the size of the largest antenna.
\end{enumerate}
The second condition has been found recently also for results related to the path loss involving \acp{LIS} and/or large antenna arrays and is defined as \textit{geometric near field} \cite{BjoSan:20, Dar:J20}.

\bigskip
\section{Examples}\label{Sec:results}
In this section, some numerical results are presented in order to discuss the proposed methods.
Results are compared with numerical solutions obtained using \ac{SVD}. In this case, the transmitting and receiving antenna are discretized into a fine mesh, then the Green function is evaluated and its \ac{SVD} decomposition computed, leading to the transmitting and receiving basis functions. The number of communication modes is obtained as the number of singular values whose sum corresponds to the $99\%$ of the overall coupling gain.

\smallskip
\subsection{Number of Communication Modes}
\begin{figure}
\centering
\includegraphics[width=0.5\columnwidth,keepaspectratio=true]{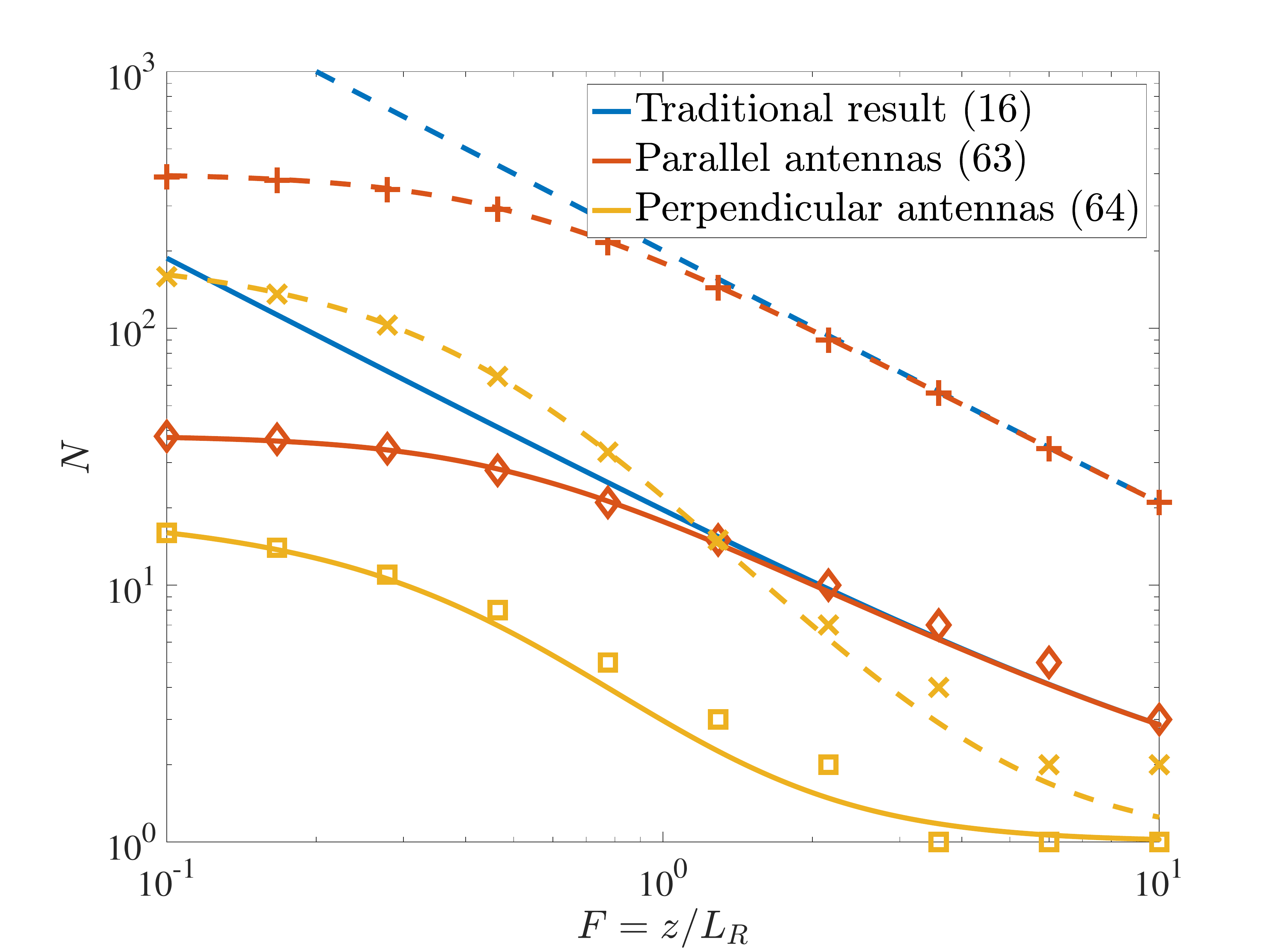}
\caption{Number of communication modes as a function of $F=z/\lr$ for parallel and perpendicular antennas, in comparison with traditional result \eqref{eq:Ngeneral}. SIS of size $\lt=20\,$cm. Continuous lines (--) refer to the millimeter-wave band ($f_0=28\,$GHz); dashed lines (-$\,$-)  refer to the terahertz band ($f_0=300\,$GHz). Markers refer to the number of communication modes obtained with SVD.}
\label{fig:Ncurve}
\end{figure}

Figure~\ref{fig:Ncurve} reports the number of communication modes for communication between a SIS and a LIS; results are obtained with the derived expressions \eqref{eq:NParr} and \eqref{eq:NPerp},  (or \eqref{eq:NparrF} and \eqref{eq:NperpF}). Two frequency ranges are considered, corresponding to the millimeter-wave band ($f_0=28\,$GHz) and terahertz band ($f_0=300\,$GHz). The traditional result valid for large distance and paraxial condition \eqref{eq:Ngeneral} is reported for comparison. It can be noticed that the actual number of communication modes diverges from that predicted by \eqref{eq:Ngeneral} when the size of the largest antenna becomes comparable with the link distance, thus entering the geometric near field ($F\approx 1$). In particular, for very short distance or very large LIS (small $F$), the limits $2\lr/\lambda$ and $\lr/\lambda$ arise for parallel and perpendicular antennas, respectively. Differently, at large distance or with a small antenna (large $F$) a single mode is obtained, especially for the lower frequency. Notice that a large number of communication modes can be exploited when communicating with SIS and a LIS of size comparable with the link distance (e.g., ~16 communication modes in the millimeter-wave band and ~170 communication modes in the terahertz band), thus enabling a large improvement of the channel capacity, even in \ac{LOS} conditions. In Fig.~\ref{fig:Ncurve},  the markers report the number of significative singular values obtained with the numerical \ac{SVD}. It is possible to see the good agreement between the number of practically-orthogonal beams obtained for a specific antenna size and orientation, and the number of significative singular values. This confirms that the proposed beamspace modeling, and then the expressions derived, can be adopted as general design formulae capable of overcoming the limits of the traditional results related to paraxial approximation, without resorting to numerical evaluations and optimal basis functions which might lead to implementation issues.

\begin{figure}
\centering
\includegraphics[width=0.5\columnwidth,keepaspectratio=true]{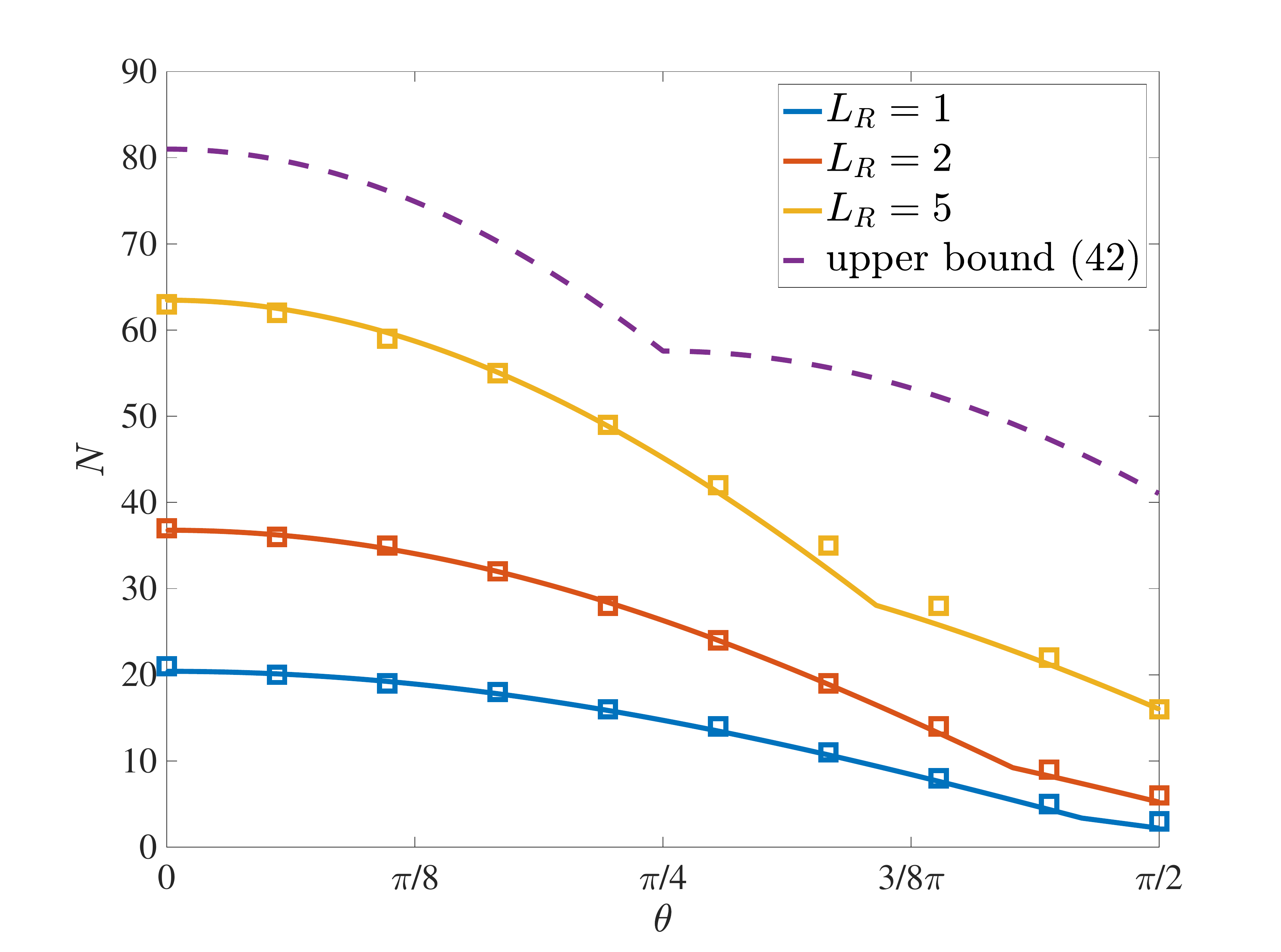}
\caption{Number of communication modes as a function of the orientation $\theta$ among the antennas for different size of the LIS, considering a SIS of size $\lt=20\,$cm, $z=2\,$m and $f_0=60\,$GHz. Markers refer to the number of communication modes obtained with SVD.}
\label{fig:NcurveGen}
\end{figure}

When a generic orientation $\theta$ among the antennas is considered, the result \eqref{eq:Ngeneric} is reported in Fig.~\ref{fig:NcurveGen}. A SIS of size $\lt=20\,$cm is considered and a frequency $f_0=60\,$GHz. It is possible to notice that the number of communication modes is minimum when antennas are perpendicular (most unfavorable condition), while it is maximum when antennas are parallel (most favorable condition).  On each curve, for a specific value of $\lr$, the left-hand limit corresponds to the value $N$ given by \eqref{eq:NParr}; differently, the right-hand limit corresponds to the value $N$ given by \eqref{eq:NPerp}. With the increasing size of $\lr$ the curves move towards the upper side of the graph (larger number of communication modes). When the size of the LIS becomes large, the limit $2\lt/\lambda$ arises for parallel antennas (left-hand limit for the upper bound curve) and the limit $\lt/\lambda$ arises for perpendicular antennas (right-hand limit). Also in this case, we have a the good agreement between the number of beams obtained for a specific antenna size and orientation, and the number of significative singular values obtained with \ac{SVD}  (square markers).\footnote{Expressions are less tight for angles close to $\pi/4$ when very large surfaces are used.} 

\subsection{TX/RX Basis Functions}

In this section, the approximate basis functions (hereinafter beams) obtained with the proposed method are compared with the optimal basis functions at transmitting and receiving antenna side computed with \ac{SVD}.

Table~\ref{tab:Scenario1} compares the beams obtained considering a transmitting SIS and a receiving LIS (first row - uplink scenario) and the beams obtained considering a transmitting LIS and a receiving SIS (second row - downlink scenario), constructed according to the method described in Sec.~\ref{sec:BasisGen}. The two left-hand columns report the functions $\phi_n(\eta)$ and $\psi_n(y)$ (in amplitude and phase) at the SIS side of size $20\,$cm; the two right-hand columns report the functions $\psi_n(y)$ and $\phi_n(\eta)$ (in amplitude and phase) at the LIS side of size $1\,$m. As discussed, the antenna at transmitting side operates phase-tapering only, so the amplitude function is not reported in this case (constant functions). In this table, a paraxial condition is considered, with $z=5\,$m. 
It is possible to see that the SIS uses a phase profile (i.e., functions $\phi_n(\eta)$) almost linear as expected (1b), corresponding to beam steering, and a small curvature can be noticed due to the relative short distance of the focal points on the receiving LIS, which is in the near-field of the SIS (corresponding to $7.5\,$m thus larger than $z$). In fact, $N=3$ beams (functions $|\psi_n(y)|$ in 1c) are realized on the receiving antenna exploiting the proposed method.\footnote{We can notice that, at the LIS side, a parabolic phase profile is obtained when beam steering is operated at the SIS side, which is not accounted by the analytic expression \eqref{eq:basisRX_SISLIS} and it is due to the first order approximation for both the phase profile at the SIS side and the Green function in \eqref{eq:distribution}.} At the transmitting antenna side, the worst-case cross-correlation among the three functions $\phi_n(\eta)$ is $-65\,$dB; at the receiving antenna side it is $-25\,$dB, thus interesting for practical applications.
When changing the role between transmitter and receiver (i.e., downlink), the transmitting LIS realizes focusing (functions $\phi_n(\eta)$ in 2d) and $N=3$  beams (functions $|\psi_n(y)|$ in  2a) are realized on the receiving SIS (same number of the uplink, as expected). At the transmitting antenna side, the worst-case cross-correlation among the three three functions $\phi_n(\eta)$ is $-43\,$dB; at the receiving antenna side it is $-25\,$dB.
The third row of figures shows the exact solutions (optimal basis functions) obtained with \ac{SVD} (3a-3d). In this case, the functions tend to occupy all the region of space of antennas, and the number of beams for each communication mode corresponds to the mode index. Moreover, it can be noticed that the same shape (in amplitude) is obtained at the SIS and LIS sides, despite the different size of the antennas. Both amplitude and phase tapering are exploited both at transmitting and receiving antenna side, and perfect orthogonality is realized. 

Then, Table~\ref{tab:Scenario2}  compares the beams obtained using the proposed method for uplink and downlink in a non-paraxial scenario. In particular, the same size of antennas of Table~\ref{tab:Scenario1} are considered, but with an orientation $\theta=\pi/4$ and at $z=2\,$m. The two left-hand columns report the functions $\phi_n(\eta)$ and $\psi_n(y)$ (in amplitude and phase) at the SIS side of size $20\,$cm; the two right-hand columns report the functions $\psi_n(y)$ and $\phi_n(\eta)$ (in amplitude and phase) at the LIS side of size $1\,$m. In this case, it is possible to see that the $N=6$ beams at the receiving antenna side (1c) are obtained using linear phase profiles at transmitting antenna side (1b). In fact, the receiving LIS is not located on the antenna boresight, resulting in a less pronounced focusing capability. It is also evident the stretching of the lobe widths, which changes in the different parts of the LIS, resulting in a number of communication modes that cannot be predicted by traditional formulations related to paraxial approximation and small antennas, assuming intrinsically beams of constant width.
At the transmitting antenna side the worst-case cross-correlation among the six functions $\phi_n(\eta)$ is $-32\,$dB; at the receiving antenna side it is $-25\,$dB.
When changing the role between transmitter and receiver (i.e., downlink), the transmitting LIS realizes focusing (2d) and $N=7$ beams (2a) are realized on the receiving SIS.\footnote{The number is slightly different from the $N=6$ of the uplink due to the already-discussed border effects.} At the transmitting antenna side the worst-case cross-correlation among the seven functions $\phi_n(\eta)$ is $-21\,$dB; at the receiving antenna side it is $-14\,$dB.
The third row of figures shows the solutions obtained with \ac{SVD} (3a-3d). Interestingly, in this case of non-paraxial scenario, the beams obtained on the LIS (3c) are similar to that realized with the proposed method, and they experience the same stretching of the widths.
Although partially overlapped, every communication mode is characterized by a single beam on a specific point of the antenna, which is similar to the approach proposed.  

\begin{table*}[]
\caption{Example of TX/RX basis functions for uplink and downlink, considering a SIS of $20\,$cm, a LIS of $100\,$cm, $z=5\,$m, $\theta=0$ (paraxial case), $f_0=28\,$GHz. Comparison with optimal basis functions obtained with SVD.}
\label{tab:Scenario1}
\centering
\begin{tabular}{ c | c c l c c }
& \multicolumn{2}{c}{Small Intelligent Surface} &  \multicolumn{2}{c}{Large Intelligent Surface}\\
\hline
\\
\rotatebox{90}{UPLINK }&
(constant amplitude)&
\includegraphics[width=0.2\textwidth,keepaspectratio=true]{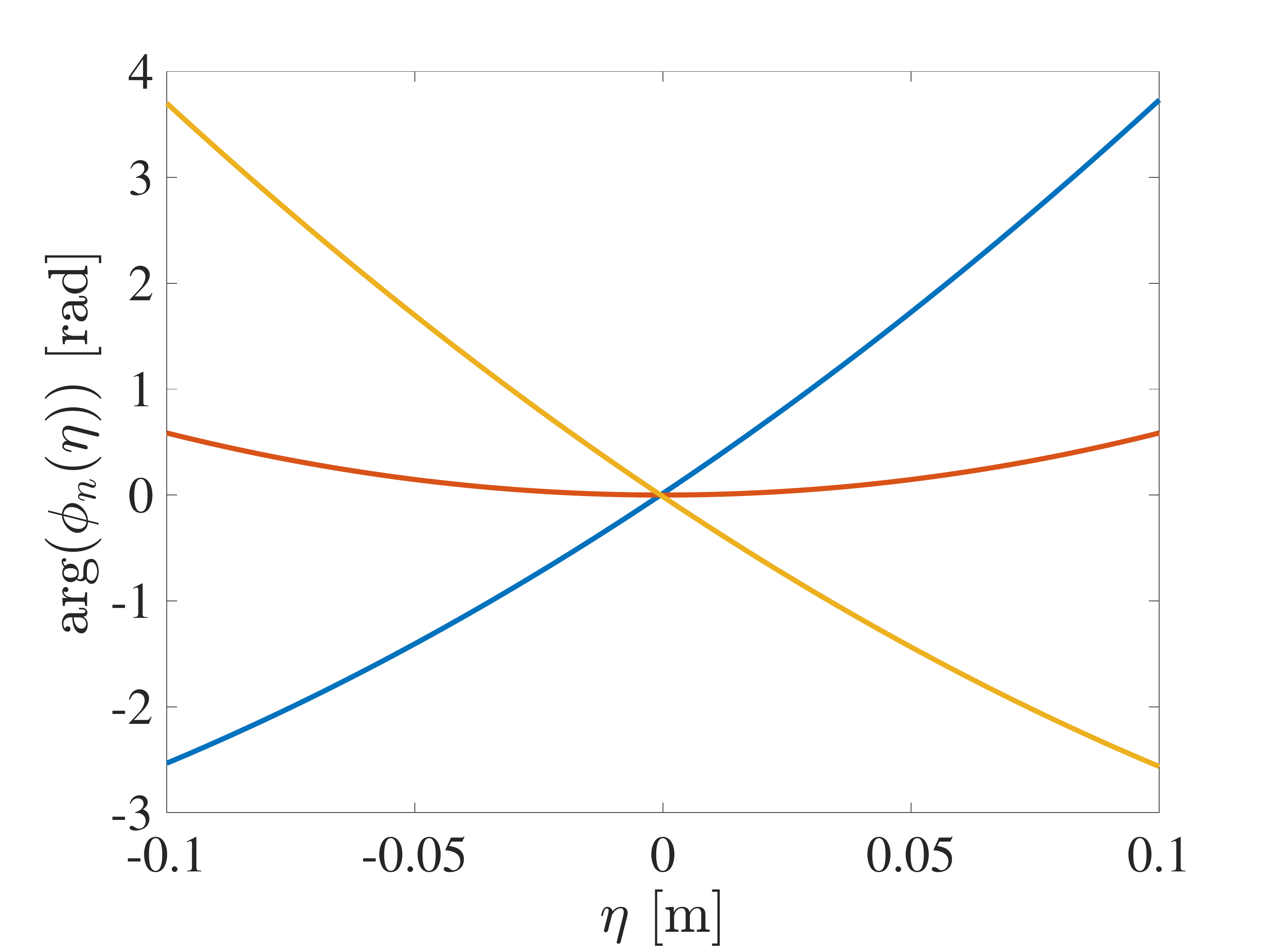}& 
\includegraphics[width=0.2\textwidth,keepaspectratio=true]{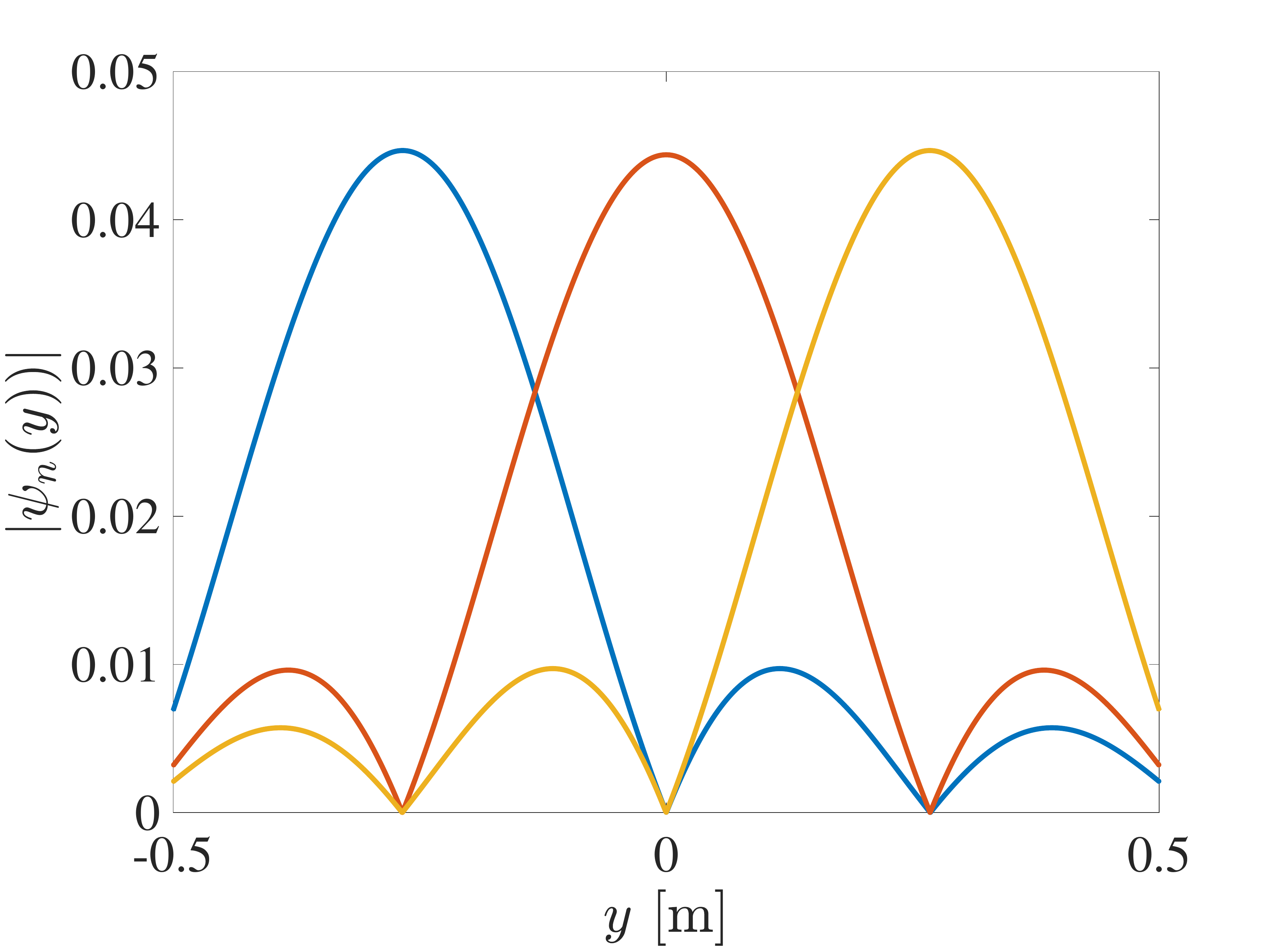}&
\includegraphics[width=0.2\textwidth,keepaspectratio=true]{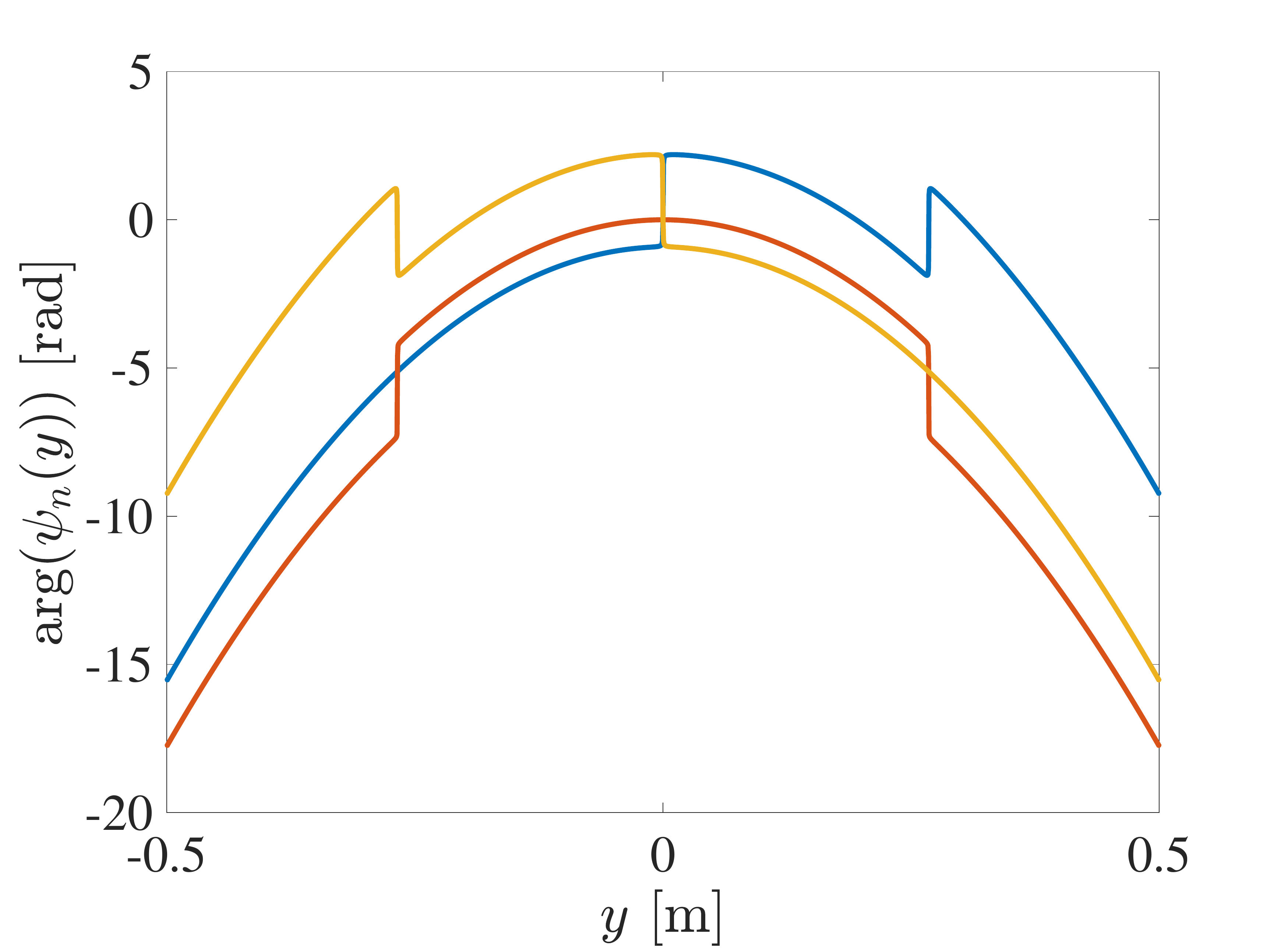}&\\
& (1a) & (1b) & $\quad\quad\quad\quad\quad\quad$(1c) &(1d)\\
\hline\\
\rotatebox{90}{DOWNLINK }&
{\includegraphics[width=0.2\textwidth,keepaspectratio=true]{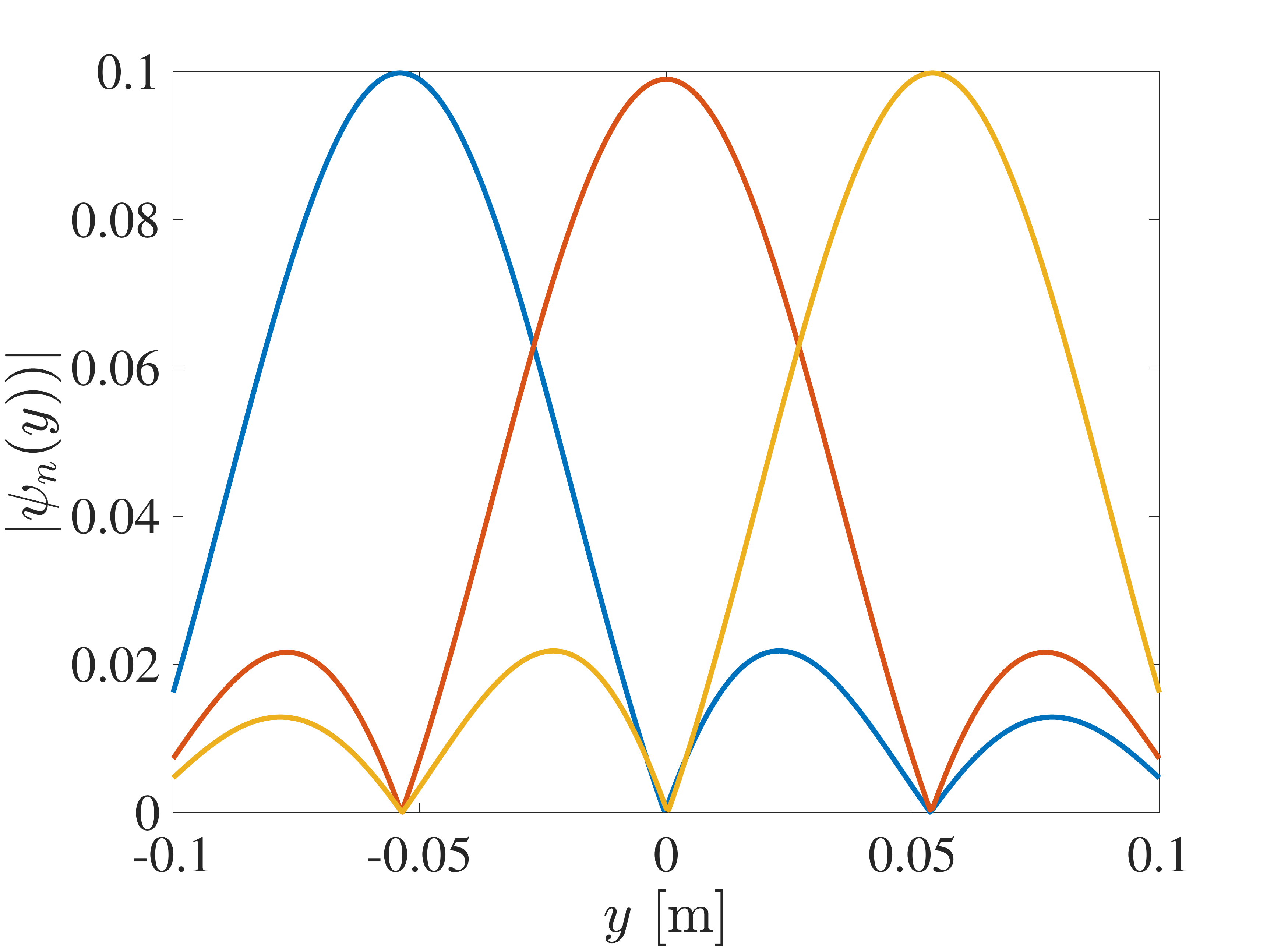}}&
{\includegraphics[width=0.2\textwidth,keepaspectratio=true]{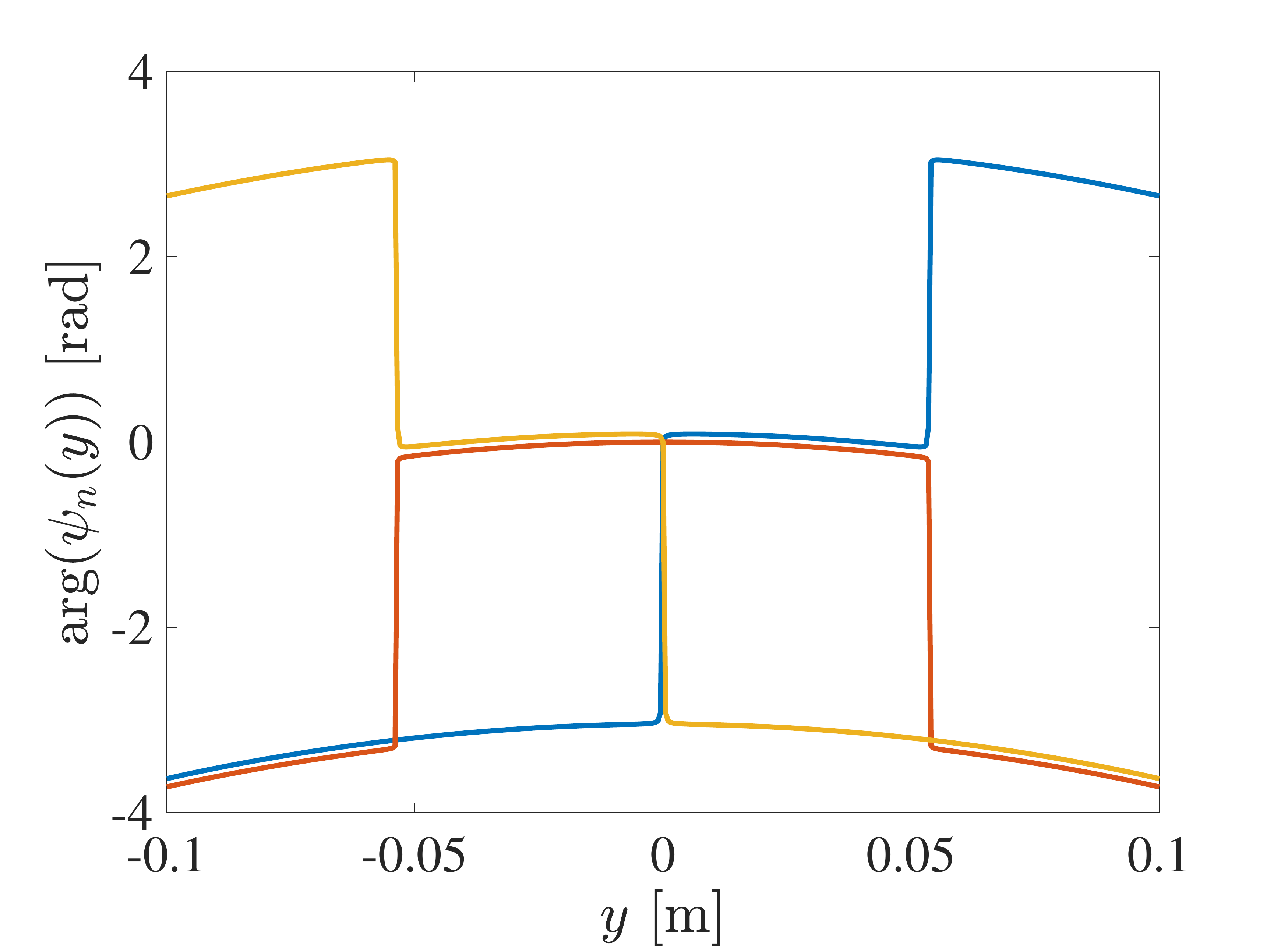}}&
\quad\quad(constant amplitude)&
{\includegraphics[width=0.2\textwidth,keepaspectratio=true]{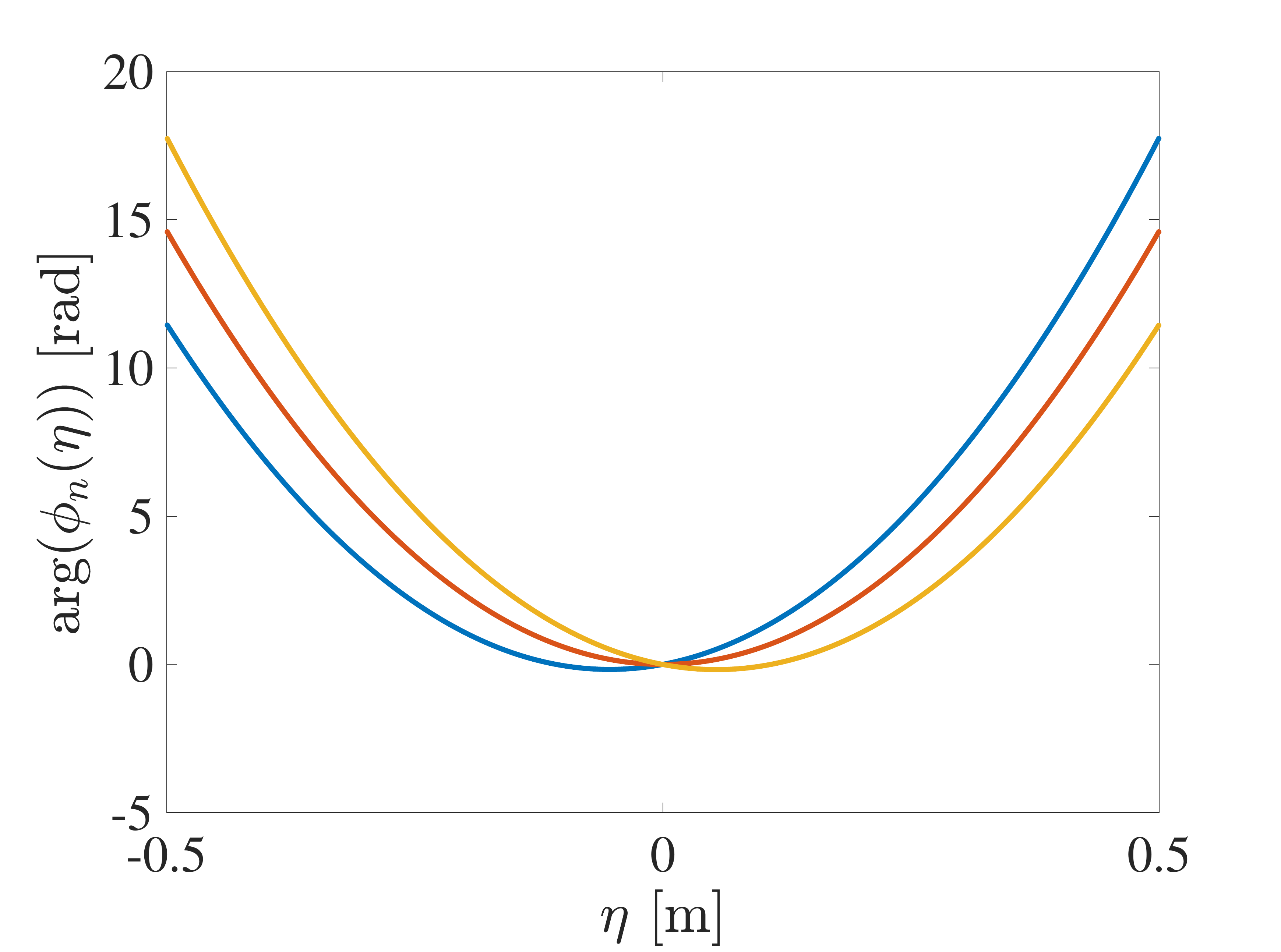}}\\
& (2a) & (2b) & $\quad\quad\quad\quad\quad\quad$(2c) &(2d)\\
\hline\\
\rotatebox{90}{SVD }&
{\includegraphics[width=0.2\textwidth,keepaspectratio=true]{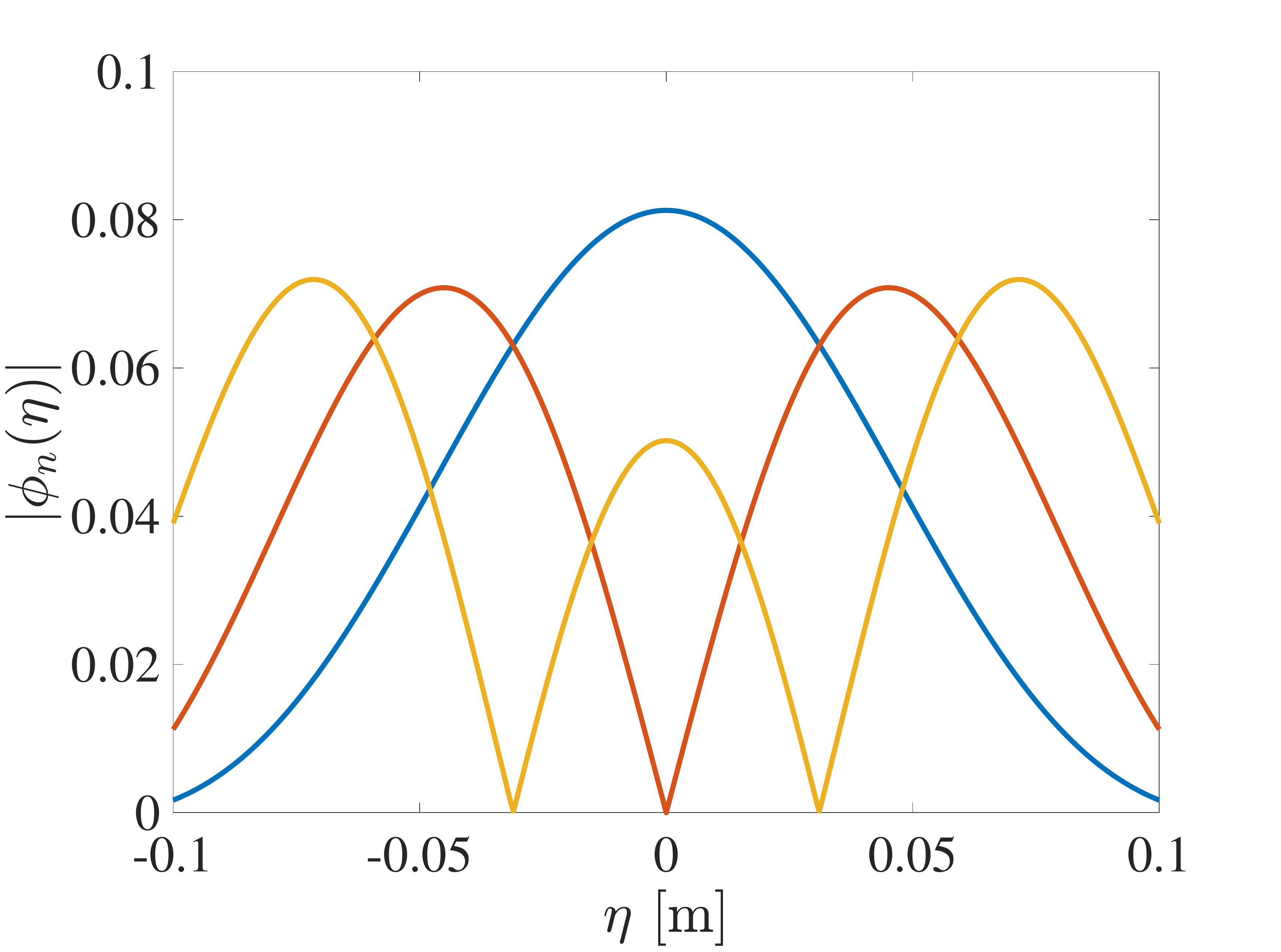}}&
{\includegraphics[width=0.2\textwidth,keepaspectratio=true]{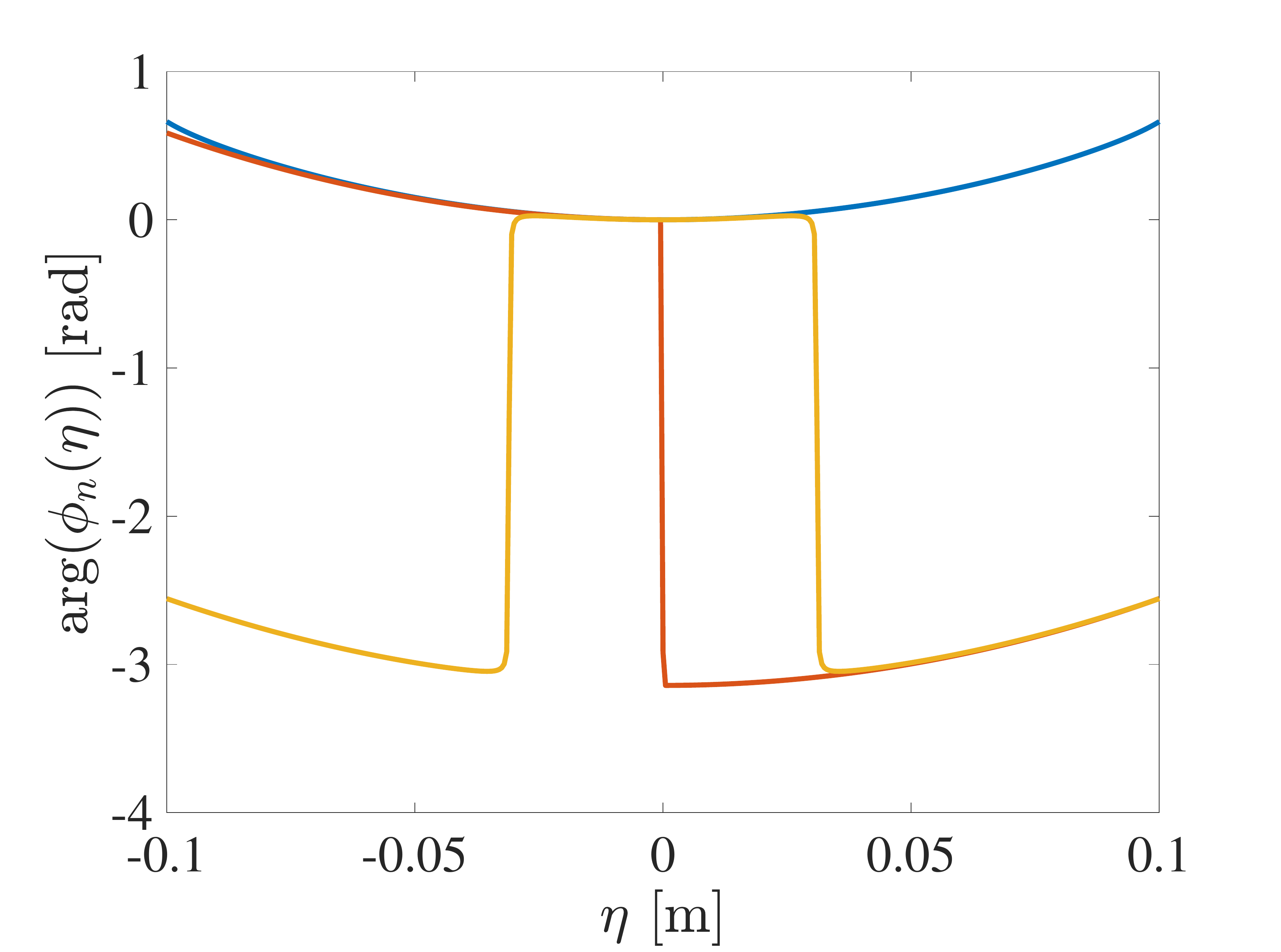}}&
{\includegraphics[width=0.2\textwidth,keepaspectratio=true]{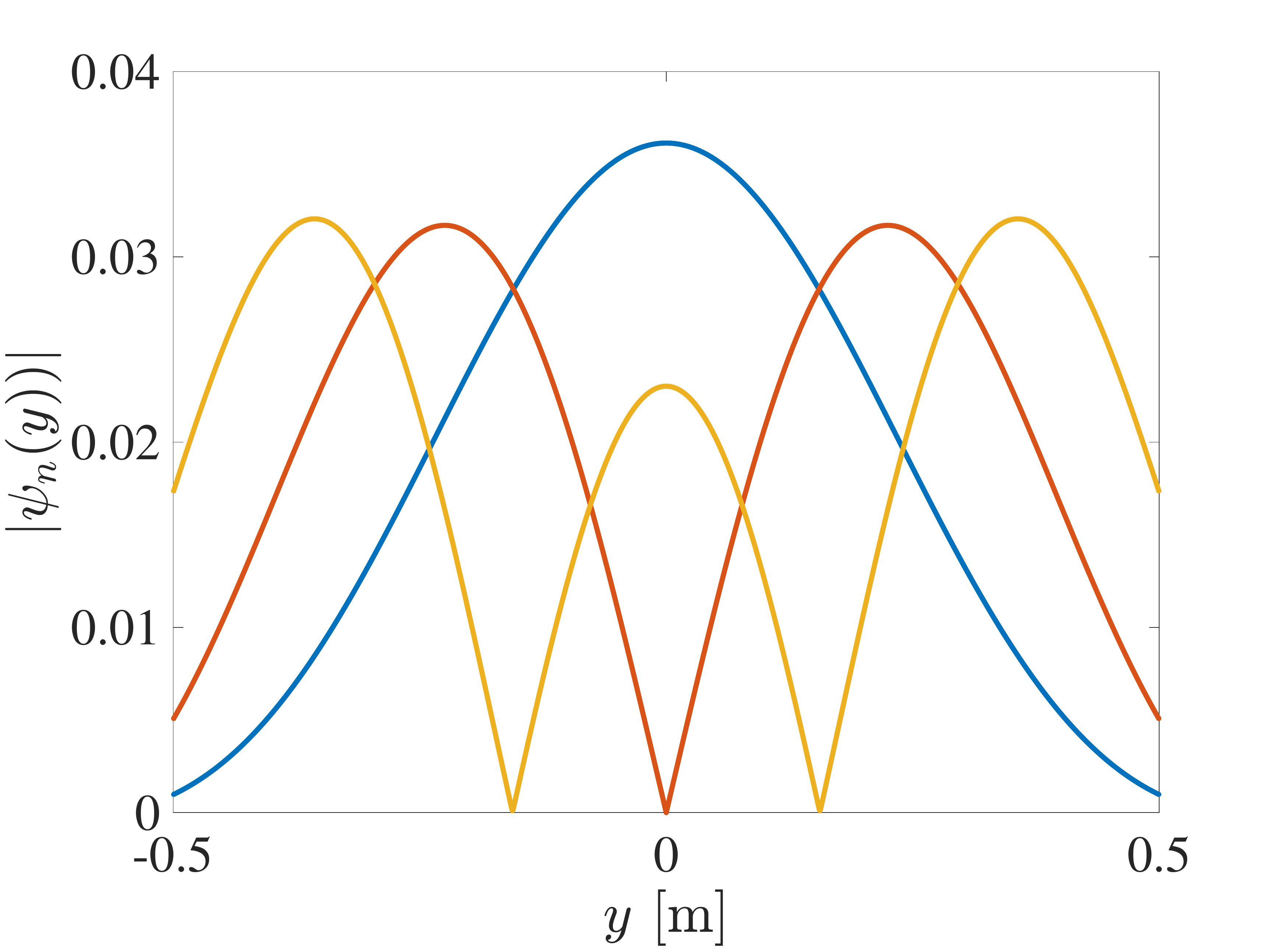}}&
{\includegraphics[width=0.2\textwidth,keepaspectratio=true]{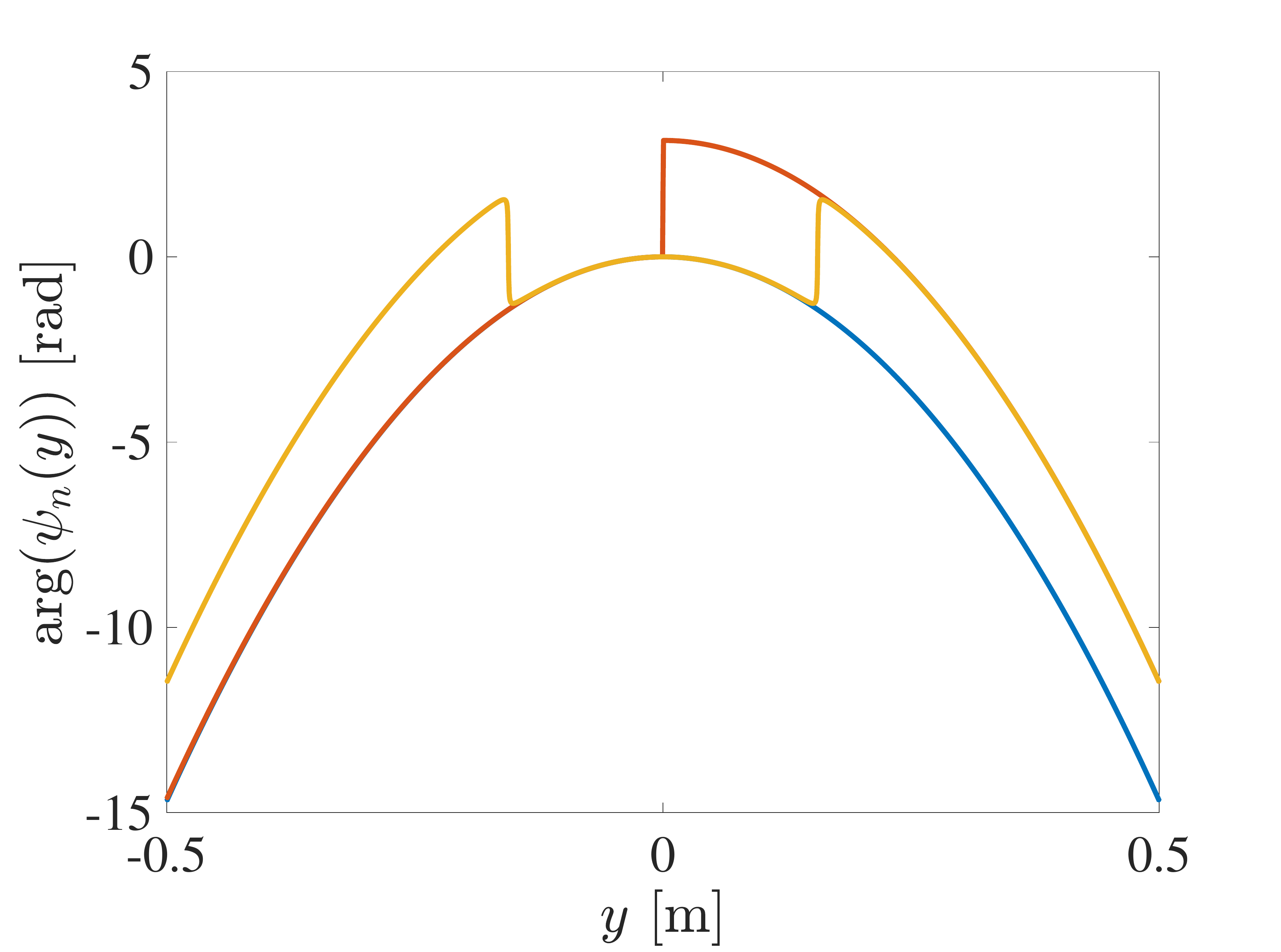}}\\
& (3a) & (3b) & $\quad\quad\quad\quad\quad\quad$(3c) &(3d)\\
\hline
\end{tabular}
\end{table*}

\begin{table*}[h!]
\caption{Example of TX/RX basis functions for uplink and downlink considering a SIS of $20\,$cm, a LIS of $100\,$cm, $z=2\,$m and $\theta=\pi/4$ (non-paraxial case), $f_0=28\,$GHz. Comparison with optimal basis functions obtained with SVD.}
\label{tab:Scenario2}
\vspace{-0.5cm}
\begin{center}
\begin{tabular}{ c | cc l cc }
& \multicolumn{2}{c}{Small Intelligent Surface} &  \multicolumn{2}{c}{Large Intelligent Surface}\\
\hline
\\
\rotatebox{90}{UPLINK }&
(constant amplitude)&
{\includegraphics[width=0.2\textwidth,keepaspectratio=true]{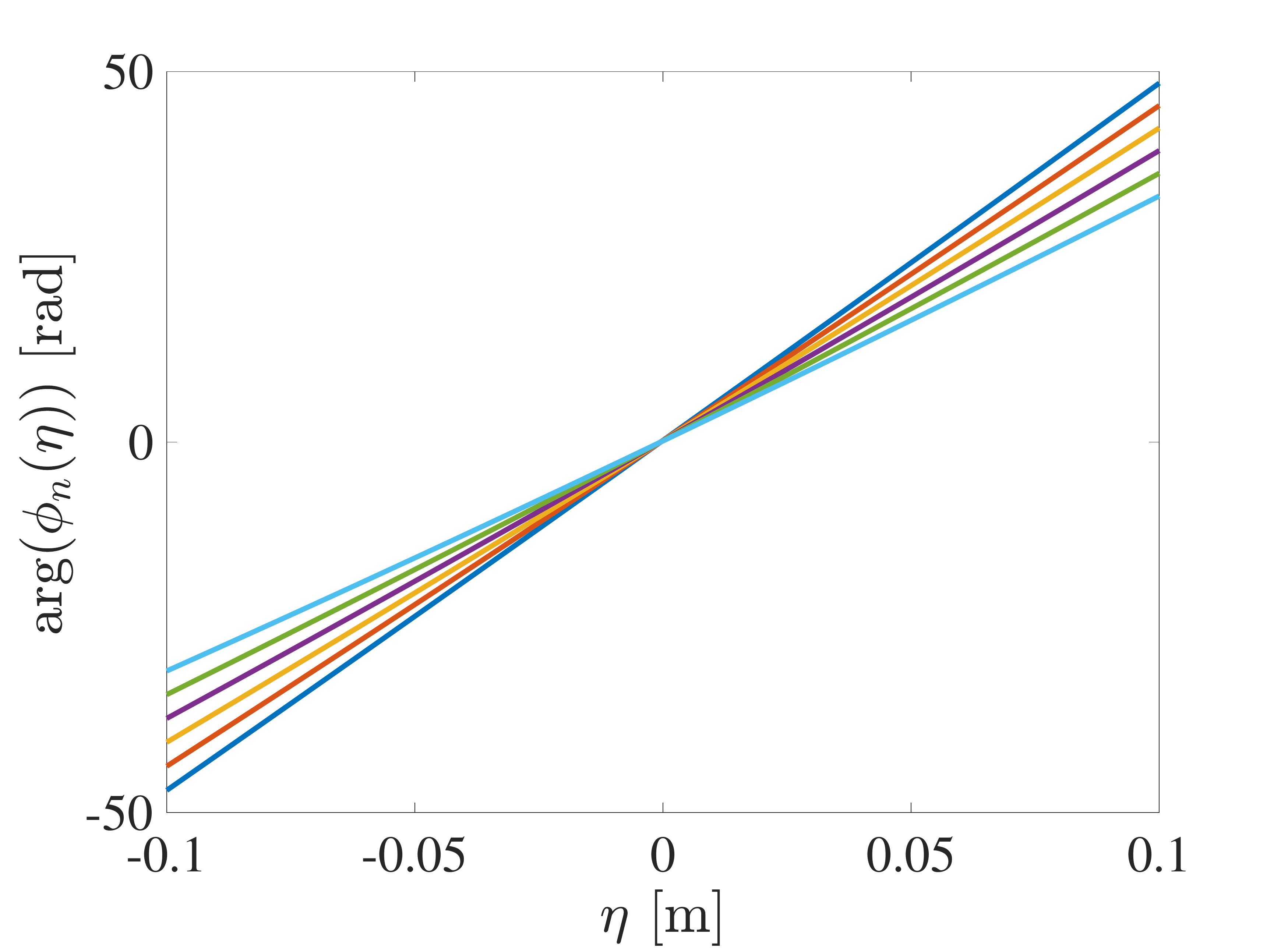}}& 
{\includegraphics[width=0.2\textwidth,keepaspectratio=true]{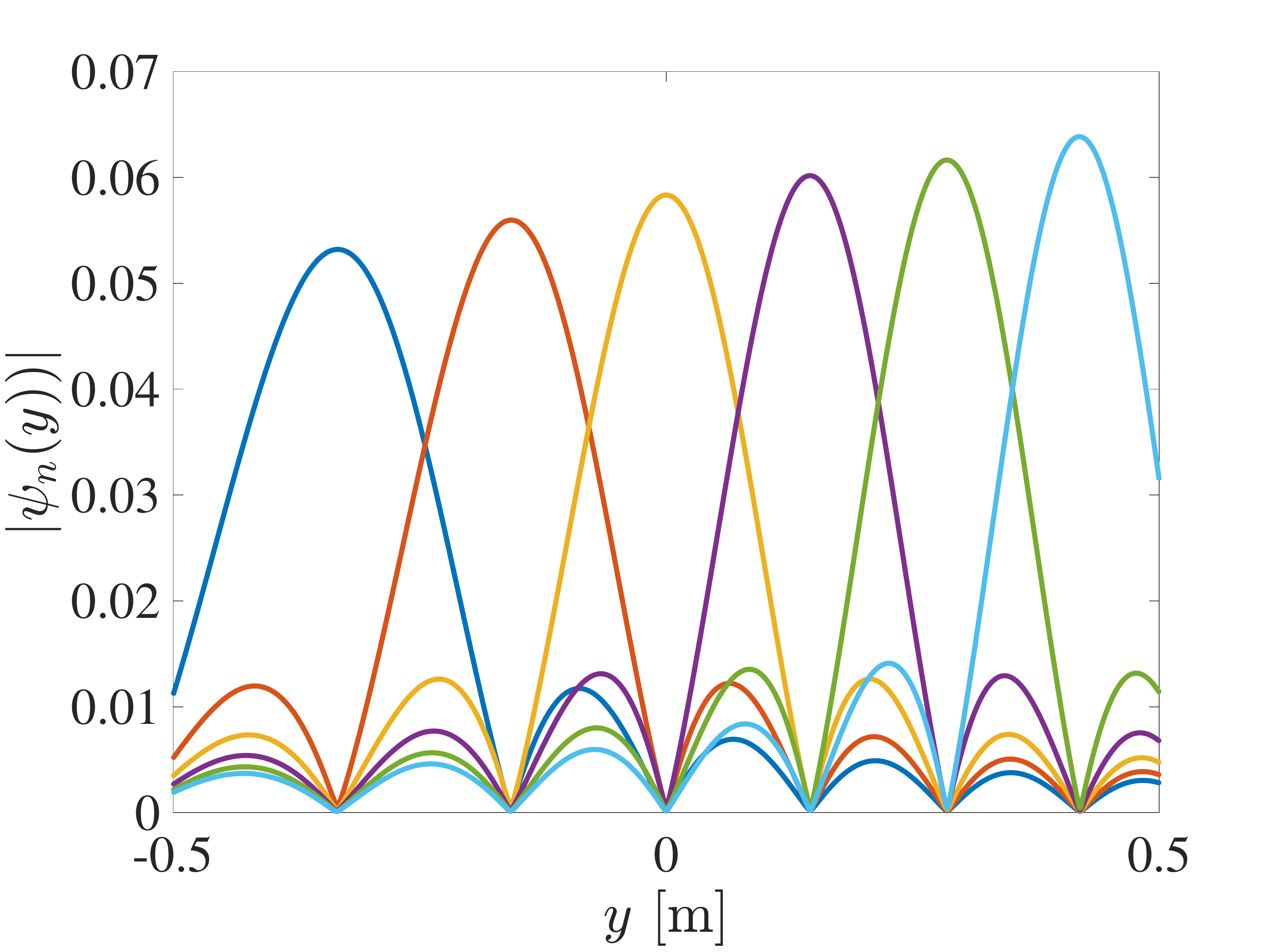}}&
{\includegraphics[width=0.2\textwidth,keepaspectratio=true]{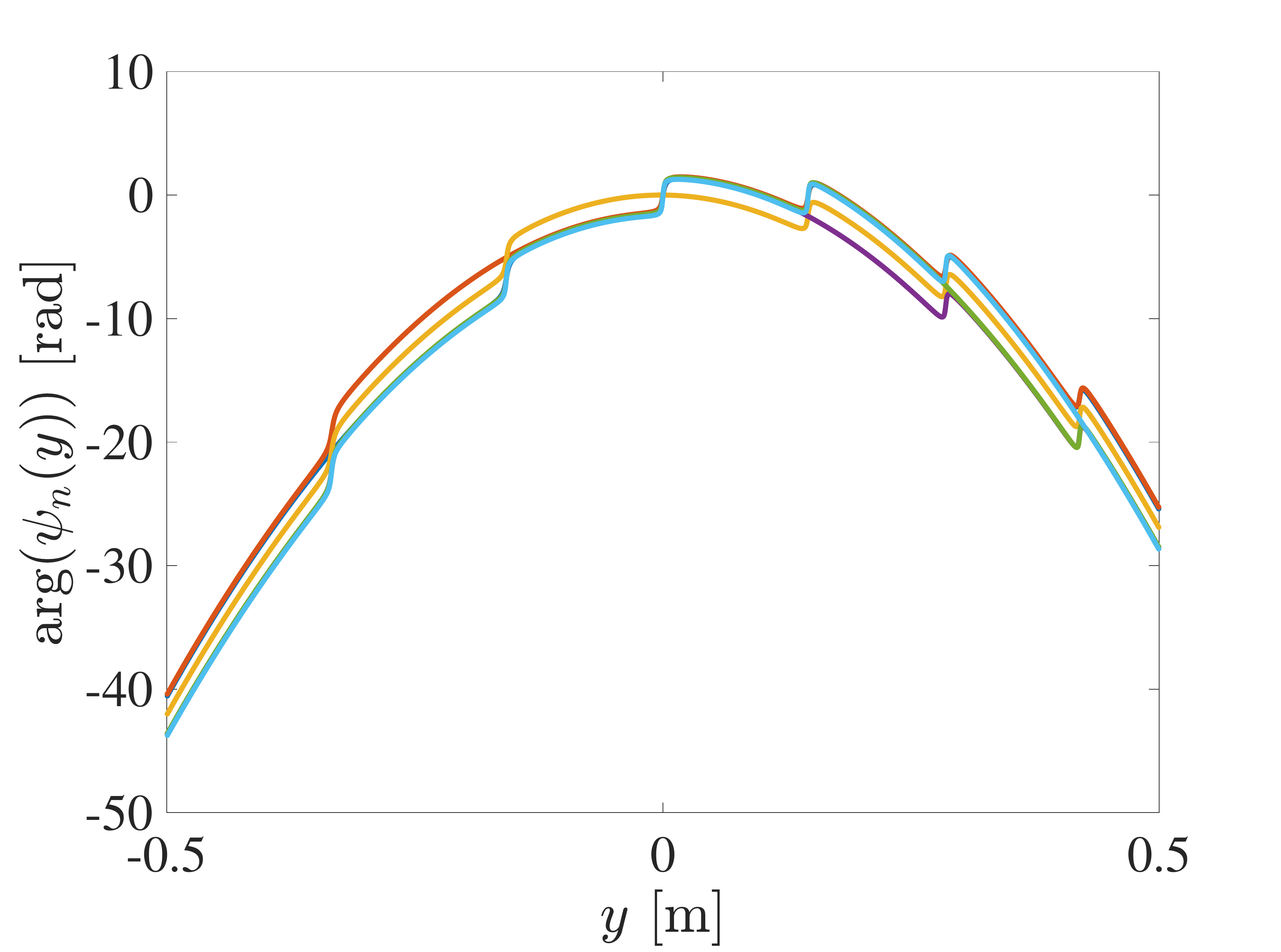}}\\
& (1a) & (1b) & $\quad\quad\quad\quad\quad\quad$(1c) &(1d)\\
\hline
\\
\rotatebox{90}{DOWNLINK }&
{\includegraphics[width=0.2\textwidth,keepaspectratio=true]{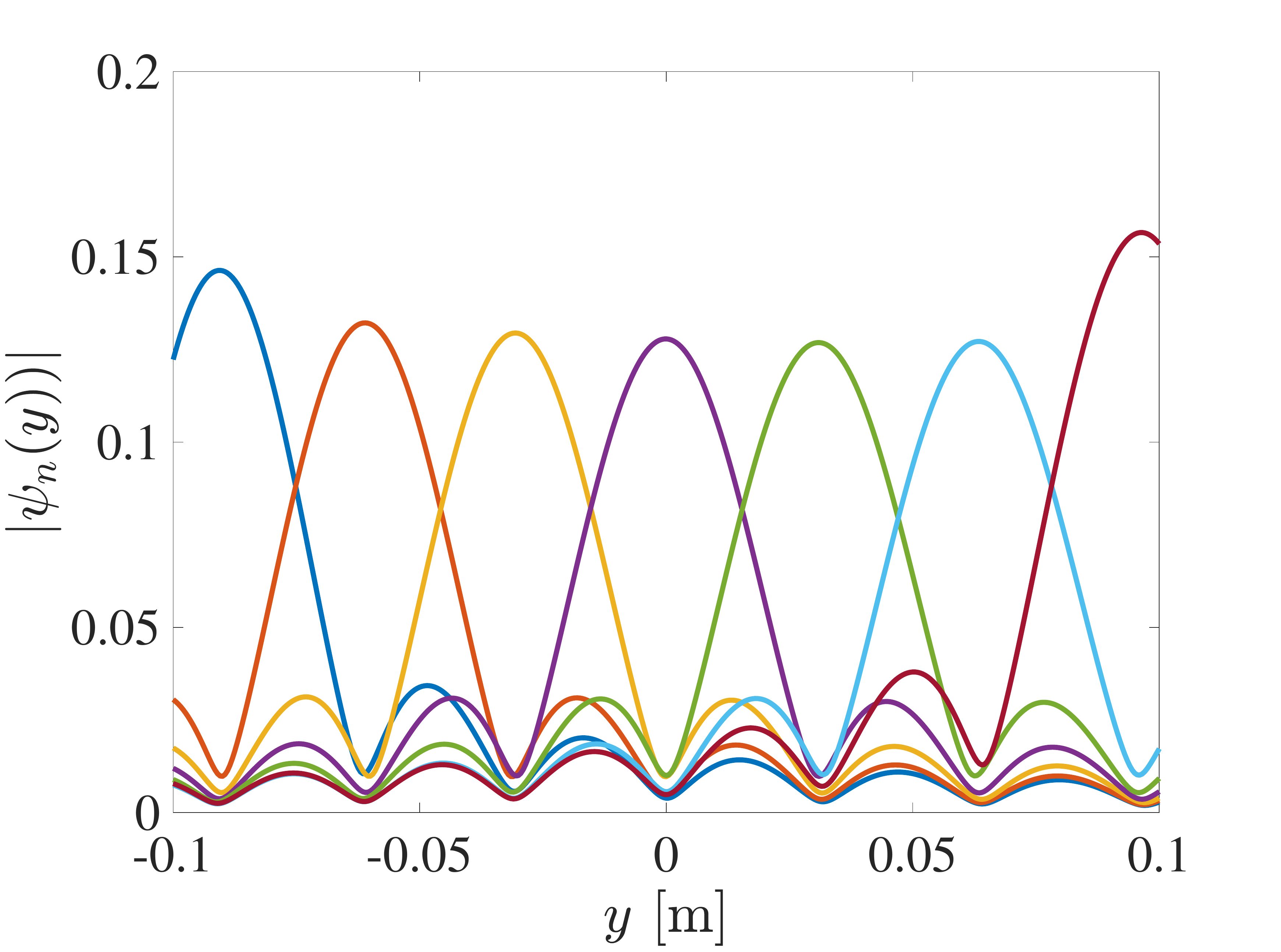}}&
{\includegraphics[width=0.2\textwidth,keepaspectratio=true]{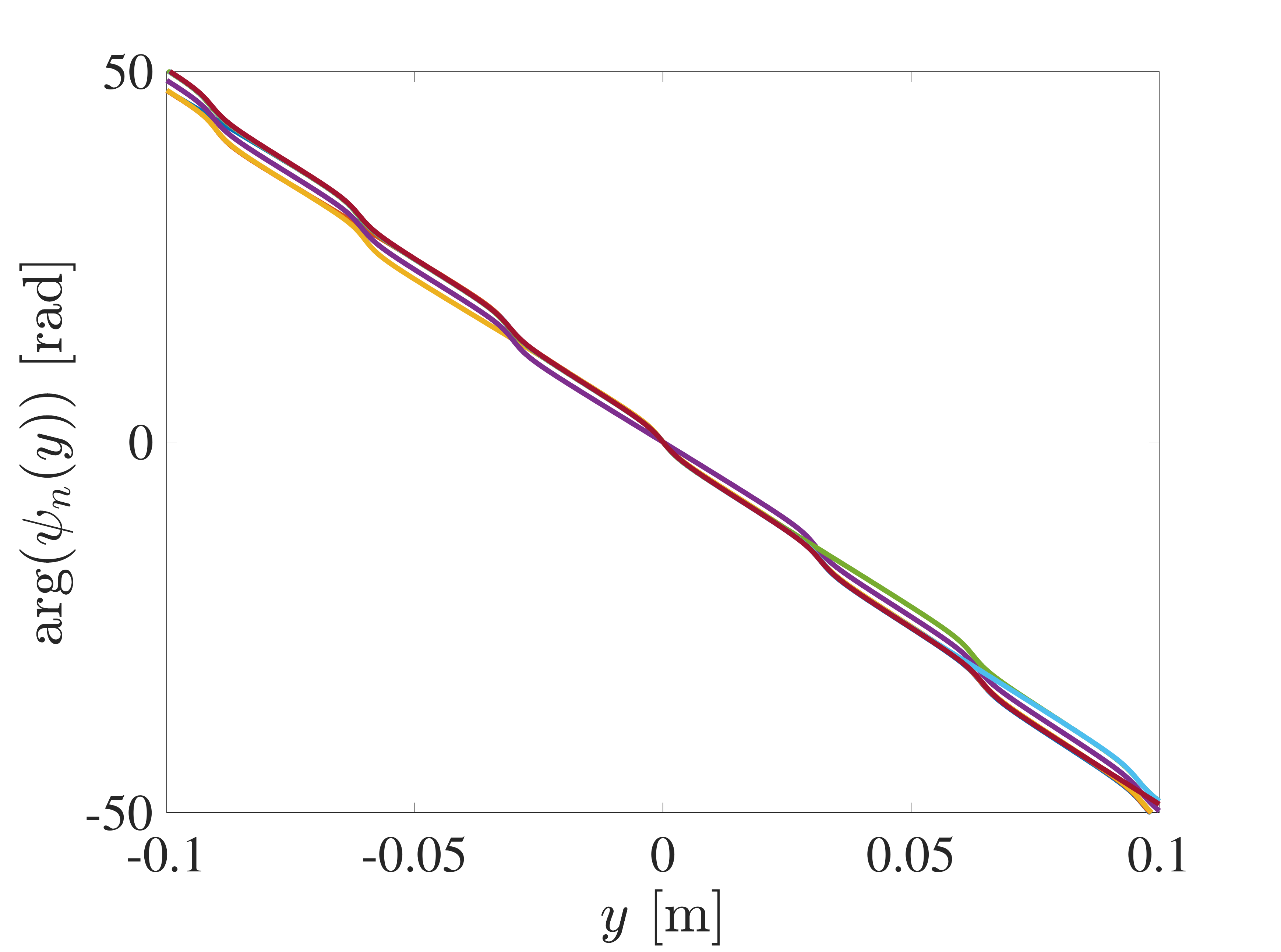}}&
\quad\quad(constant amplitude)&
{\includegraphics[width=0.2\textwidth,keepaspectratio=true]{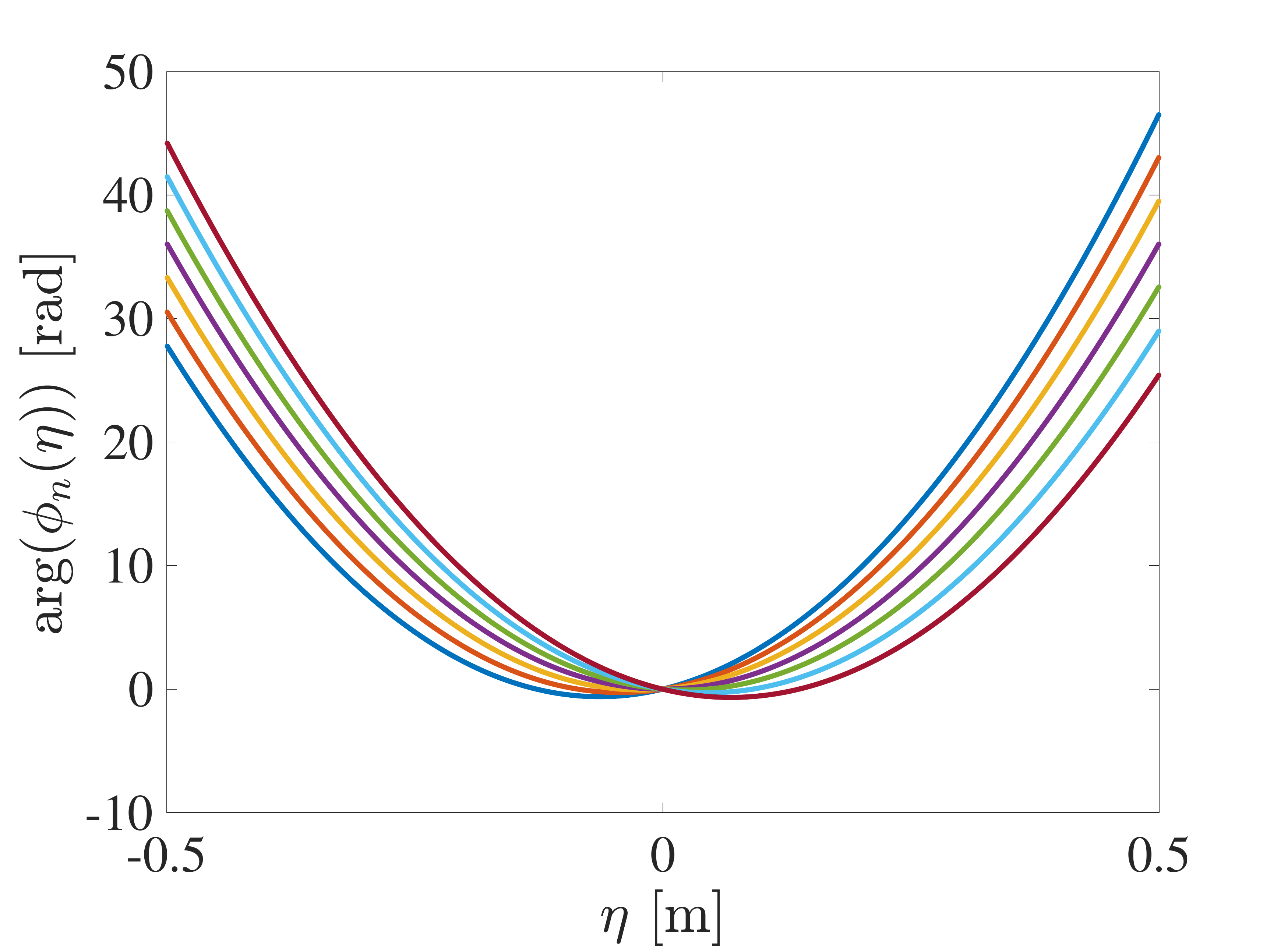}}\\
& (2a) & (2b) & $\quad\quad\quad\quad\quad\quad$(2c) &(2d)\\
\hline
\\
\rotatebox{90}{SVD }&
{\includegraphics[width=0.2\textwidth,keepaspectratio=true]{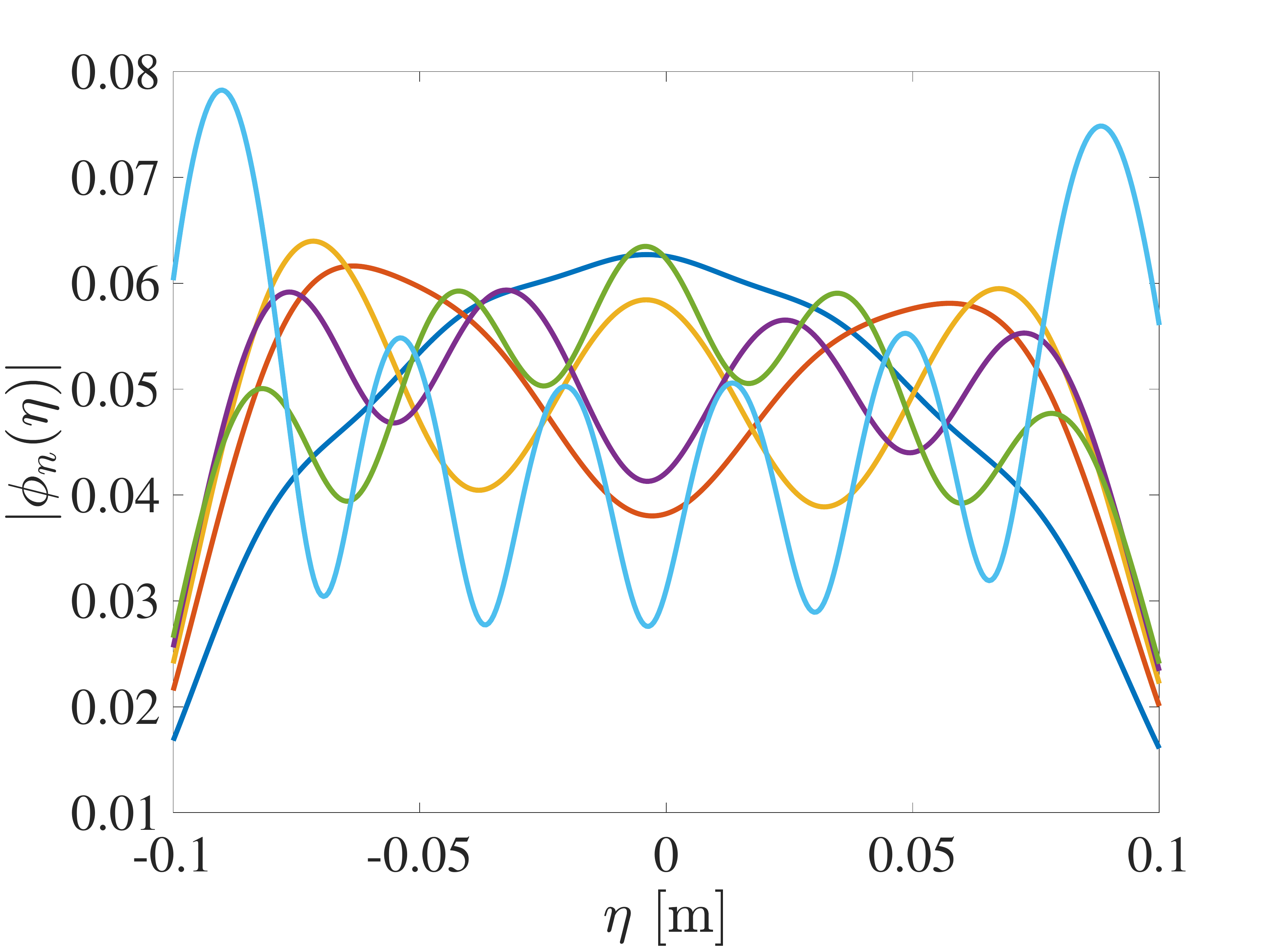}}&
{\includegraphics[width=0.2\textwidth,keepaspectratio=true]{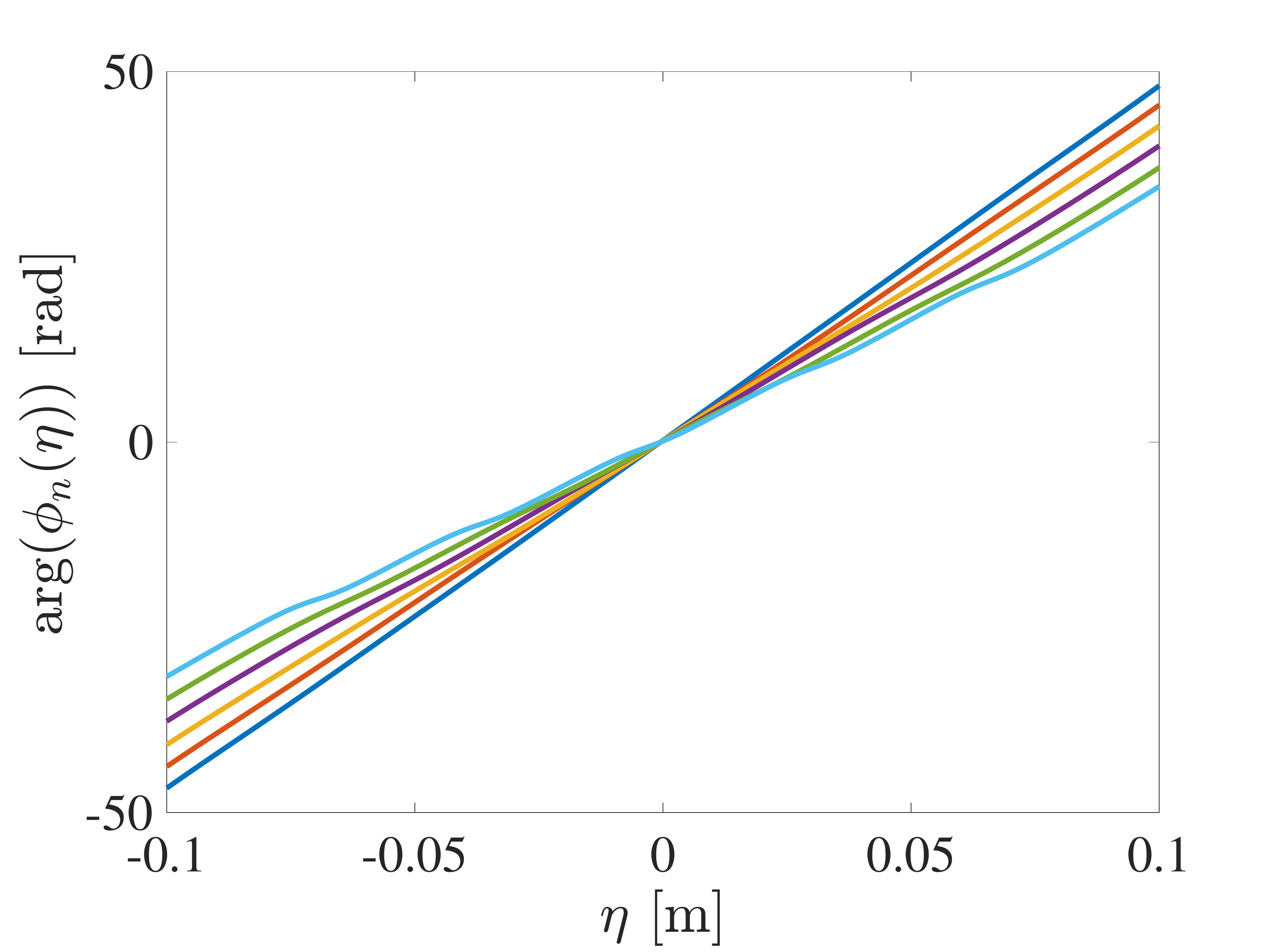}}&
{\includegraphics[width=0.2\textwidth,keepaspectratio=true]{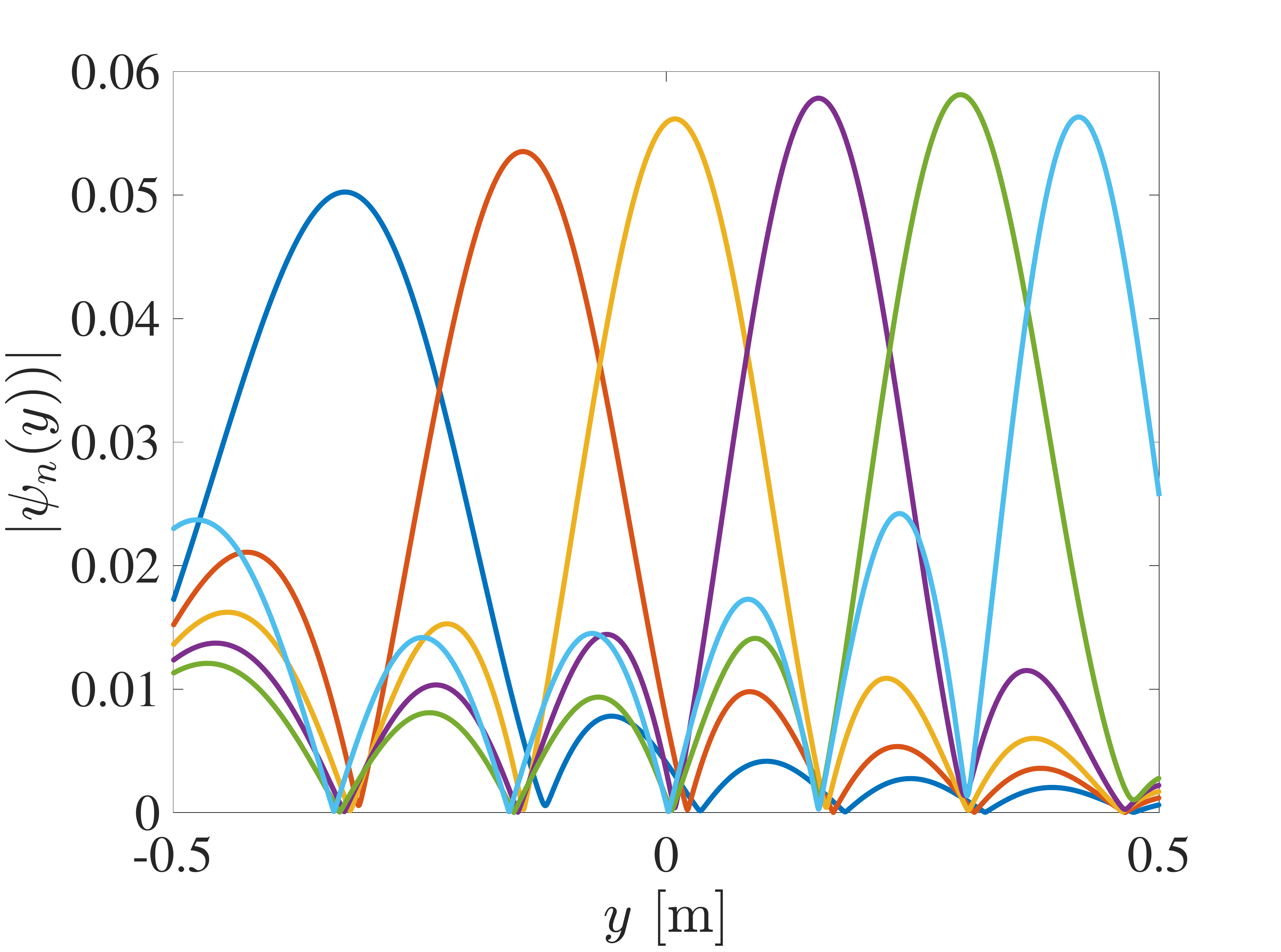}}&
{\includegraphics[width=0.2\textwidth,keepaspectratio=true]{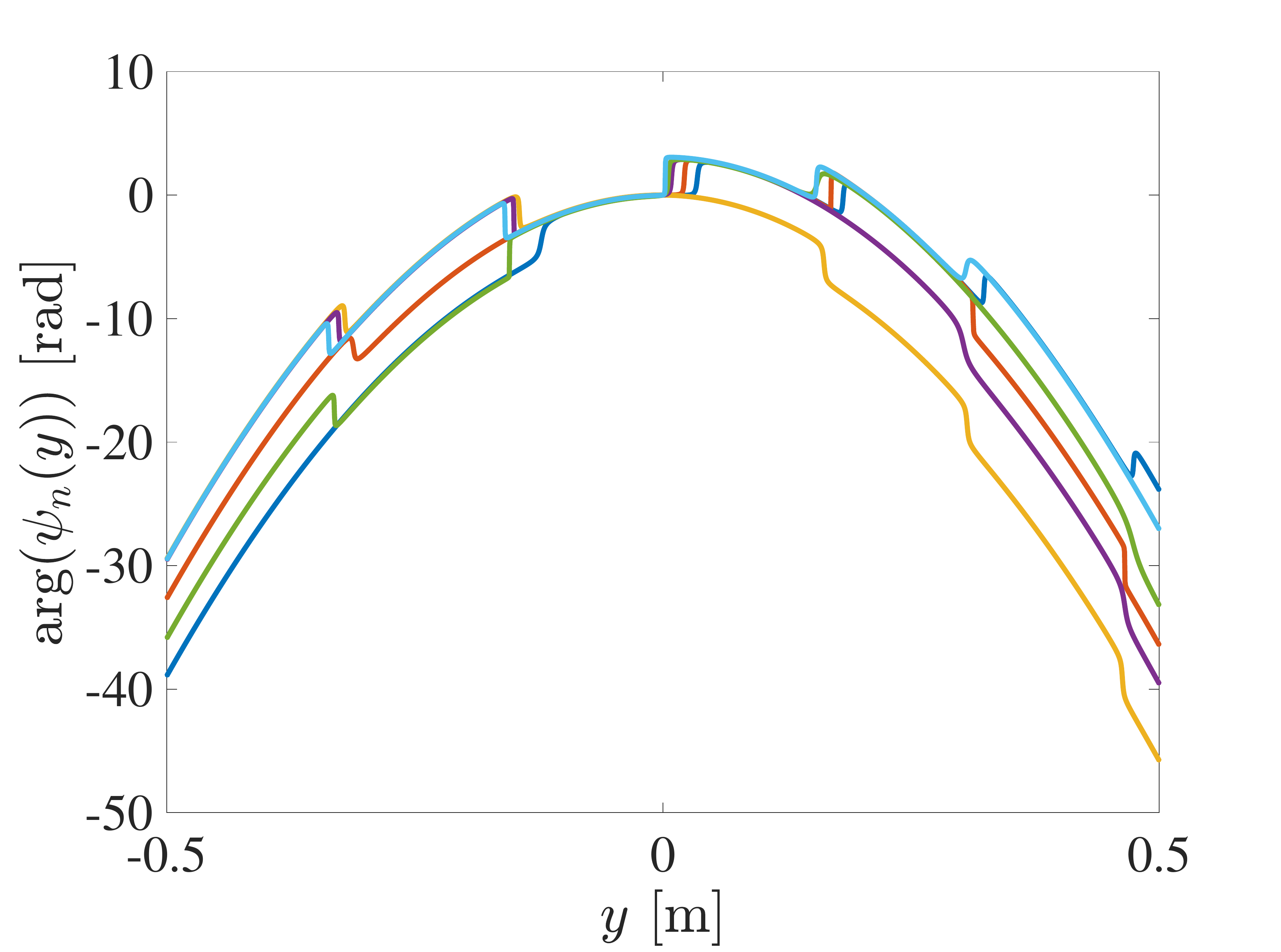}}\\
& (3a) & (3b) & $\quad\quad\quad\quad\quad\quad$(3c) &(3d)\\
\hline
\end{tabular}
\end{center}
\end{table*}

\section{Conclusion}\label{sec:conclusion}

A simple and practical method for the definition of communication modes in the near field has been proposed in this paper, exploiting concepts borrowed by diffraction theory and multi-focusing capability of large antennas. Starting from a beamspace modeling of communication modes, novel expressions for their number have been derived, together with closed-form definitions for the basis sets at the transmitting and receiving antennas for several cases of interest, such as the communication between a LIS and a SIS. It has been shown that almost-optimal basis sets can be designed starting from focusing functions, thus avoiding numerical evaluations and the implementation of complex amplitude/phase antenna profiles.  The expressions derived for the number of communication modes are valid in a generic setup, beyond the traditional paraxial approximation, and account also for the intrinsic limits arising when the link distance becomes comparable to the antenna size (geometric near field). Traditional results valid under paraxial approximation have been revised in light of the proposed method, showing that the beamspace modeling leads to the same number of communication modes. The discussion on the operating zones around a transmitting antenna, depending on the geometry and operating frequency conditions, has shown the relationship among near-far field boundary definition (Fraunhofer distance) and the possibility of obtaining multiple communication modes, thus enabling \ac{MIMO}-like communication even in \ac{LOS} channel conditions without exploiting multi-path propagation.

\appendices

\section{Evaluation of the Phase Profile in the Uplink Scenario Between SIS and LIS}\label{App:SISLIS}
We have
\begin{equation}
r(0) =\sqrt{z^2+y^2} \, 
\end{equation}
and
\begin{align}
 \frac{\partial}{\partial \eta} r(\eta) = \frac{1}{2\sqrt{(z+\eta\sin\theta)^2+(y-\eta\cos\theta)^2}}  \left[\frac{\partial}{\partial \eta}(z+\eta\sin\theta)^2 +\frac{\partial}{\partial \eta}(y-\eta\cos\theta)^2   \right]
\end{align}
which gives 
\begin{align}
 \frac{\partial}{\partial \eta} r(\eta) = \frac{(z+\eta\sin\theta)\sin\theta-(y-\eta\cos\theta)\cos\theta}{\sqrt{(z+\eta\sin\theta)^2+(y-\eta\cos\theta)^2}} \, .
\end{align}
By posing $\eta=0$ and discarding all the phase terms independent of $\eta$, since the addition of a constant phase shift would not change the focusing behavior, \eqref{eq:focusingSIS} is obtained.

\section{Number of Communication Modes for Parallel Antennas}\label{App:NParr}
We have
\begin{equation}
\frac{\lambda}{\lt}n<\sin\arctan{\frac{\lr}{2z}}\, .
\end{equation}
By exploiting $\sin\arctan(x)=x/\sqrt{1+x^2}$, it is obtained
\begin{equation}
n^+=\frac{\lt\lr}{2\lambda z \sqrt{1+\frac{\lr^2}{4z^2}}}\, 
\end{equation}
then \eqref{eq:NParr}.

\section{Number of Communication Modes for Perpendicular Antennas}\label{App:NPerp}

We have
\begin{equation}
\frac{\lambda}{\lt}n<1-\sin\arccot{\frac{\lr}{2z}}\, .
\end{equation}
By exploiting $\sin\arccot(x)=1/\sqrt{1+x^2}$, it is obtained
\begin{equation}
n^+=\frac{\lt}{\lambda}\left(1- \frac{1}{\sqrt{1+\frac{\lr^2}{4z^2}}}\right)\, 
\end{equation}
then \eqref{eq:NPerp}.

\section{Number of Communication Modes for the General Case}\label{App:NGen}
When considering a generic orientation $\theta$ and $\yc=0$, the number of beams  $n^+$ intercepted by the upper part of the receiving antenna will be different by  the number of beams  $n^-$ intercepted by the lower half of the antenna.
Without loss of generality, let us consider $0<\theta<\pi/2$.
\subsection{Positive Semi-Axis}
In this case we have $0<y_n<\lr/2$, with $n<0$. 
By inverting the relation
\begin{align}
0<z \tan{\left[\arcsin{\left(-\frac{\lambda}{\lt}n-\sin\theta\right)}+\theta\right]}<\frac{\lr}{2}
\end{align}
for $n=-1, -2,\ldots -n^+$, it can be shown that
\begin{equation}\label{eq:kplusGeneric}
n^+=\frac{\lt}{\lambda }\left[ \sin\left(\arctan{\frac{\lr}{2z}-\theta}\right)+\sin\theta\right]\, .
\end{equation}
Result \eqref{eq:kplusGeneric} gives a number of beams increasing with $\theta$ up to $\theta=\frac{1}{2}\arctan{\frac{\lr}{2z}}$, then decreasing for larger values of $\theta$. This is reasonable, since the number of communication modes related to beams in the positive semi-axis should first increase when the boresight direction of the transmitting antenna (for which the beams realized have the smallest width, so their number is larger than for other directions) is oriented towards the positive semi-axis, then decrease for larger rotation values.

\subsection{Negative Semi-Axis}
In this case we have $-\lr/2<y_n<0$, with $n>0$. Thus, by inverting the relation
\begin{align}\label{eq:tangent}
-\frac{\lr}{2}<z \tan{\left[\arcsin{\left(-\frac{\lambda}{\lt}n-\sin\theta\right)}+\theta\right]}<0
\end{align}
for $n=1, 2,\ldots n^-$, it can be shown that
\begin{equation}\label{eq:kminusGeneric}
n^-=\frac{\lt}{\lambda }\left[ \sin\left(\arctan{\frac{\lr}{2z}+\theta}\right)-\sin\theta\right]\, .
\end{equation}
In order to have $n^->0$ from \eqref{eq:kminusGeneric}, it must be satisfied
\begin{equation}
\sin\left[\arctan{\frac{\lr}{2z}-\theta}\right]+\sin\theta > 0
\end{equation}
that results in
\begin{equation}\label{ed:conditionNminus}
\theta<\frac{\pi}{2}-\frac{1}{2}\arctan{\frac{\lr}{2z}}\, .
\end{equation}

\subsection{Number of Communication Modes}

For a generic $\theta$ with $0\leq\theta\leq{\pi}/{2}$ we have
\begin{equation}
N=1+n^+ + n^- 
\end{equation}
where $n^+$ and $n^-$ are given by  \eqref{eq:kplusGeneric} and \eqref{eq:kminusGeneric}, respectively.

\bibliographystyle{IEEEtran}


\end{document}